\documentclass[pdftex,twocolumn,epjc3]{svjour3}
\pdfoutput=1

\RequirePackage[T1]{fontenc}
\RequirePackage{graphicx}
\RequirePackage{microtype}
\RequirePackage{tikz}

\RequirePackage{placeins}
\RequirePackage{capt-of}
\RequirePackage{xspace}
\RequirePackage{xcolor}
\RequirePackage[amssymb]{SIunits}
\RequirePackage{amssymb}
\RequirePackage{amsmath}
\RequirePackage{setspace,relsize}
\RequirePackage{threeparttable}
\RequirePackage{booktabs}
\RequirePackage{multirow}
\RequirePackage[small,nooneline,center]{subfigure}
\RequirePackage{mathptmx} % use Times fonts if available on your TeX system
\RequirePackage{latexsym}
\RequirePackage{flushend}
\RequirePackage[numbers,sort&compress]{natbib}
\RequirePackage{afterpage}

\usepackage{preprintcover}
\usepackage[hyperindex,breaklinks]{hyperref}

\newcommand{\invfb}{\ensuremath{\text{fb}^{-1}}\xspace}
\newcommand{\mm}{\ensuremath{\text{mm}}\xspace}
\newcommand{\eV}{\ensuremath{\text{e\kern-0.15ex{}V}}\xspace}

\newcommand{\GeV}{\ensuremath{\text{G\eV}}\xspace}
\newcommand{\TeV}{\ensuremath{\text{T\eV}}\xspace}
\providecommand{\unit}[2]{\ensuremath{#1\;#2}\xspace}
\newcommand{\alpgen}{\textsc{Alpgen}\xspace}

\newcommand{\pythiasix}{\textsc{Pythia}\,6\xspace}
\newcommand{\pythiaeight}{\textsc{Pythia}\,8\xspace}
\newcommand{\jimmy}{\textsc{Jimmy}\xspace}
\newcommand{\herwig}{\textsc{Herwig}\xspace}
\newcommand{\herwigjimmy}{\textsc{Herwig}+\allowbreak\textsc{Jimmy}\xspace}

\newcommand{\herwigpp}{\mbox{\textsc{Herwig}\plusplus}\xspace}
\newcommand{\sherpa}{Sherpa\xspace}
\newcommand{\powheg}{\textsc{Powheg}\xspace}

\def\loss{\ensuremath{\text{LO\raisebox{.3ex}{$\mspace{0.5mu}*\mspace{-1mu}*$}}}\xspace}

\newcommand{\plusplus}{\raisebox{0.2ex}{\smaller++}}

\def\pt{\ensuremath{p_\mathrm{T}}\xspace}
\def\pT{\pt}

\def\ptsum{\ensuremath{\sum\pt}\xspace}

\def\etamod{\ensuremath{|\eta|}\xspace}

\newcommand{\ptmean}{\ensuremath{\langle\pt\rangle}\xspace}
\def\ptsum{\ensuremath{\sum\pt}\xspace}
\newcommand{\Nchg}{\ensuremath{N_\text{ch}}\xspace}
\newcommand{\ZpT}{\ensuremath{p_\mathrm{{T}}^\mathrm{{Z}}}\xspace}
\newcommand{\dNchgdetadphi}{\ensuremath {\langle N_\text{ch}/\delta\eta\,\delta\phi\rangle} \xspace}
\newcommand{\dpTsumdetadphi}{\ensuremath{\langle \sum\!\pt/\delta\eta\,\delta\phi \rangle}\xspace}
\newcommand{\Nchgdetadphi}{\ensuremath {N_\text{ch}/\delta\eta\,\delta\phi} \xspace}
\newcommand{\pTsumdetadphi}{\ensuremath{\sum\!\pt/\delta\eta\,\delta\phi}\xspace}
\def\Mll{\ensuremath{m_{\mathrm{ll}}}}
\def\DMll{\ensuremath{\Delta m_{\mathrm{ll}}}}
\newcommand{\Zee}{Z \rightarrow e^+ e^-}
\newcommand{\Zmm}{Z \rightarrow \mu^+ \mu^-}
\newcommand{\Z}{$Z$-boson\xspace}

\sfcode`\.=1001
\sfcode`\?=1001
\sfcode`\!=1001
\sfcode`\:=1001

\usepackage{ifthen}
\newcommand{\EqRef}[1]{\ifthenelse{\boolean{hmode}\and\spacefactor=1001}{Equation}{eq.}~\ref{#1}\xspace}
\newcommand{\FigRef}[1]{\ifthenelse{\boolean{hmode}\and\spacefactor=1001}{Figure}{Fig.}~\ref{#1}\xspace}
\newcommand{\FigsRef}[1]{\ifthenelse{\boolean{hmode}\and\spacefactor=1001}{Figures}{Figs.}~\ref{#1}\xspace}
\newcommand{\TabRef}[1]{\ifthenelse{\boolean{hmode}\and\spacefactor=1001}{Table}{Tab.}~\ref{#1}\xspace}
\newcommand{\TabsRef}[1]{\ifthenelse{\boolean{hmode}\and\spacefactor=1001}{Tables}{Tabs.}~\ref{#1}\xspace}
\newcommand{\SecRef}[1]{\ifthenelse{\boolean{hmode}\and\spacefactor=1001}{Section}{Sect.}~\ref{#1}\xspace}
\newcommand{\SecsRef}[1]{\ifthenelse{\boolean{hmode}\and\spacefactor=1001}{Sections}{Sects.}~\ref{#1}\xspace}
\newcommand{\AppRef}[1]{\ifthenelse{\boolean{hmode}\and\spacefactor=1001}{Appendix}{App.}~\ref{#1}\xspace}
\newcommand{\AppsRef}[1]{\ifthenelse{\boolean{hmode}\and\spacefactor=1001}{Appendices}{Apps.}~\ref{#1}\xspace}

\newcommand{\resultsimg}[3][0.8]{\subfigure[]{\includegraphics[width=#1\textwidth]{#2}\label{#3}}}

\PreprintCoverPaperTitle{Measurement of distributions sensitive to the underlying event in inclusive Z-boson production in $pp$
collisions at $\sqrt{s}=7$ TeV with the ATLAS detector} 

\PreprintIdNumber{CERN-PH-EP-2014-162}  

\PreprintCoverAbstract{
A measurement of charged-particle distributions
  sensitive to the properties of the underlying event is presented for an inclusive sample of events containing 
  a \Z , decaying to an electron or muon pair.
  The measurement is based on data 
  collected using the ATLAS detector at the LHC in proton--proton collisions at a centre-of-mass energy 
  of $7$ TeV with an integrated luminosity of $4.6$ fb$^{-1}$.
  Distributions of the charged particle multiplicity
  and of the charged particle transverse momentum
  are measured
  in regions of azimuthal angle defined with respect to
  the \Z direction.     
  The measured distributions are compared to similar distributions measured in jet events,
  and to the predictions of various Monte Carlo generators implementing
  different underlying event models.
}

\PreprintJournalName{EPJC}

\begin{document}

\title{Measurement of distributions sensitive to the underlying event in inclusive Z-boson production in $pp$
collisions at $\sqrt{s}=7$ TeV with the ATLAS detector}
\titlerunning{ATLAS underlying event in 7\;\TeV Z boson events}
\author{The ATLAS Collaboration}

\institute{CERN, CH-1211 Geneva 23. Switzerland}

\date{Received: 12 September 2014 / Accepted: 23 November 2014}

\maketitle

\begin{abstract}
  A measurement of charged-particle distributions
  sensitive to the properties of the underlying event is presented for an inclusive sample of events containing 
  a \Z , decaying to an electron or muon pair.
  The measurement is based on data 
  collected using the ATLAS detector at the LHC in proton--proton collisions at a centre-of-mass energy 
  of $7$ TeV with an integrated luminosity of $4.6$ fb$^{-1}$.
  Distributions of the charged particle multiplicity
  and of the charged particle transverse momentum
  are measured
   in regions of azimuthal angle defined with respect to
   the \Z direction.     
  The measured distributions are compared to similar distributions measured in jet events,
  and to the predictions of various Monte Carlo generators implementing
  different underlying event models.  
\end{abstract}

%%%%%%%%%%%%%%%%%%%%%%%%%%%%%%%%%%%%
%        Introduction
%%%%%%%%%%%%%%%%%%%%%%%%%%%%%%%%%%%%
\section{Introduction}
\label{sec:intro}

In order to perform precise Standard Model measurements or to search for new physics
phenomena at hadron colliders, it is important to have a good understanding of not
only the short-distance \textit{hard} scattering process, but also of the
accompanying 
activity
-- collectively termed the
\textit{underlying event} (UE). 
This includes partons not participating in the hard-scattering 
process (beam remnants), and 
additional hard scatters in the same proton-proton collision, 
termed multiple parton interactions (MPI).
Initial and final state gluon radiation (ISR, FSR) also 
contribute to the UE activity.
It is impossible to unambiguously separate the UE from the
hard scattering process on an event-by-event basis.  However, distributions can be
measured that are sensitive to the properties of the UE.

The soft interactions contributing to the UE cannot be calculated reliably using
perturbative quantum chromodynamics (pQCD) methods, and
are generally described using different phenomenological models, usually implemented
in Monte Carlo (MC) event generators. These models contain many parameters whose
values and energy dependences are not known \textit{a priori}. Therefore, 
the model parameters must be tuned to experimental data
to obtain insight into the nature of
soft QCD processes and to optimise the description of UE contributions for
studies of hard-process physics.

Measurements of distributions sensitive to the properties of the UE have been performed 
in proton-proton ($pp$) collisions at $\sqrt{s}=\unit{900}{\GeV}$ and \unit{7}{\TeV} in 
ATLAS~\cite{Aad:2010fh, Aad:2011qe, :2012fs, Aad:2013bjm, Aad:2014hia}, 
ALICE~\cite{ALICE:2011ac} and 
CMS~\cite{Khachatryan:2010pv,Collaboration:2012tb}. 
They have also been performed  
in $p\bar{p}$ collisions in events with jets and in Drell-Yan events at 
CDF ~\cite{PhysRevD.70.072002, PhysRevD.82.034001} at centre-of-mass energies of
$\sqrt{s}=\unit{1.8}{\TeV}$ and \unit{1.96}{\TeV}.

This paper reports a measurement of distributions sensitive to the UE, performed with the ATLAS
detector~\cite{:2008zzm} at the LHC in $pp$ collisions at
a centre-of-mass energy of $7$ TeV.
The full dataset acquired during 2011 
is used, corresponding to an integrated luminosity of 
 \unit{4.64 \pm 0.08}{\femto\reciprocal\barn}.
Events with a \Z candidate decaying into an electron or muon pair were selected,
and observables constructed from 
the final state charged particles
(after excluding the lepton pair) 
were studied 
as a function of the transverse momentum~\footnote
  {The ATLAS
  reference system is a Cartesian right-handed coordinate system, with the
  nominal collision point at the origin. The anti-clockwise beam direction
  defines the positive $z$-axis, while the positive $x$-axis is defined as
  pointing from the collision point to the center of the LHC ring and the
  positive $y$-axis points upwards. The azimuthal angle $\phi$ is measured
  around the beam axis, and the polar angle $\theta$ is measured with
  respect to~the $z$-axis. The pseudorapidity is given by $\eta = -\ln
  \tan\mspace{-0.1mu}( \theta/2 )$. Transverse momentum is defined relative to the beam axis.}
of the \Z candidate, \ZpT .

This paper is organised as follows: 
the definitions of the underlying event observables are given in \SecRef{sec:obs}.
The ATLAS detector is described in \SecRef{sec:atlasdet}.
In \SecRef{sec:mc}, the MC models used in this analysis are discussed.
\SecRef{sec:sel} and \SecRef{sec:pu} ~describe the event selection, and the correction for 
the effect of multiple proton-proton interactions in the same bunch crossing (termed pile-up).
The correction of the data to the particle level, and the combination of the electron and muon channel results
are described in \SecRef{sec:corr}. % and \SecRef{sec:comb}.
\SecRef{sec:syst} contains the estimation of the systematic uncertainties.
The results are discussed in \SecRef{sec:res} and finally the conclusions are presented in \SecRef{sec:concl}.

%%%%%%%%%%%%%%%%%%%%%%%%%%%%%%%%%%%%
%         Observables
%%%%%%%%%%%%%%%%%%%%%%%%%%%%%%%%%%%%
\section{Underlying event observables}
\label{sec:obs}

\begin{figure}[tbp]
  \begin{center}
    \begin{tikzpicture}[>=stealth, very thick, scale=1.1]
      %% Make figure text smaller than for the body
      %\smaller
      \small

      %% Circle and region dividers
      \draw[color=blue!80!black] (0, 0) circle (3.0);
      \draw[rotate= 30, color=gray] (-3.0, 0) -- (3.0, 0);
      \draw[rotate=-30, color=gray] (-3.0, 0) -- (3.0, 0);

      %% Delta{phi} definition arcs
      \draw[->, color=black, rotate=-2] (0, 3.5) arc (90:47:3.5) node[right] {$\Delta{\phi}$};
      \draw[->, color=black, rotate=2] (0, 3.5) arc (90:133:3.5) node[left] {$-\Delta{\phi}$};

      %% Tracks
      \draw[->, color=red, ultra thick] (0, 2) -- (0, 4) node[above] {\textcolor{black}{\Z}};
      \draw[->, color=green!70!black, ultra thick] (0, -2) -- ( 0.0, -4);
      \draw[->, color=green!70!black, ultra thick, rotate around={-15:(0,-0.3)}] (0, -2) -- ( 0.0, -4);
      \draw[->, color=green!70!black, ultra thick, rotate around={ 15:(0,-0.3)}] (0, -2) -- ( 0.0, -4);

      %% Region definition labels
      \draw (0,  1.4) node[text width=2cm] {\begin{center} Toward \goodbreak $|\Delta\phi| < 60^\circ$ \end{center}};
      \draw (0, -1.1) node[text width=2cm] {\begin{center} Away \goodbreak $|\Delta\phi| > 120^\circ$ \end{center}};
      \draw ( 1.7, 0.3) node[text width=3cm] {\begin{center} Transverse \goodbreak $60^\circ < |\Delta\phi| < 120^\circ$ \end{center}};
      \draw (-1.7, 0.3) node[text width=3cm] {\begin{center} Transverse \goodbreak $60^\circ < |\Delta\phi| < 120^\circ$ \end{center}};
    \end{tikzpicture}
    \caption{Definition of UE regions as a function of the azimuthal angle with respect to the \Z.}
    \label{fig:ueregions}
  \end{center}
\end{figure}
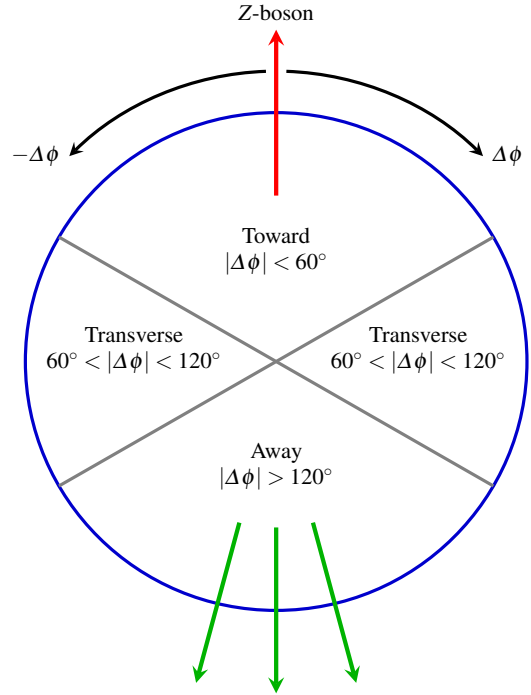

Since there is no final-state gluon radiation associated with a \Z , 
lepton-pair production consistent with \Z decays provides a cleaner final-state environment than jet production
for measuring the characteristics of the underlying event in certain regions of phase space.
The direction of the \Z candidate is used to define regions in the azimuthal plane
that have different sensitivity to the UE, a concept first used in~\cite{Field:2002vt}.
As illustrated in \FigRef{fig:ueregions}, the azimuthal angular difference between charged tracks and the \Z,
$|\Delta\phi|=|\phi-\phi_\text{\Z}|$, is used to define the following three azimuthal UE regions:
\begin{itemize}
\item $|\Delta\phi| < 60^{\circ}$, the \textit{toward} region,
\item $60^{\circ} < |\Delta\phi| < 120^{\circ}$, the \textit{transverse} region, and
\item $|\Delta\phi| > 120^{\circ}$, the \textit{away} region.
\end{itemize}

These regions are well defined only when
the measured \ZpT\ is large enough that, taking into account detector resolution, it can be used to define a direction.
The away region is dominated by particles balancing the momentum of the \Z except at low values of \ZpT .
The transverse region is sensitive to the underlying event, since it is by construction
perpendicular to the direction of the \Z and hence it is expected to have a lower
level of activity from the hard scattering process compared to the away region.
The two opposite transverse regions may be distinguished on an event-by-event basis  through their amount of activity, 
as measured by the sum of the charged-particle transverse momenta 
in each of them.
The more or
less-active transverse regions are then referred to as \textit{trans-max}
and\textit{ trans-min}, respectively, with the difference between them
on an event-by-event basis for a given observable defined as \textit{trans-diff}~\cite{PhysRevD.38.3419, PhysRevD.57.5787}.
The activity in the toward region, which is similarly 
unaffected by additional activity from the hard
scatter,
is measured in this analysis, in contrast to the underlying event analysis in dijet events~\cite{Aad:2014hia}.

The observables measured in this analysis
are derived from the number, \Nchg , and transverse momenta, \pT ,
of stable charged particles in each event.
They have been 
studied both as one-dimensional distributions, inclusive in the
properties of the hard process, and as \textit{profile} histograms which present the
dependence of the mean value of each observable (and its uncertainty)
on \ZpT .
The observables are summarised in~\TabRef{tab:obs}. 
The mean charged-particle transverse momentum
is constructed 
on an event-by-event basis and is then averaged over all events
to calculate the observable mean \pT .

\begin{table}[h]
   \caption{Definition of the measured observables. These are defined for each azimuthal region under consideration except for \ZpT .}
  \begin{center}
    \begin{tabular}{ll}
     \toprule
      Observable & Definition \\
     \midrule
     \ZpT & Transverse momentum of the \Z  \\
     &\\  
     \Nchgdetadphi & \multirow{2}{50mm}{Number of stable charged particles per unit $\eta$--$\phi$} \\
     &\\
     &\\
     \pTsumdetadphi & \multirow{2}{50mm}{Scalar \pt\, sum of stable charged particles per unit $\eta$--$\phi$} \\
     &\\
     &\\
     Mean \pT & \multirow{2}{50mm}{Average \pt\, of stable charged particles}  \\
     &\\         
   \bottomrule
    \end{tabular}
    \label{tab:obs}
  \end{center}

\end{table}

%%%%%%%%%%%%%%%%%%%%%%%%%%%%%%%%%%%%
%        ATLAS Detector
%%%%%%%%%%%%%%%%%%%%%%%%%%%%%%%%%%%%
\section{The ATLAS detector}
\label{sec:atlasdet}

The ATLAS detector~\cite{:2008zzm} covers almost the full solid angle around the collision point.
The components that are relevant for this analysis are the tracking
detectors, the liquid-argon (LAr) electromagnetic sampling calorimeters and 
the muon spectrometer.

The inner tracking detector (ID) has full coverage in azimuthal angle $\phi$ and covers the pseudorapidity range $\etamod~ < 2.5$.
It consists of a silicon pixel detector (pixel), a semiconductor tracker (SCT) and a straw-tube transition radiation tracker
(TRT). These detectors are located at a radial distance from the beam line of
50.5--\unit{150}{\mm}, 299--\unit{560}{\mm} and 563--\unit{1066}{\mm},
respectively, and 
are contained within a 2~T
axial magnetic field. The inner detector barrel (end-cap) consists of
3 ($2 \times 3$) pixel layers, 4 ($2 \times 9$) layers of double-sided silicon strip modules, and 73 ($2 \times 160$) layers of
TRT straw-tubes. These detectors have position resolutions typically of \unit{10}{\micro\metre}, \unit{17}{\micro\metre} and
\unit{130}{\micro\metre} for the $r$--$\phi$ coordinates (only for TRT barrel), respectively.
The pixel and SCT detectors provide measurements of the $r$--$z$  coordinates with typical resolutions of 
\unit{115}{\micro\metre} and \unit{580}{\micro\metre}, respectively. 
The TRT acceptance is $\etamod< 2.0$.
A track traversing the barrel typically has 11 silicon hits (3 pixel clusters and 8 strip clusters) and more than 30 straw-tube hits.

A high-granularity lead, liquid-argon  electromagnetic sampling calorimeter~\cite{2010LAr}
covers the pseudorapidity range $|\eta| < 3.2$.
Hadronic calorimetry in the range $|\eta| < 1.7$ is provided
by an iron scintillator-tile calorimeter, consisting of a central barrel and two smaller extended barrel cylinders, one on either
side of the central barrel. 
In the end-caps ($|\eta| > 1.5$), the acceptance of the LAr hadronic calorimeters matches
the outer $|\eta|$ limits of the end-cap electromagnetic calorimeters.
The LAr forward calorimeters provide both electromagnetic and hadronic energy 
measurements, and extend the coverage to $|\eta| < 4.9$.

The muon spectrometer (MS) measures the deflection of muon tracks in the large superconducting air-core toroid magnets
in the pseudorapidity range $|\eta| < 2.7$. It is instrumented with separate trigger and high-precision tracking chambers. Over
most of the $\eta$-range, a precision measurement of the track coordinates in the principal bending direction of the magnetic
field is provided by monitored drift tubes. At large pseudorapidities, cathode strip chambers with higher granularity
are used in the innermost plane over the range $2.0 < |\eta| < 2.7$.

The ATLAS trigger system consists of a
hardware-based Level-1 (L1) trigger and a software-based High Level Trigger, subdivided into 
the Level-2 (L2) and Event-Filter (EF)~\cite{Aad:2012xs} stages.
In L1, electrons are selected by requiring adjacent electromagnetic (EM) trigger towers exceed a certain $E_\mathrm{T}$
threshold, depending on the detector $\eta$. The EF uses the offline
reconstruction and identification algorithms to apply the final electron selection in the trigger.
The $\Zee$ events are selected in this analysis by using a dielectron trigger
in the region $|\eta| < 2.5$ with an electron transverse energy, $E_\mathrm{T}$, threshold of $12$ GeV.
The muon trigger system, which covers the pseudorapidity range
$|\eta| < 2.4$, consists of resistive plate chambers in the barrel ($|\eta| < 1.05$) and thin gap chambers in the end cap regions
($1.05 < |\eta| < 2.4$). 
Muons are reconstructed in the EF combining L1 and L2 information.
The $\Zmm$ events in this analysis are selected with a 
first-level trigger that requires the presence of a muon candidate reconstructed in the muon spectrometer 
with transverse momentum of at least $18$~GeV.
The trigger efficiency for the events selected as described in \SecRef{sec:sel} is very close to $100\%$.

%%%%%%%%%%%%%%%%%%%%%%%%%%%%%%%%%%%%
%         QCD Monte Carlo Models
%%%%%%%%%%%%%%%%%%%%%%%%%%%%%%%%%%%%

\section{Monte Carlo simulations}
\label{sec:mc}

Monte Carlo event samples including a simulation of the ATLAS detector response are used to
correct the measurements for
detector effects, and to estimate systematic uncertainties. 
In addition, predictions of different phenomenological models implemented 
in the MC generators are compared to the data corrected to the particle level.
Samples of inclusive $\Zee$ and $\Zmm$ events were produced using the
leading order (LO) \pythiasix~\cite{Sjostrand:2006za}, \pythiaeight~\cite{Sjostrand:2007gs}, 
\herwigpp~\cite{Bahr:2008pv,Gieseke:2011na}, \sherpa~\cite{Gleisberg:2008ta}, \alpgen~\cite{Mangano:2002ea} 
and next to leading order (NLO) \powheg~\cite{Alioli:2008gx} event generators,
including various parton density function (PDF) parametrisations.
The \alpgen and \sherpa matrix elements are generated for up to five additional partons,
thereby filling the phase space with sufficient statistics for the full set of measured observables.
It should be noted, that since the measurements are all reported in bins of \ZpT , 
the results presented in this paper are not sensitive to the predicted shape of the \ZpT\ spectrum, 
even though they are sensitive to jet activity in the event. 
\TabRef{tab:mc} lists the different MC models used in this paper.

\pythiasix, \pythiaeight and \herwigpp are 
all leading-logarithmic parton shower (PS) models matched to 
leading-order matrix element (ME) calculations, but with different ordering algorithms for parton showering,
and different hadronization models.
In scattering processes modelled by lowest-order perturbative QCD
two-to-two parton scatters, with a sufficiently low \pT threshold, the partonic jet cross-section
exceeds that of the total hadronic cross-section.
This can be interpreted in terms of MPI.
In this picture, the ratio of the partonic jet cross-section to the
total cross-section is interpreted as the mean number of parton
interactions per event. 
This is implemented using phenomenological models~\cite{Buckley:2011ms},
which include (non-exhaustively) further low-\pT\ screening of the
partonic differential cross-section, and use of phenomenological transverse matter-density 
profiles inside the hadrons. 
The connection of colour lines 
between partons, and the rearrangement of the colour structure of an event by reconnection of the colour
strings, are implemented in different ways in these phenomenological models.

The \pythiasix and \pythiaeight generators both use \pT-ordered parton showers,
and a hadronisation model based on the fragmentation of colour strings.
The \pythiaeight generator adds to the \pythiasix MPI model by interleaving not
only the ISR emission sequence with the MPI scatters, but also the FSR emissions.
The \herwigpp generator implements a cluster hadronization scheme 
with parton showering ordered by emission angle.
The \sherpa generator uses LO matrix elements 
with a model for MPI similar to that of \pythiasix and a cluster hadronisation model similar to that of \herwigpp .
In \alpgen the showering is performed with the \herwig 
generator. The original Fortran \herwig~\cite{Corcella:2002jc} generator %by itself
 does not simulate multiple partonic interactions;  these are added by the
\jimmy~\cite{Butterworth:1996zw} package.
The \alpgen generator provides leading-order multi-leg matrix element
events: it includes more complex hard process topologies than those used by
the other generators, but does not include loop-diagram contributions. The \alpgen
partonic events are showered and hadronised by the \herwigjimmy generator
combination, making use of MLM matching\,\cite{Mangano:2002ea} between the
matrix element and parton shower to avoid double-counting of jet production
mechanisms. A related matching process is used to interface \pythiasix to
the next-to-leading-order (NLO) \powheg generator, where the matching scheme
avoids both double-counting and NLO subtraction singularities~\cite{Aad:2013lpa, Nason:2013uba}.

Different settings of model parameters, tuned to 
reproduce existing experimental data, have been used for the MC generators. The 
\pythiasix, \pythiaeight , \herwig + \jimmy , \herwigpp and \sherpa tunes have been performed using
mostly Tevatron and early LHC data. 
The parton shower generators used with \alpgen and \powheg do not
use optimised tunes specific to their respective parton shower matching schemes.

\begin{table*}[ht]
  \singlespacing
  \caption{Main features of the Monte-Carlo models used. The abbreviations ME, PS, MPI, LO and NLO respectively stand for
  matrix element, parton shower, multiple parton interactions, leading order and next to leading order in QCD.}
  \begin{center}
    \begin{tabular}{lllll}  
      \toprule
      Generator &  Type &  Version &  PDF & Tune  \\
      \midrule        
       \pythiasix & LO PS & 6.425 & CTEQ6L1~\cite{Pumplin:2002vw}  & Perugia2011C~\cite{Skands:2009zm} \\ 
       &&&&\\
       \pythiaeight & LO PS & 8.165 & CTEQ6L1  & AU2~\cite{py8mc12}   \\       
              &&&&\\
       \herwigpp  &  LO PS & 2.5.1 & MRST \loss~\cite{Sherstnev:2007nd}   &  UE-EE-3~\cite{Gieseke:2012ft} \\
              &&&&\\
       \sherpa &  LO multi-leg & 1.4.0 & CT10~\cite{Lai:2010vv}  & Default \\     
                    & ME + PS &  /1.3.1 && \\  
              &&&&\\
       \alpgen &  LO multi-leg ME   &  2.14   & CTEQ6L1 &  \\ 
       + \herwig & + PS & 6.520  &  MRST \loss & AUET2~\cite{atlastunesto2010data}  \\
       +\jimmy & (adds MPI) & 4.31 & &\\
              &&&&\\
       \powheg & NLO ME &  -  & CT10 &  \\       
        + \pythiaeight & + PS & 8.165 & CT10 & AU2 \\
       \bottomrule
    \end{tabular}
    \label{tab:mc}
  \end{center}
\end{table*}

For the purpose of correcting the data for detector effects, samples generated with
\sherpa (with the CTEQ6L1 PDF and the corresponding UE tune), and \pythiaeight tune 4C~\cite{Corke:2010yf}
were passed through ATLFAST2~\cite{ATLAS:1300517}, a fast detector simulation software package,
which used full simulation in the ID and MS and a fast simulation of the calorimeters.
Comparisons between MC events at the reconstructed and particle level are then used to correct the data for detector effects. 
Since the effect of multiple proton-proton interactions is   
corrected using a data-driven technique (as described in \SecRef{sec:pu}), 
only single proton-proton interactions are simulated in these MC samples.

%%%%%%%%%%%%%%%%%%%%%%%%%%%%%%%%%%%%
%       Selections
%%%%%%%%%%%%%%%%%%%%%%%%%%%%%%%%%%%%

\section{Event selection}
\label{sec:sel}

The event sample was collected during stable beam conditions, with
all detector subsystems
operational.
To reject contributions from cosmic-ray muons and other non-collision backgrounds,
events are required to have 
a primary vertex (PV).
The PV is defined as the reconstructed vertex in the event with the highest $\sum\pt^2$
of the associated tracks, 
consistent with the beam-spot position
(spatial region inside the detector where collisions take place)
and with at least two associated tracks with $\pT > 400$~MeV.

Electrons are reconstructed from energy deposits measured in the EM calorimeter and associated to ID tracks. 
They are required to satisfy $\pT > 20$ GeV and $|\eta| < 2.4$, excluding 
the transition region $1.37 < |\eta| < 1.52$ between the barrel and end-cap
electromagnetic calorimeter sections. Electron identification uses
shower shape, track-cluster association and TRT criteria~\cite{Aad:2014fxa}. 
Muons are reconstructed from 
track segments in the MS associated to ID tracks~\cite{Aad:2012mu}.
They are required to have $\pT > 20$ GeV and $|\eta| < 2.4$.
Both electrons and muons are required to have longitudinal impact 
parameter multiplied by $\sin\theta$ of the ID track, $|z_0|\sin \theta < 10$~mm with respect to the PV.
The dilepton invariant mass of oppositely charged leptons, \Mll , is 
required to be in the region $66 < \Mll< 116$~GeV at this stage.
No explicit isolation requirement is applied to the muons, but in the case of electrons,
some isolation is implied
by the identification algorithm. 
The correction for this effect is discussed in \SecRef{sec:comb}.

The tracks in the calculation of UE observables satisfy the following criteria~\cite{Collaboration:2010ir}:
  \begin{itemize}
  \item $\pt > 0.5$~GeV and $|\eta| < 2.5$;
  \item a minimum of one pixel and six SCT hits; 
  \item a hit in the innermost pixel layer, 
   if the corresponding pixel module was active;
  \item transverse and %weighted-
   longitudinal impact parameters with respect to the PV,
  $|d_0| < 1.5$~mm and $|z_0|\sin \theta < 1.5$~mm, respectively;
  \item for tracks with $\pt > 10$~GeV, a goodness of fit probability greater than 
       $0.01$ in order to remove mis-measured tracks.
   \end{itemize}

The tracks corresponding to the leptons forming the \Z candidate are excluded.

%%%%%%%%%%%%%%%%%%%%%%%%%%%%%%%%%%%%
%       Pileup
%%%%%%%%%%%%%%%%%%%%%%%%%%%%%%%%%%%%

\section{Correction for pile-up} 
\label{sec:pu}

The average expected number of pile-up events per hard-scattering interaction ($\mu$) 
 was typically in the range $3-12$ in the 2011 dataset. 
Of the tracks selected by the procedure described above and 
compatible with the PV of
the hard-scattering event, up to $15\%$ originate from pile-up, as described below.
Due to the difficulty in modelling accurately the soft interactions in $pp$ collisions and 
the fact that pile-up conditions vary significantly over the data-taking period,
a data-driven procedure has been derived to
correct the measured observables for the pile-up contribution.

The measured distribution of any track-based observable can be expressed as the convolution
of the distribution of this variable for the tracks originating from the \Z production vertex,
with the distribution resulting from the superimposed pile-up interactions.
The pile-up contribution is estimated from data by sampling tracks originating
from a vertex well separated from the hard-scattering PV.
In each event, the pile-up contribution to a given observable is derived from tracks selected with 
the same longitudinal and transverse impact parameter requirements 
as the PV tracks, but with respect to two points located at 
$z$ distances of $+2$~cm and $-2$~cm from the hard-scattering PV. 
The shift of $2$~cm relative to the PV introduces 
a bias in the density of the pile-up interactions.
This is corrected on the basis of
the shape of the distribution of the $z$ distance between pairs of interactions in the same bunch crossing.
This distribution is well approximated by a Gaussian with variance $\sigma = \sqrt{2}\sigma_{BS}$, 
where $\sigma_{BS} \approx 6$ cm is the effective 
longitudinal variance of the interaction region 
averaged over all events.
Pile-up distributions are thus obtained for each observable and are deconvoluted from the corresponding measured
distributions at the hard-scattering PV.

\begin{figure*}[ht]
  \centering
     \includegraphics[width=0.95\textwidth]{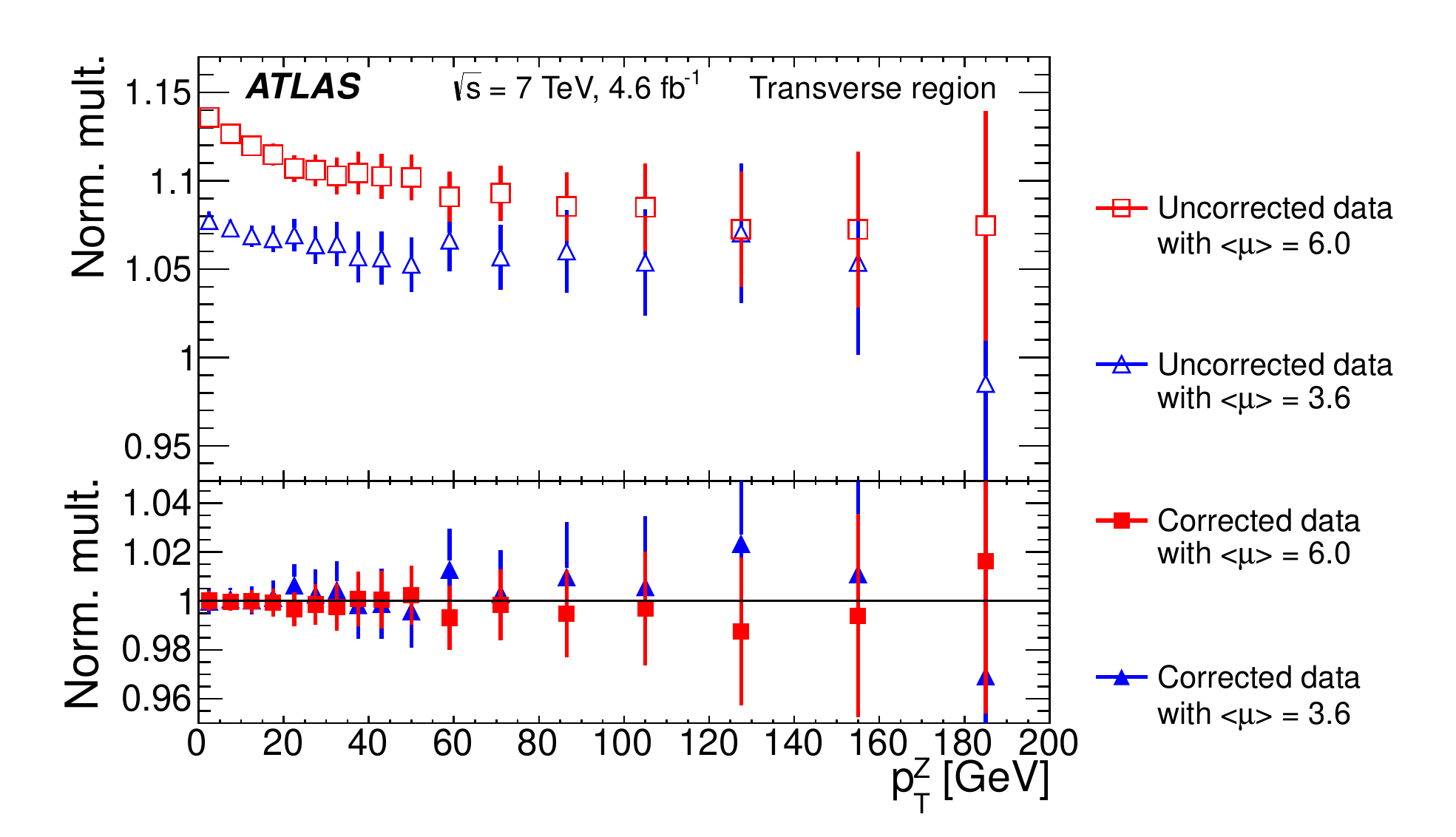} 
   \caption{
   Average charged particle multiplicity density, \dNchgdetadphi in the transverse region
   for two samples with different average numbers of interactions, $\langle \mu \rangle$, 
   normalised to the average density in
   the full sample after pile-up correction, before (top) and after (bottom) pile-up correction.
   The data are shown    
   as a function of the transverse momentum of the \Z, \ZpT .
   Only statistical uncertainties are shown.  
}
  \label{fig:pup-trv}
\end{figure*}

The stability of the pile-up correction for different beam conditions 
is demonstrated in~\FigRef{fig:pup-trv}. 
The figure compares the distributions of the average charged particle multiplicity 
density, \dNchgdetadphi as a function of \ZpT ,
before and after pile-up correction, for two sub-samples with an average of
$3.6$ and $6$ interactions per bunch crossing ($\langle \mu \rangle $), respectively.
Each distribution is normalised to that obtained for the full sample after pile-up correction.
The dependence of the normalised charged multiplicity distributions on \ZpT
which can be seen before correction in~\FigRef{fig:pup-trv} reflects the fact that actual contributions to this observable 
depend on \ZpT , while the pile-up contribution is independent of \ZpT .
The pile-up corrected results agree to better than $2\%$, a value much smaller than the size
of the correction, which may be as large as $20\%$ for this 
observable in low \ZpT\ bins for 
the data-taking periods with the highest values of $\langle\mu\rangle$.
The systematic uncertainty arising from this procedure is discussed in \SecRef{sec:syst}.

%PU Fig

%%%%%%%%%%%%%%%%%%%%%%%%%%%%%%%%%%%%
%       Corrections
%%%%%%%%%%%%%%%%%%%%%%%%%%%%%%%%%%%%

\section{Unfolding to particle level, background corrections and channel combination} 
\label{sec:corr}

After correcting for
pile-up, an iterative Bayesian unfolding~\cite{D'Agostini:1994zf}
of all the measured observables to the particle level is performed.
This is followed by a correction of the unfolded distributions for the small amount of background from 
other physics processes. At this point, the electron and muon measurements are combined to produce the final results.

%%%%%%%%%%%%%%%%%%%%%%%%%%%%%%%%%%%%
%       Unfolding
%%%%%%%%%%%%%%%%%%%%%%%%%%%%%%%%%%%%

\subsection{Unfolding} 
\label{sec:unfolding}

The measurements are presented in the 
fiducial region defined by the \Z reconstructed from a pair of
oppositely charged electrons or muons 
each with $\pT > 20$~GeV and $|\eta| < 2.4$ and with a lepton pair invariant mass in the range  $66 < \Mll < 116$~GeV.

The results in \SecRef{sec:res} are presented in the Born approximation, 
using the leptons before QED FSR to reconstruct the \Z . These results are also provided 
in HEPDATA~\cite{Buckley:2010jn} using \textit{dressed} leptons.
These are defined by adding vectorially to the $4$-momentum of each lepton after 
QED FSR the $4$-momenta of any photons not produced in hadronic decays and found within a cone of 
$\Delta R = 0.1$ around the lepton, where the angular separation $\Delta R$ is 
given by $\sqrt{(\Delta \eta)^2 + (\Delta \phi)^2}$.

The UE observables are constructed from stable charged particles 
with $\pT > 0.5$~GeV and $|\eta| < 2.5$, excluding \Z decay products. 
Stable charged particles are
defined as those with a proper lifetime $\tau > 0.3 \times
\unit{\ensuremath{10^{-10}}}{\second}$, either directly produced in $pp$ interactions 
or from the subsequent decay of particles with a shorter
lifetime.

Bayesian iterative unfolding was used to correct for
residual detector resolution effects. This method requires two inputs: an input
distribution of the observable (the MC generator-level distribution is used for
this), and a detector response matrix which relates the uncorrected measured distribution in this
observable to that defined at the event generator level, also termed the particle level.
The detector response matrix element,
$S_{ij}$
is the probability that a particular event from bin~$i$ of the particle-level
distribution is found in bin~$j$ of the corresponding reconstructed
distribution, and is obtained using simulation.
For the profile histogram
observables in this paper, a two-dimensional (2D) histogram was created with a fine binning for the
observable of interest, such that each unfolding bin
corresponds to a region in the 2D space.

The unfolding process is iterated to avoid dependence on the input
distribution: the corrected data distribution produced in each iteration is used
as the input for the next. In this analysis, four iterations were performed since
this resulted only in a small residual bias when tested on MC samples while keeping the
statistical uncertainties small. 
The unfolding uses the \sherpa simulation for the input distributions and unfolding matrix.
In the muon channel, the 
MC events are reweighted at the particle level 
in terms of a multi-variable distribution constructed for each distribution of interest
using the ratio of data to detector-level MC, 
so that the detector-level MC closely matches the data.
This additional step is omitted in the electron channel for the reasons discussed in \SecRef{sec:comb}.

The dominant correction to the data is that related to track reconstruction and 
selection efficiencies, in particular at low-\pT. 
 After the selection described in \SecRef{sec:sel}, the rate of fake
 tracks (those constructed from tracker noise and/or hits which were not
 produced  by a single particle)
 is found to be very small.
 This, as well as a small contribution of  secondaries
 (\textit{i.e.} tracks arising from hadronic interactions, photon
 conversions to electron-positron pairs, and decays of long-lived particles)
 is corrected for by the unfolding procedure.

%%%%%%%%%%%%%%%%%%%%%%%%%%%%%%%%%%%%
%       Backgrounds
%%%%%%%%%%%%%%%%%%%%%%%%%%%%%%%%%%%%

\subsection{Backgrounds}
\label{sec:bg}

\begin{figure*}[htb]
  \centering
     \includegraphics[width=0.85\textwidth]{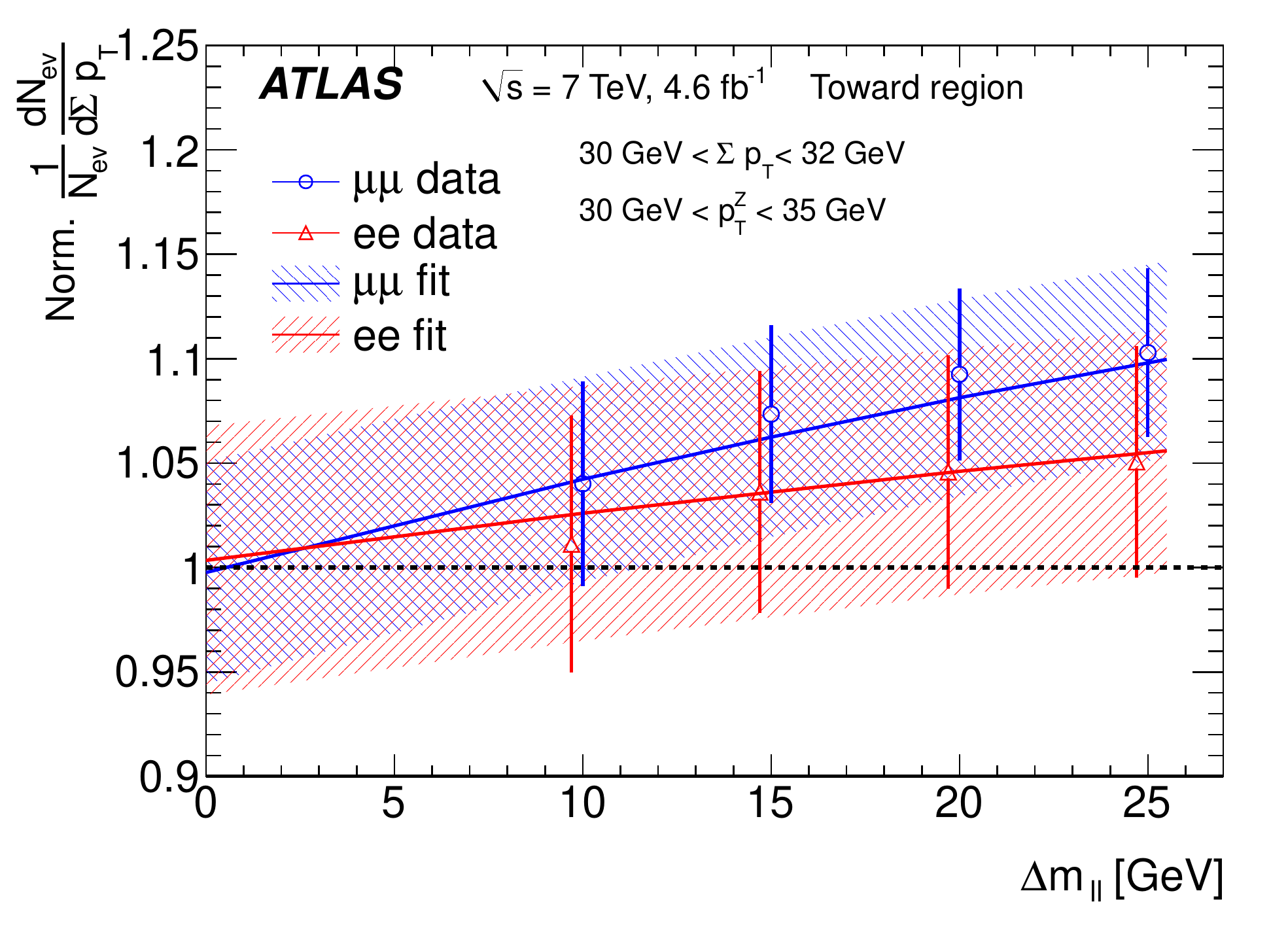} 
   \caption{   
   Impact of non-resonant backgrounds on the measurement of  \ptsum
   in the bin $30$~GeV $< \Sigma \pT < 32$~GeV 
   and in the toward region for $30$~GeV $< \ZpT < 35$~GeV.
   This is shown separately for the electron and muon channels   
   as a function of the window applied to the dilepton mass
   $|\Mll - M_{\mathrm{Z}}| < \Delta\Mll$.   
   The unfolded value for each channel is normalised to the corrected combined result.   
   The statistical uncertainties at individual $\Delta\Mll$ points are strongly
   correlated within each channel. 
   The uncertainty range of the linear fit is shown by hatched bands
   for each channel.    
   This includes the statistical and systematic uncertainties from the fit itself, as well as the relevant correlations. 
   The vertical line at
   $\Delta m_{\ell\ell} = 0$ marks the points to which the extrapolations are made.   
   }
  \label{fig:bglin}
\end{figure*}

The background to the \Z signal decaying into a lepton pair 
consists of a dominant component from multijet production, 
smaller components from other physics sources, and a very small component from non-collision backgrounds.
A fully data-driven correction procedure has been developed 
and applied directly to the unfolded distributions to take into account the influence of the backgrounds.

The primary vertex requirement removes almost all of the beam-induced non-collision background events.
Similarly, the impact parameter requirements on the leptons reduce the cosmic-ray 
background to a level below $0.1\%$ of the signal. 
These residual backgrounds were considered as negligible in the analysis.

The $pp$ collision backgrounds to $\Zee$ or $\Zmm$ decays
were found to be of the order of a few percent of the signal in the mass window~\cite{Aad:2013ysa}.
The \textit{resonant} backgrounds from $WZ$, $ZZ$ and $Z\gamma$ pair production with a $Z$ boson decaying into leptons 
were estimated from simulated samples and found to amount to less than 0.2\% of the selected events.
Their impact on the underlying event observables is negligible and they were not considered further here.

The contribution from the \textit{non-resonant} backgrounds (\textit{i.e.} from all other $pp$ collision processes)
is larger, typically between $1\%$ and $2\%$ of 
the signal, depending on the \ZpT\ range considered, and is dominated by multijet production 
with a combination of light-flavour jets misidentified as electrons and heavy-flavour jets 
with a subsequent semileptonic decay of a charm or beauty hadron. 
This contribution is estimated 
to correspond to 0.5\% of the signal for $\Zee$ decays and to $1-2\%$ of the signal for $\Zmm$ decays.
The background in the electron channel is somewhat lower
because of the implicit isolation requirement imposed 
on the electrons through the electron identification requirements. 
Smaller contributions to the non-resonant background arise from diboson, $t\bar t$ and single top production and 
amount to less than 0.3\% of the signal, increasing to $1\%$ at $\ZpT > 50$~GeV.
The still smaller contributions from processes such as $W$ or $Z$
production with jets, where a jet is misidentified as a lepton, are treated in the same way 
as the multijet background. These contributions amount to less than $0.1\%$ of the signal sample.

The non-resonant background is corrected for by studying the UE observables as a function of $\Delta\Mll$, 
the half-width of the mass window around the \Z signal peak.
Since the distributions of UE observables in non-resonant 
background processes are found to be
approximately constant as a function of the dilepton mass 
and the background shape under the \Z mass peak is approximately linear, the 
background contribution to any UE observable is approximately proportional to \DMll.
Thus, the background contribution
can be corrected for by calculating the UE observables for different values of $\Delta\Mll$, 
chosen here to be between $10$ and $25$~GeV, 
and extracting the results which could be measured for a pure signal with $\DMll \to 0$.
This procedure is performed separately for each bin of the distributions of interest.

  The validity of the linear approximation for the \DMll\ dependence of
   the background contribution was checked for all observables
   studied in this analysis. An example is presented
   in~\FigRef{fig:bglin}, where the $\Delta\Mll$ dependence is shown for
   one bin of the $\ptsum$ differential distribution, as obtained in the
   toward region for  $30 < \ZpT < 35$~GeV and shown separately for the
   electron and muon channels.
   The values plotted in~\FigRef{fig:bglin} are normalised to the
   corrected combined value.
   The values of the observables in the muon channel increase linearly
   with $\Delta\Mll$.
   The difference in the slope observed between the muon and the electron
   samples is due to the larger background in the muon channel, as
   discussed above. A straight line is fitted through the points obtained
   for the various $\Delta\Mll$ values shown in~\FigRef{fig:bglin} for
   each channel.
   For each bin in the observable and \ZpT\ , the muon and electron channels values agree with each other after
   extrapolating to $\Delta\Mll = 0$
   within the uncertainties of the fit procedure, which are
   represented by the shaded areas
   and include the statistical and systematic uncertainties from the fit itself (as discussed 
   in \SecRef{sec:syst}, as well as the relevant correlations.

The effect of the background on the unfolded 
distributions 
can be summarised as follows: in the case of the 
electron channel, which has less background than the muons, the background 
in the average values of \ptsum and \Nchg is below $1\%$.
 The absence of any
 isolation requirement applied to the muons leads to significantly higher background levels in 
 certain regions, with corrections ranging from as high as $6-8\%$
for the average values of \ptsum in the toward region at high \ZpT ,
to about $1\%$ for the average values of \Nchg .
The background correction is done after unfolding to avoid resolution issues present at the detector level.

%%%%%%%%%%%%%%%%%%%%%%%%%%%%%%%%%%%%%%%%%%%%%%%%

\subsection{Combination of the electron and muon channels}  
\label{sec:comb}

Before combining the electron and muon channels, the analysis must correct for a bias over 
a limited region of the phase space which affects the measurements in the electron channel when one 
of the electrons is close to a jet produced in association with the $Z$ boson. 
This bias is observed at high \ZpT ,
mostly in the toward region and to a lesser extent in the transverse region, and affects the $\ptsum$
distribution for high values of $\ptsum$, typically $\ptsum > 30$~GeV. 
It arises from the imperfect modelling of the electron shower shape variables in the simulation,
 which leads to an underestimate of the 
electron identification efficiency for electrons close to jets.
The bias on the observable can be as large as $50\%$ for $\ptsum = 100$~GeV. 
Since it is not reproduced precisely enough by the simulation
of the electron shower, in the relevant narrow regions of phase space a
tightened isolation criterion was applied to electrons to exclude the
mismodelled event configurations and the proper geometric correction was
deduced from the muon channel unaffected by jet overlap.
The combined results for electrons and 
muons in the affected bins are assigned a larger uncertainty, since 
the contribution of events from the electron-decay channel is 
significantly reduced leading to a larger overall uncertainty.
The most significant effect is observed for the \ptsum $>100$~GeV in the toward and transverse region.

As discussed in~\SecRef{sec:obs} and in \SecRef{sec:unfolding}, the electron and muon results are unfolded and then combined, 
both as Born-level lepton pairs and as dressed lepton pairs, and accounting for the uncorrelated and correlated terms 
in the systematic uncertainties between the channels (as described in~\SecRef{sec:syst}). 
Combining the dressed electron and muon pairs induces $< 0.1\%$ additional systematic uncertainty on the UE observables
compared to the Born level results.

\begin{figure*}[tbp]
  \centering
  \vspace{0.5cm}
   \resultsimg{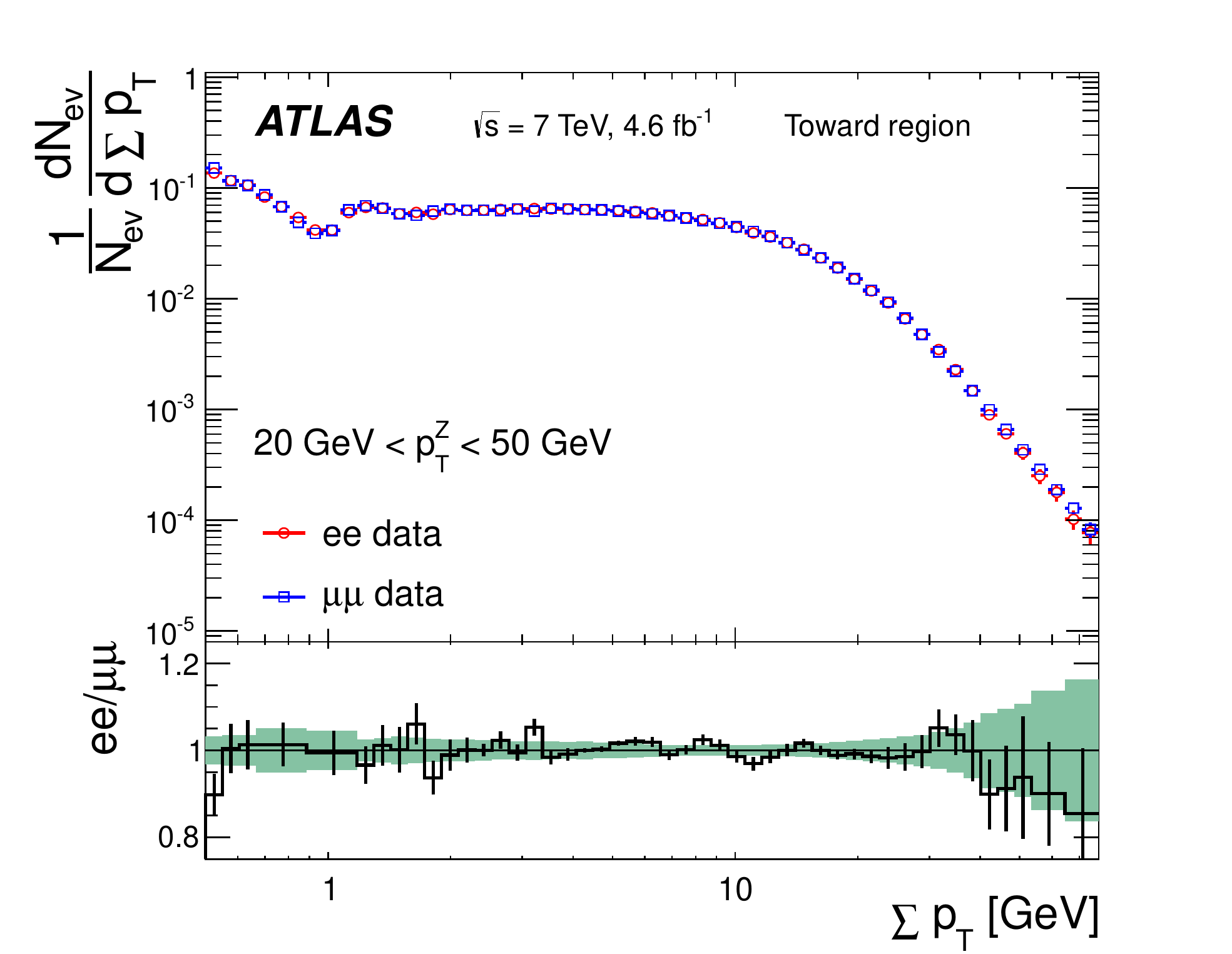}{fig:bc-ptsum}\\
   \resultsimg{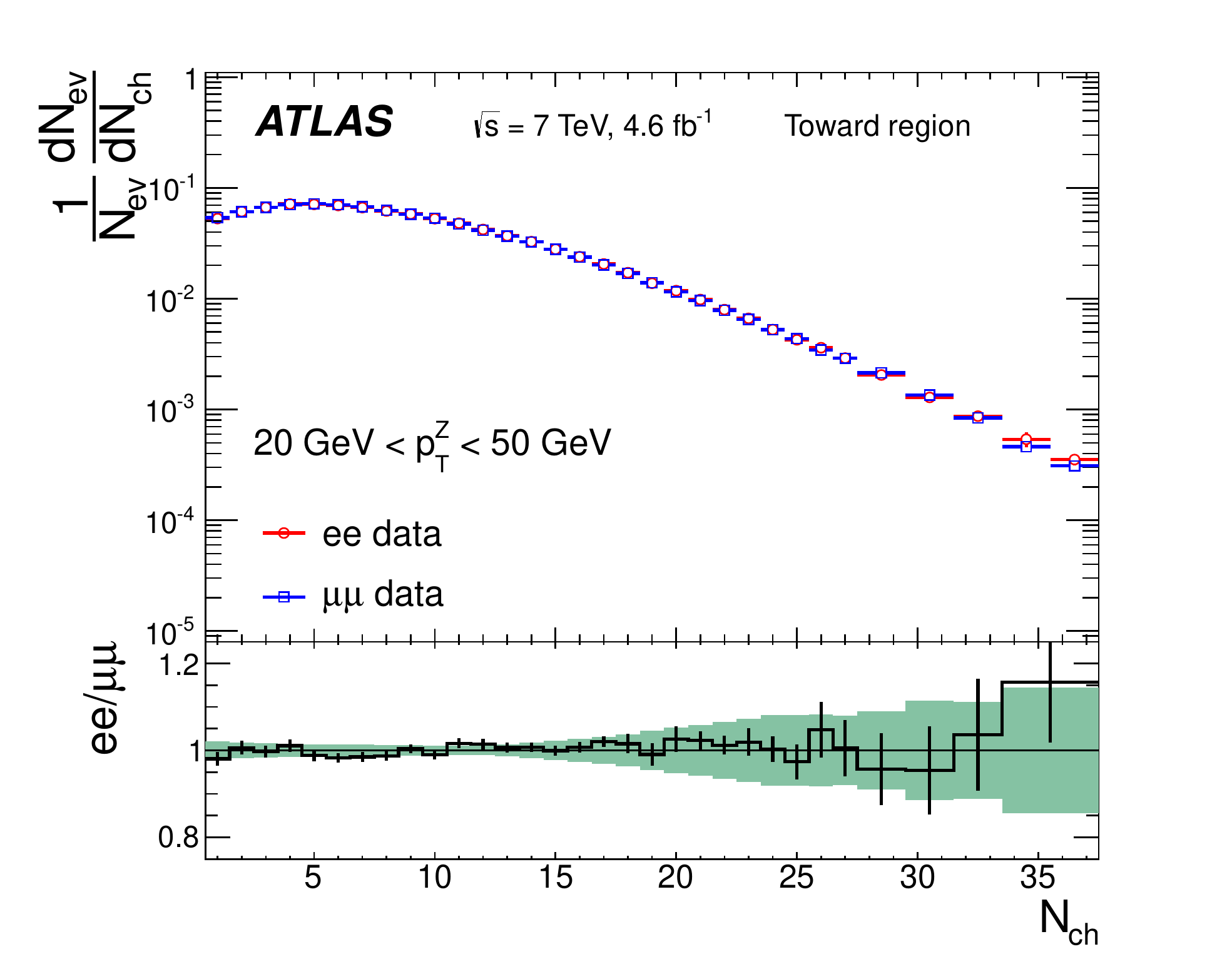}{fig:bc-nchg}
   \caption{
   Unfolded and corrected distributions of charged particle $\ptsum$ (a)  and $\Nchg$ (b)  for $20 < \ZpT < 50$~GeV 
   shown separately for the $\Zee$ and $\Zmm$ samples after all corrections have been applied.
   The bottom panels show the ratios between the electron and the  muon 
distributions where the error bars are purely statistical and the shaded areas represent the total uncertainty,
including systematic, on the combined result.}
   \label{fig:born-comb}
\end{figure*}

Figure~\ref{fig:born-comb} illustrates the excellent agreement between the fully unfolded and corrected UE observables for 
the electron and muon channels, once the specific correction procedure 
described above has been applied to the electron channel in the limited phase space regions where significant 
hadronic activity occurs close to one of the electrons. 
As shown for the specific region $20 < \ZpT < 50$~GeV in \FigRef{fig:born-comb}, the differential distributions for \ptsum
and \Nchg agree within statistical uncertainties over most of the range of relevance, 
except for high values of \ptsum , where the electron bias has been corrected as described above,
and where the total uncertainty on the combined measurement has been enlarged as shown by the 
shaded error band in the ratio plot. 
The shape of the \ptsum distribution in the region around $1$~GeV reflects the \pT\ threshold of $0.5$~GeV 
applied in the track selection.

%%%%%%%%%%%%%%%%%%%%%%%%%%%%%%%%%%%%
%         Systematics
%%%%%%%%%%%%%%%%%%%%%%%%%%%%%%%%%%%%

\section{Systematic Uncertainties} 
\label{sec:syst}

The following sources of uncertainty have been assessed for the measured distributions after 
all corrections and unfolding.
Table~\ref{tab:sysSummry} summarises the typical sizes of the
systematic uncertainties for the UE observables as a
function of \ZpT.

\begin{table*}[ht]
  \singlespacing
  \caption[Summary of systematic uncertainties]{
  Typical contributions to the systematic uncertainties (in \%) 
  on the unfolded and corrected distributions of interest in the  
  toward and transverse regions for the profile distributions.
  The range of values in the columns $3-5$ indicate the variations as a function
   of \ZpT , while those in the last column indicate the variations as a function of \Nchg.   
  The column labelled \textit{Correlation} indicates whether
  the errors are treated as correlated or not between the electron and muon channels.}
  \begin{center}
    \begin{tabular}{lccccc}
      \toprule
      Observable           & Correlation              & \Nchg vs \ZpT      & \ptsum vs \ZpT     &  Mean \pT\  vs \ZpT     &  Mean \pT\  vs \Nchg  \\
      \midrule
       Lepton selection       & No       & $0.5 - 1.0$   & $0.1 - 1.0$ & $<0.5$      & $0.1 - 2.5$       \\ 
      Track reconstruction    & Yes        & $1.0 - 2.0$   & $0.5 - 2.0$ & $<0.5$      & $<0.5$           \\ 
      Impact parameter requirement  & Yes      & $0.5 - 1.0$   & $1.0-2.0$   & $0.1 - 2.0$ & $<0.5$     \\                  
      Pile-up removal        & Yes             & $0.5 - 2.0$   & $0.5 - 2.0$ & $<0.2$      & $0.2 - 0.5$   \\
      Background correction   & No          & $0.5 - 2.0$   & $0.5 - 2.0$ & $<0.5$      & $<0.5$          \\        
      Unfolding       & No                  & $0.5 - 3.0$   & $0.5 - 3.0$ & $<0.5$      & $0.2-2.0$     \\ 
      Electron isolation      & No             & $0.1 - 1.0$   & $0.5 - 2.0$ & $0.1 - 1.5$ & $<1.0$           \\
      \midrule
      {\bf Combined systematic uncertainty}  &    & $1.0 - 3.0$   & $1.0 - 4.0$ & $<1.0$      & $1.0 - 3.5$     \\   
      \bottomrule          
    \end{tabular}
    \label{tab:sysSummry}
  \end{center}
\end{table*}

\begin{description}

\item [Lepton selection -] 
systematic uncertainties due to the lepton selection efficiencies have been assessed using MC simulation.
The data are first unfolded using the nominal MC samples, then with samples
corresponding to a $ \pm 1 \sigma$ variation of the efficiencies~\cite{Aad:2013ysa}.
These uncertainties are assumed to be uncorrelated between the electron and muon channels. 
The resulting uncertainty is less than $1\%$ 
for all observables over most of the kinematic range.

\item [Track reconstruction -]
the systematic uncertainty on the track reconstruction 
efficiency originating from uncertainties on the detector 
material description is estimated as in Ref.~\cite{Collaboration:2014zz}
for particles with $|\eta| < 2.1$ and as in Ref.~\cite{Collaboration:2010ir} for $|\eta| > 2.1$.
The typical value for $|\eta| < 2.1$ is $\pm 1\%$ while it is approximately $5\%$ for $|\eta| > 2.1$.
The effect of this uncertainty on the final results is less than $2\%$. 
This uncertainty is fully correlated between the electron and muon channels.

\item [Impact parameter requirement -] 
the fraction of secondary particles 
(i.e. those originating from decays and interactions in the inner detector material) 
in data is reproduced by the MC simulation to an accuracy of $\sim 10-20\%$,
obtained by comparing $d_0$ distributions in MC and in the data corrected for pile-up.
To assess the corresponding systematic uncertainty, 
the track impact parameter requirements on $|d_0|$ and $|z_0|\mathrm{sin}\theta$
are varied from the nominal values of $1.5$~mm to $1.0$~mm and $2.5$~mm, 
resulting in fractions of secondaries varying between $0.5\%$ to $4.0\%$,
and the resulting distributions are unfolded using MC samples selected with the same impact parameter requirements.
The maximum residual difference of $2\%$ or less 
between these unfolded distributions
and the nominal unfolded distribution 
is taken as the uncertainty arising from this requirement. This uncertainty is also fully correlated 
between the electron and muon channels.

\item [Pile-up correction -] the pile-up correction uncertainty 
 originates from the uncertainty in
 the pile-up density fitted along with the spatial distribution of tracks originating from pile-up,  
 and the difference between the pile-up densities measured for \Z and for randomly triggered events. 
 In addition to these, the stability of the correction method with respect to the instantaneous 
 luminosity was estimated by performing
     the correction procedure independently on datasets with
     different average numbers of reconstructed vertices, as
     shown in~\FigRef{fig:pup-trv}. 
    The total uncertainty due to the pile-up correction is taken
     to be the quadratic combination of the uncertainties from these sources,
     and it is at most $2\%$ for the average underlying event observables.
     The overall uncertainty is fully correlated between the electron and muon channels.

\item [Background correction -] the uncertainty is evaluated by comparing the
results of the linear fit to those obtained using a second-order
polynomial.
This uncertainty is at most $2\%$ 
for the maximum background uncertainty on \ptsum , which is the most strongly affected variable, and 
is assumed to be uncorrelated between the electron and muon channels.
Any potential correlation arising from the common $t\overline{t}$ and diboson backgrounds is neglected
because they become sizable only for $\ZpT > 100$~GeV,
where the total uncertainty is dominated by the statistical uncertanty on the background.

\item [Unfolding -] 
 the uncertainty due to the model-dependence of the unfolding procedure is
 taken from the degree of non-closure between the \pythiaeight initial particle-level distributions 
 and the corresponding detector-level \pythiaeight distributions unfolded and corrected
 using the \sherpa sample, which was reweighted to agree 
 with \pythiaeight at the detector level.
 This uncertainty varies between $0.5\%$ and $3\%$ for the profile distributions, and is assumed
  to be uncorrelated between the electron and muon channels.

 \item [Bias due to implicit isolation -]
  this uncertainty is estimated by varying the electron isolation requirement used to derive 
  the correction discussed in~\SecRef{sec:comb}. 
  The uncertainty is assigned to the electron channel and does not exceed $\sim 1$\% for 
  the profile distributions.

\end{description}

Other potential sources of systematic uncertainty have been
found to be negligible.
The total uncertainty in each measured bin is 
obtained by propagating the systematic component of the error matrix through the channel combination.
For the differential distributions in \SecRef{sec:distr}, the unfolding model dependent uncertainty
increases to about $5\%$, resulting in slightly larger overall systematic uncertainties.

%%%%%%%%%%%%%%%%%%%%%%%%%%%%%%%%%%%%
%        Results
%%%%%%%%%%%%%%%%%%%%%%%%%%%%%%%%%%%%
\section{Results}
\label{sec:res}

%%%%%%%%%%%%%%%%%%%%%%%%%%%%%%%%%%%%%%%%
\subsection{Overview of the results}
\label{sec:ovw}

The results are shown in \SecRef{sec:distr}, first for the differential distributions of 
charged particle \ptsum and \Nchg
in intervals of \ZpT , and then for the same distributions for a representative \ZpT\ range
compared to MC model predictions. 
The normalised quantities, \Nchgdetadphi and \pTsumdetadphi, 
are obtained by dividing \Nchg or \ptsum by the angular area in $\eta$--$\phi$ space.
This allows for direct comparisons between the total transverse and trans-min/max 
quantities, and between the current result
and experiments with different angular acceptances.
The angular areas for the  transverse, toward, and away region observables are 
$\delta\phi \, \delta\eta = (2\times \pi/3) \times (2 \times 2.5) = 10\pi/3$, 
while for trans-max/min/diff, $\delta\phi \, \delta\eta  =5\pi/3$.

Since the away region is dominated by the
jets balancing the  \ZpT~\cite{Aad:2013ysa}, 
the focus will be on the toward, transverse, trans-max 
and trans-min regions.
In the transverse region, 
the extra jet activity is more likely to be assigned to the trans-max region.
Assuming the same flat UE activity in trans-min and trans-max regions,
the trans-diff region,
the difference between the observables measured in trans-max and trans-min regions, is expected to be 
dominated by the hard scattering component.
In ~\SecRef{sec:pro} profile histograms are shown. 
Finally, in~\SecRef{sec:jet}, the results are
compared to previous measurements from ATLAS where distributions 
sensitive to the underlying event were measured as a function of the kinematics 
of either the leading charged particle~\cite{Aad:2010fh}, or the leading jet~\cite{Aad:2014hia}.

%%%%%%%%%%%%%%%%%%%%%%%%%%%%%%%%%%%%%%%%
\subsection{Differential distributions}
\label{sec:distr}

\begin{figure*}[tbp]
  \centering
  \vspace{0.5cm}
   \resultsimg{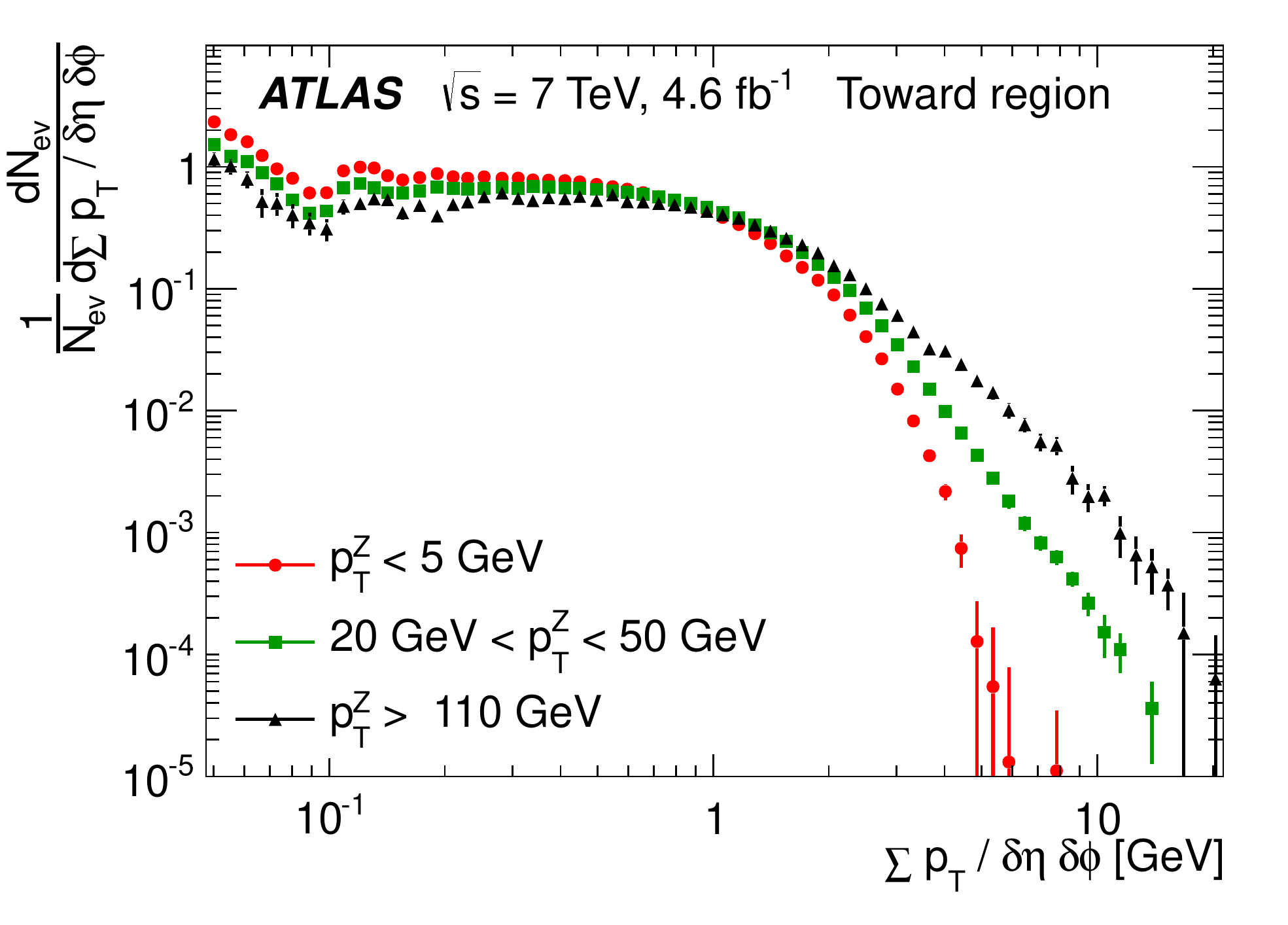}{fig:tow-ptsum}\\
   \resultsimg{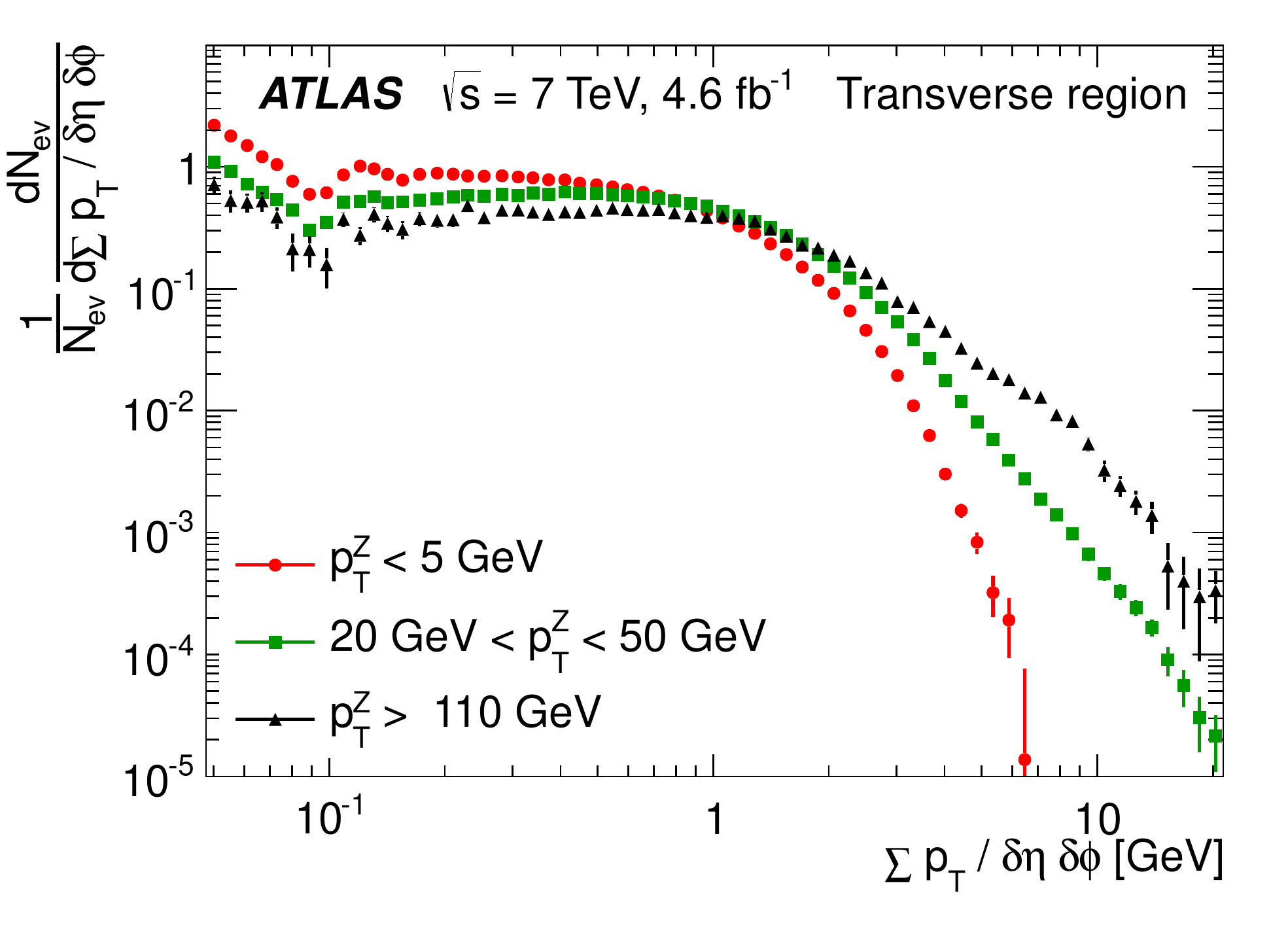}{fig:trv-ptsum}
   \caption{Distributions of the scalar \pT\ sum density of charged particles, \pTsumdetadphi , in three different
  \Z transverse momentum, \ZpT , intervals, in the toward (a) and transverse (b) regions.
   The error bars depict combined statistical and systematic uncertainties.}
  \label{fig:1DcptsumAB}
\end{figure*}

\begin{figure*}[tbp]
  \centering
  \vspace{0.5cm}
   \resultsimg{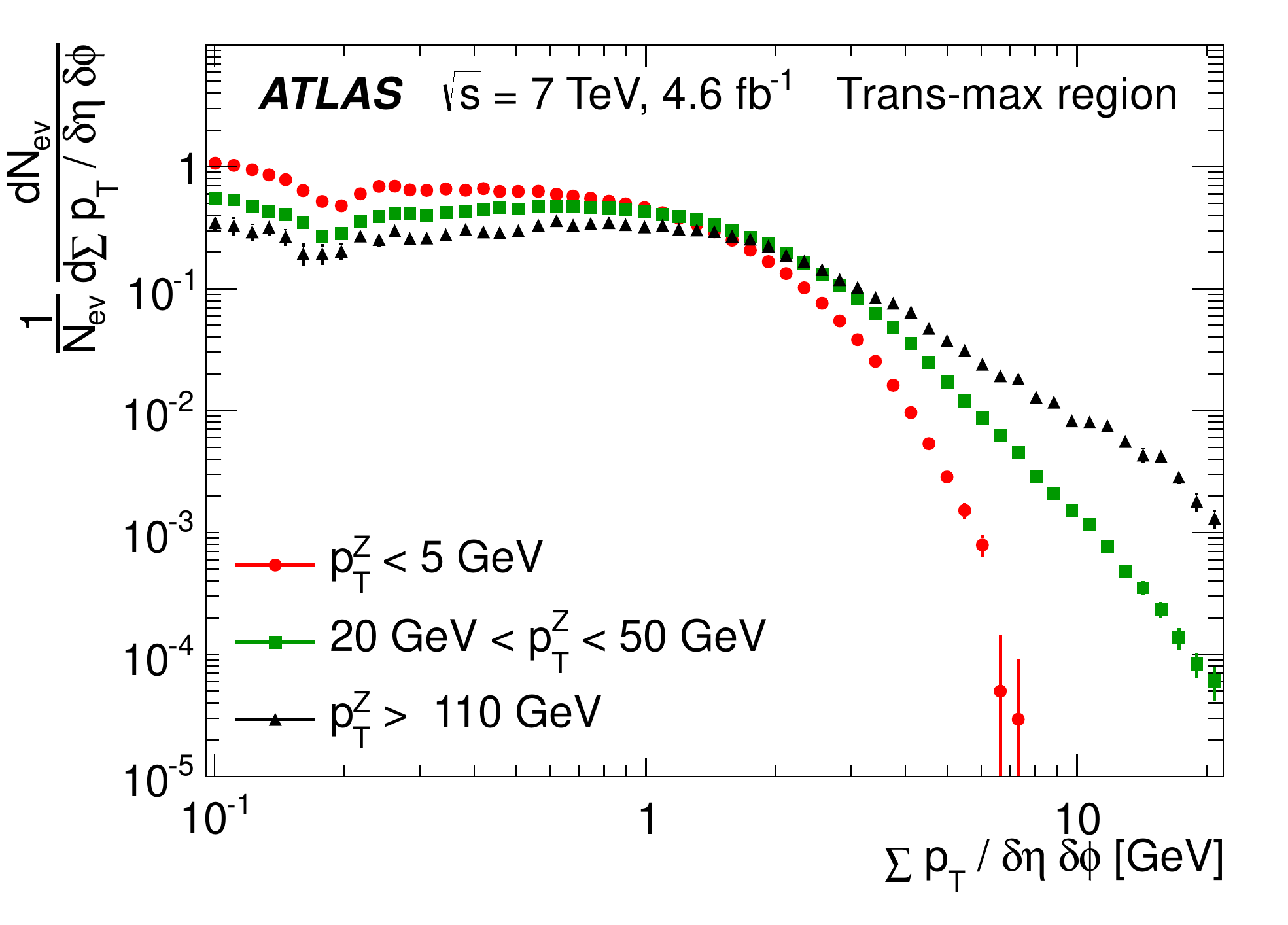}{fig:max-ptsum}\\
   \resultsimg{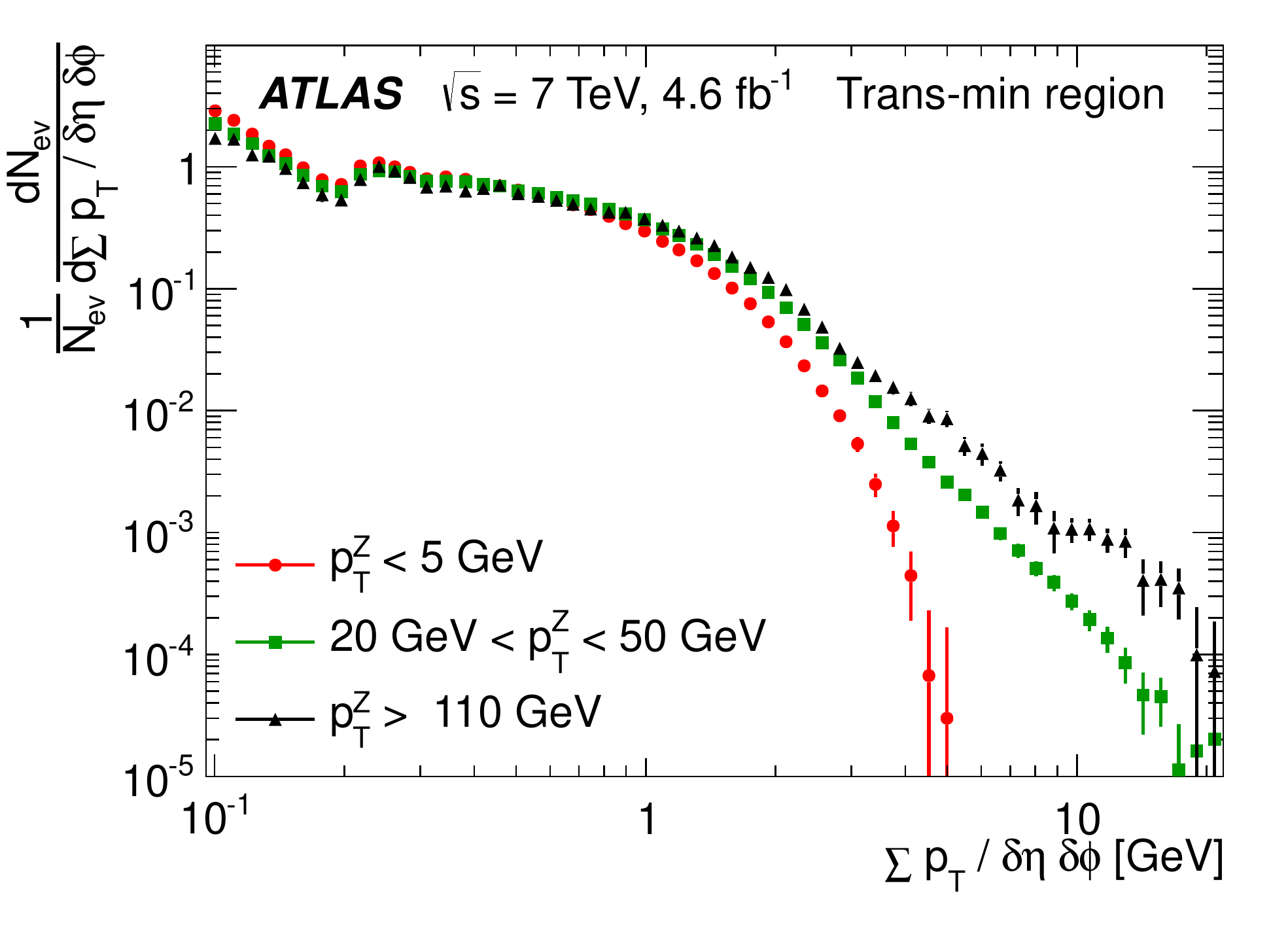}{fig:min-ptsum}
   \caption{Distributions of the scalar \pT\ sum density of charged particles, \pTsumdetadphi , in three different
  \Z transverse momentum, \ZpT , intervals, in the trans-max (a) and trans-min (b) regions.
   The error bars depict combined statistical and systematic uncertainties.}
  \label{fig:1DcptsumCD}
\end{figure*}

The distributions of the charged-particle \pTsumdetadphi and \Nchgdetadphi  in intervals of \ZpT\
show the dependence of the event activity on the hard scale.
The distributions of \pTsumdetadphi in three different \ZpT ranges 
are shown in~\FigRef{fig:1DcptsumAB} and in~\FigRef{fig:1DcptsumCD}.
At values below \pTsumdetadphi of $0.1$~GeV, the
distributions exhibit a decrease, which is independent of \ZpT.
This is followed by a sharp increase at higher \pTsumdetadphi ,
which is an artifact 
of requiring at least two tracks with \pT\ of at least $0.5$~GeV in every event.
Then a broad distribution can be seen extending to \pTsumdetadphi of about $1$~GeV, followed by a
steep decrease, the rate of which depends on the \ZpT\ interval.
For lower \ZpT\ values, the decrease is faster.
These features are fairly independent of the UE regions, with the exception of 
the trans-min region, in which the \pTsumdetadphi distribution
is approximately independent of \ZpT\ up to \pTsumdetadphi of $1$~GeV.
If there were no hard scattering contributions in the trans-min region and the remaining underlying event activity were
independent of the hard scattering scale then this \ZpT\ independence of the \pTsumdetadphi 
distribution would be expected~\cite{Frankfurt:2010ea}.

\begin{figure*}[tbp]
  \centering
  \vspace{0.5cm}
   \resultsimg{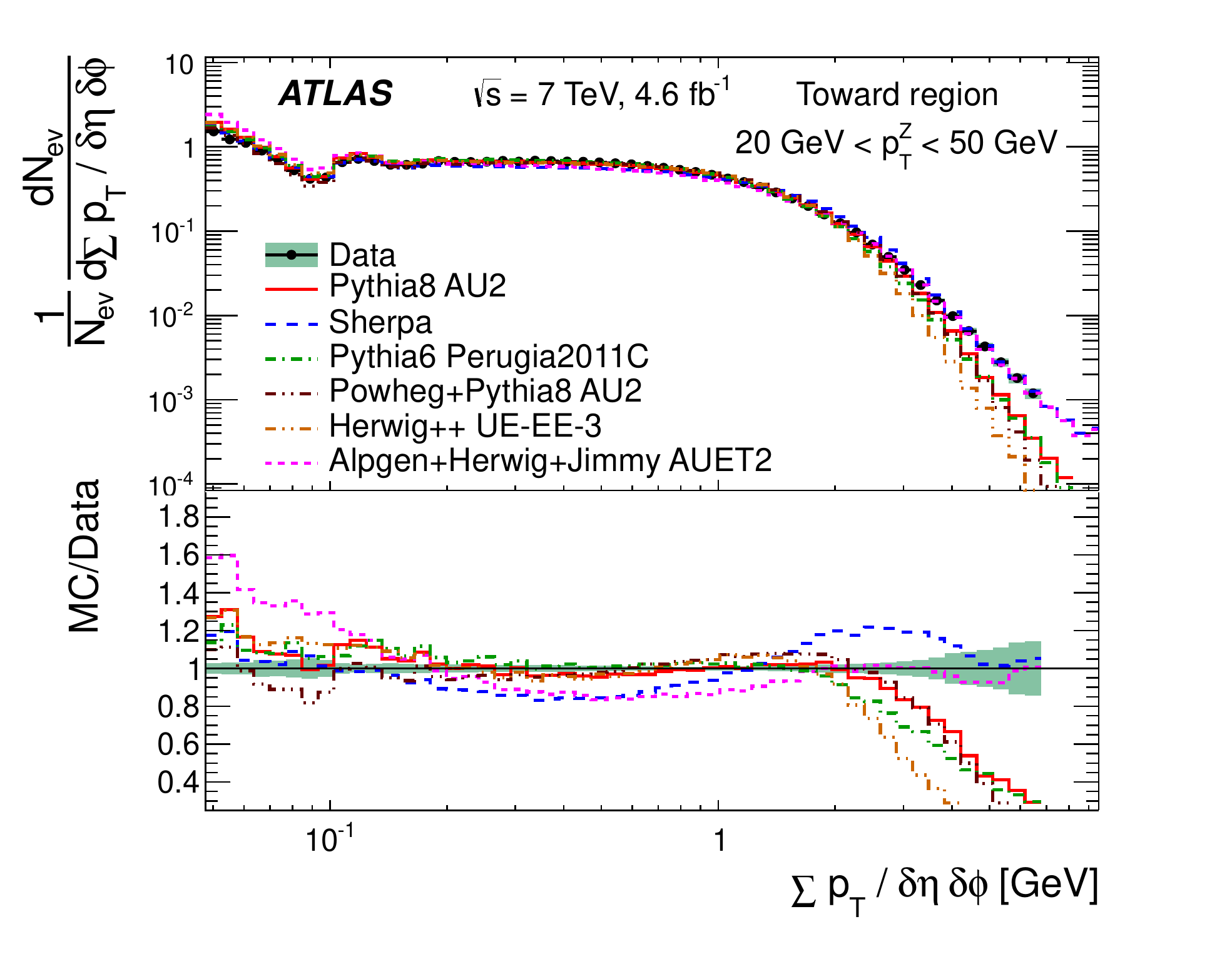}{fig:tow-ptsum-mc}\\
   \resultsimg{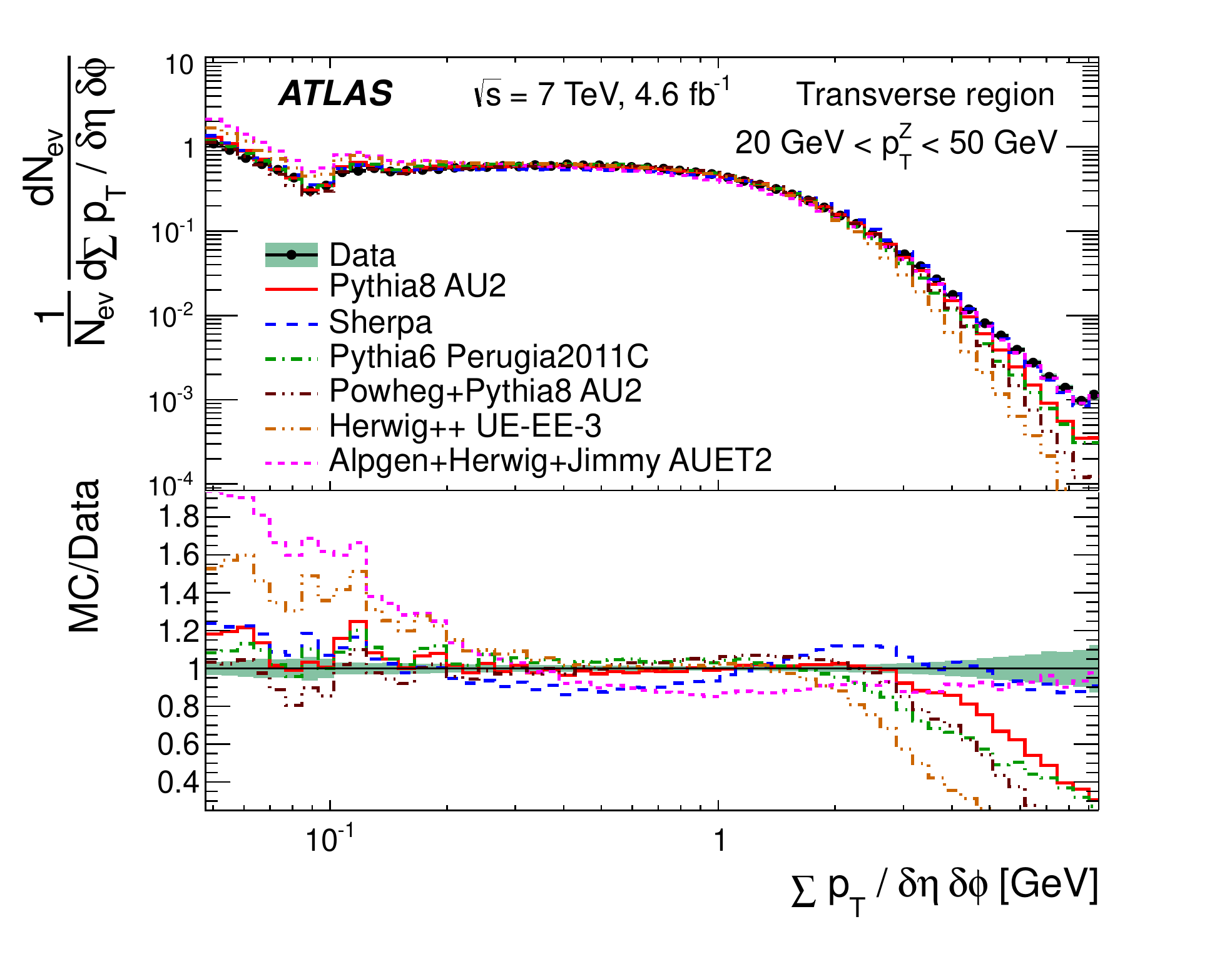}{fig:trv-ptsum-mc}
   \caption{Comparisons of data and MC predictions for the scalar \pT\ sum  density of charged particles, \pTsumdetadphi , for 
  \Z transverse momentum, \ZpT , in the interval $20-50~\GeV$, in the 
   the toward (a) and transverse (b) regions.  
   The bottom panels in each plot show the ratio of MC predictions to data.
   The shaded bands represent the combined statistical and systematic uncertainties, while
   the error bars show the statistical uncertainties.}
  \label{fig:mc-1DcptsumAB}
\end{figure*}

\begin{figure*}[tbp]
  \centering
  \vspace{0.5cm}
   \resultsimg{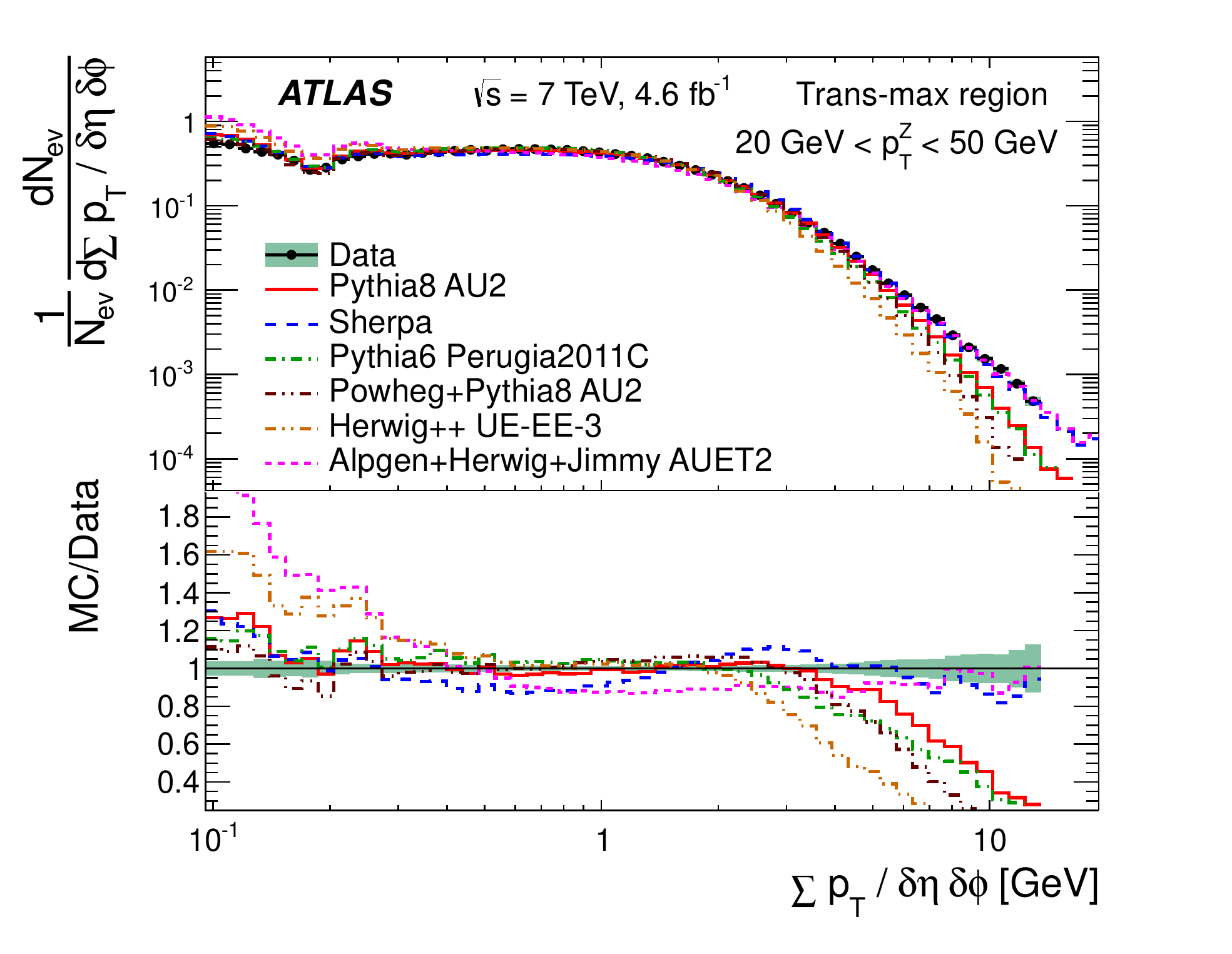}{fig:min-ptsum-mc} \\
   \resultsimg{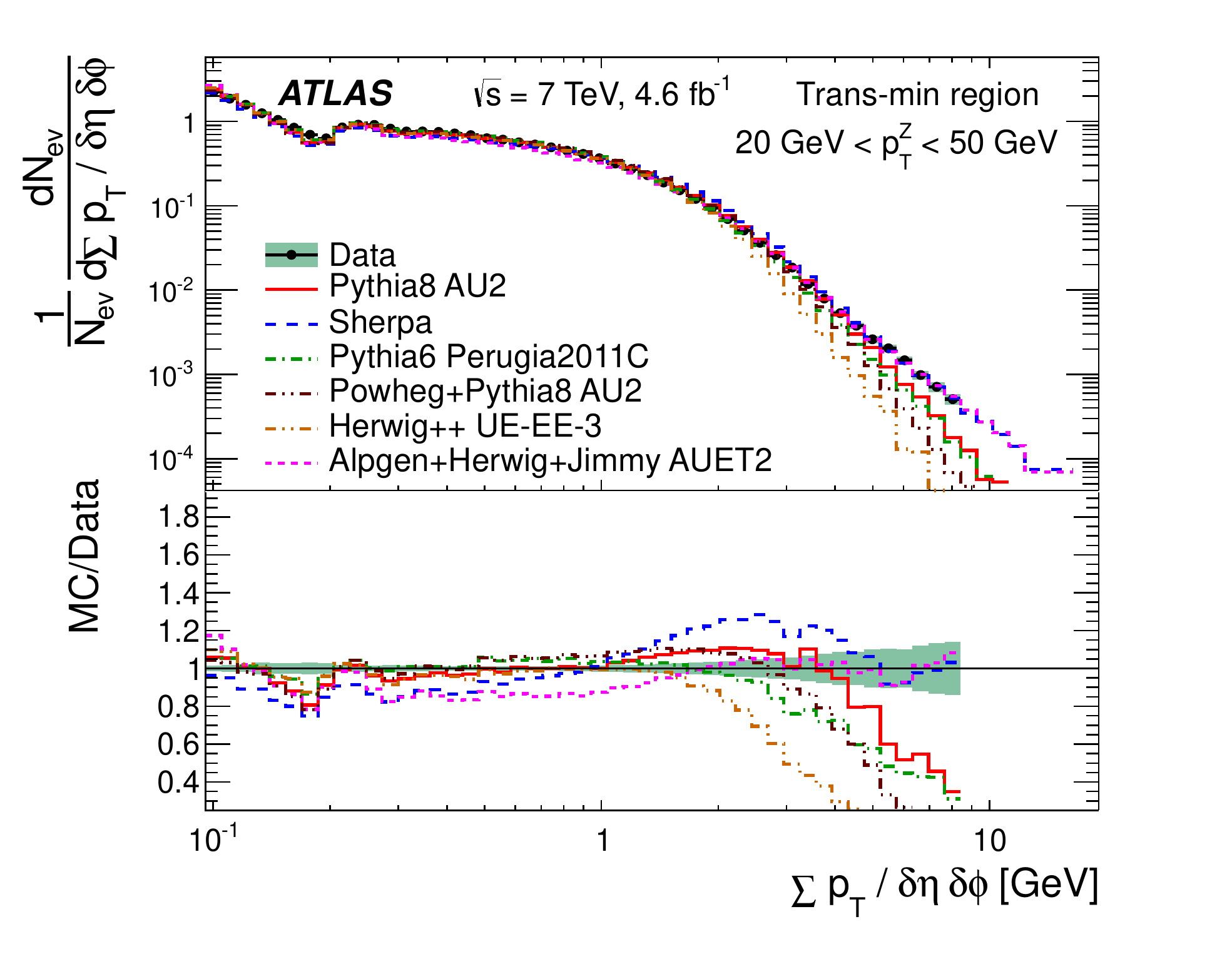}{fig:max-ptsum-mc}
   \caption{Comparisons of data and MC predictions for the scalar \pT\ sum density of charged particles, \pTsumdetadphi , for 
  \Z transverse momentum, \ZpT , in the interval $20-50~\GeV$, in the trans-max (a) and trans-min (b) regions.
   The bottom panels in each plot show the ratio of MC predictions to data.
   The shaded bands represent the combined statistical and systematic uncertainties, while
   the error bars show the statistical uncertainties.}
  \label{fig:mc-1DcptsumCD}
\end{figure*}

In~\FigRef{fig:mc-1DcptsumAB} and~\FigRef{fig:mc-1DcptsumCD}, for a selected interval of \ZpT , between $20 - 50$~GeV, 
the \pTsumdetadphi distributions in all the UE regions
are compared to various MC model predictions (as described in~\TabRef{tab:mc}).
For $\pTsumdetadphi  < 0.1$~GeV, there is a large spread in the predictions of the MC models
relative to the data, with \powheg providing the best description.
The intermediate region with $0.1 < \pTsumdetadphi < 1$~GeV, is well reproduced by most of the MC models.
For the higher \pTsumdetadphi ranges, most of the MC models underestimate 
the number of events, with the exception of \sherpa and \alpgen, 
which have previously been shown to provide good models of 
multi-jet produced in association with a \Z~\cite{Aad:2013ysa}.
This observation may indicate that even the 
trans-min region is not free of additional jets coming from the hard scatter.

\begin{figure*}[tbp]
  \centering
  \vspace{0.5cm}
   \resultsimg{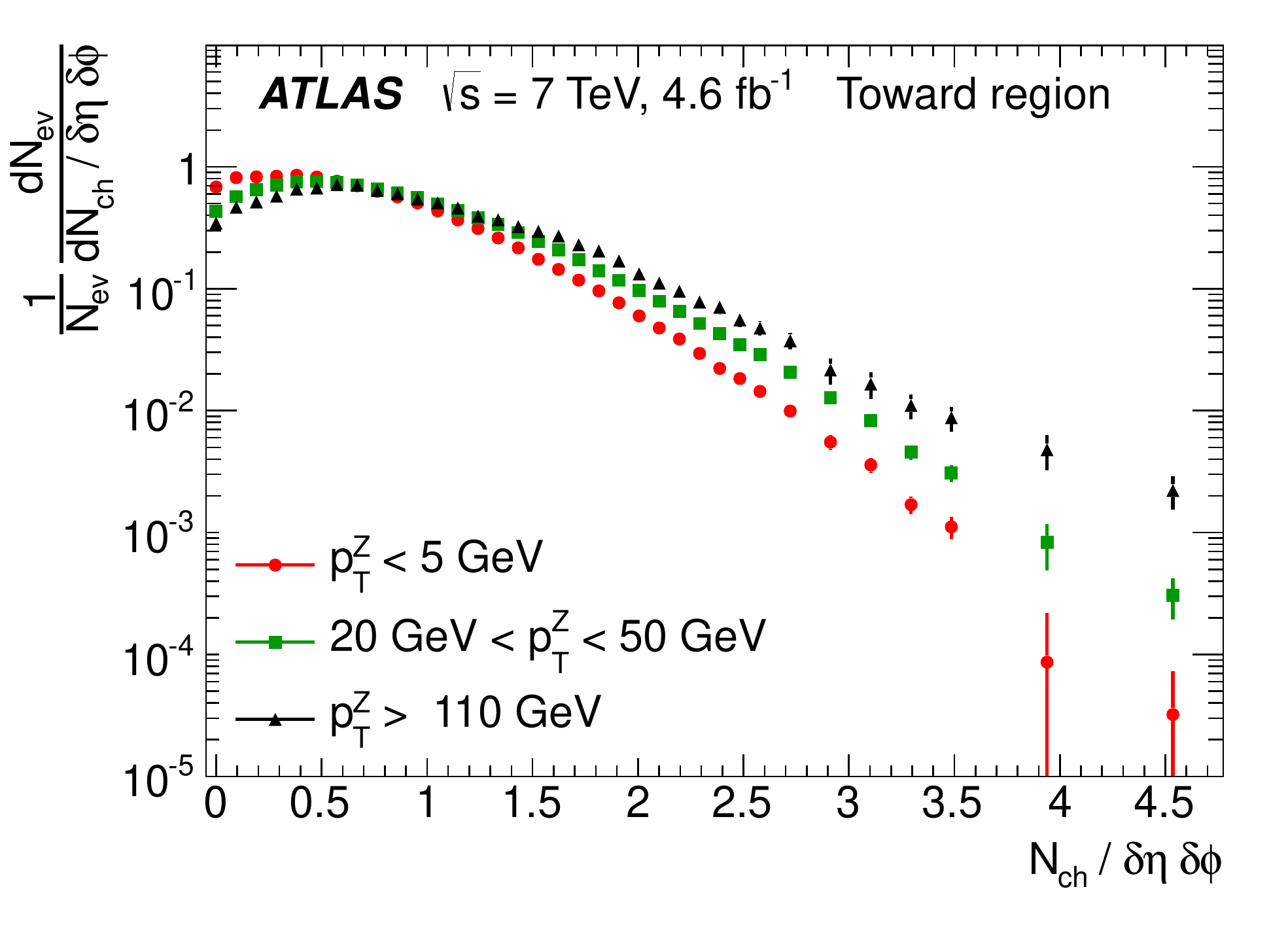}{fig:tow-nchg}\\
   \resultsimg{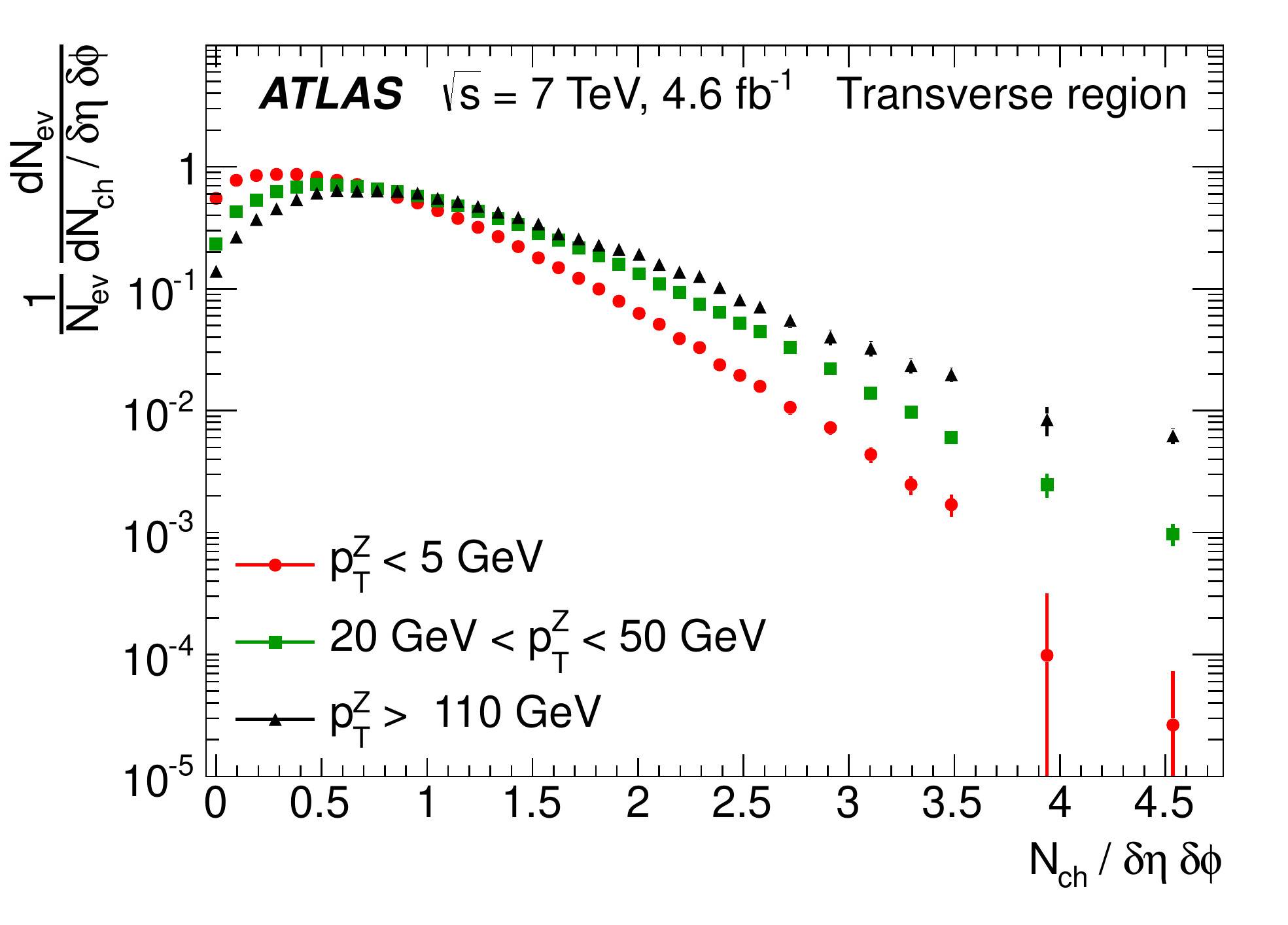}{fig:trv-nchg}
   \caption{Distributions of charged particle multiplicity density, \Nchgdetadphi , in three different
  \Z transverse momentum, \ZpT , intervals, in the toward (a) and transverse (b) regions. 
 The error bars depict combined statistical and systematic uncertainties.}
   \label{fig:1DnchgAB}
\end{figure*}

\begin{figure*}[tbp]
  \centering
  \vspace{0.5cm}
   \resultsimg{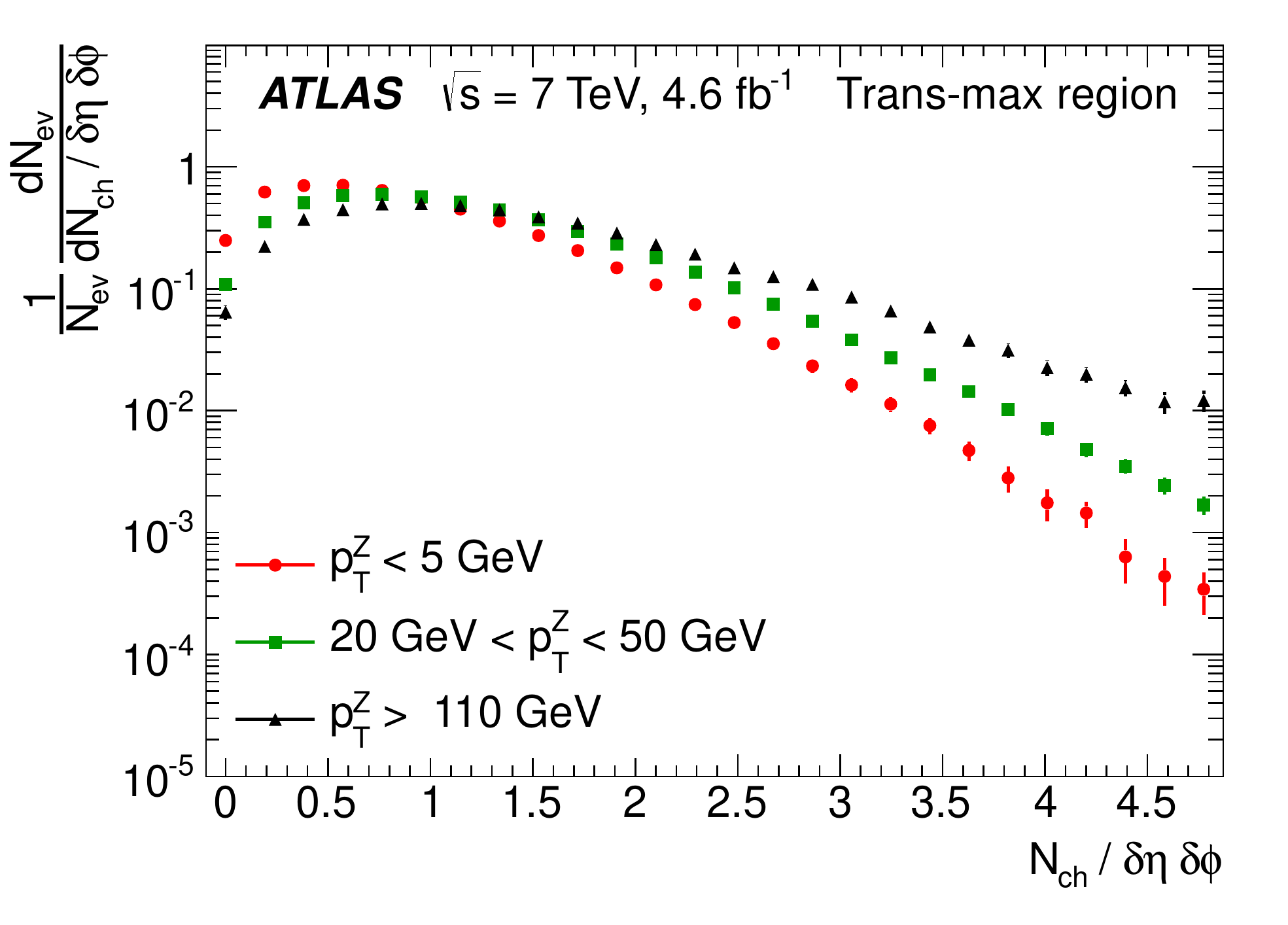}{fig:min-nchg}\\
   \resultsimg{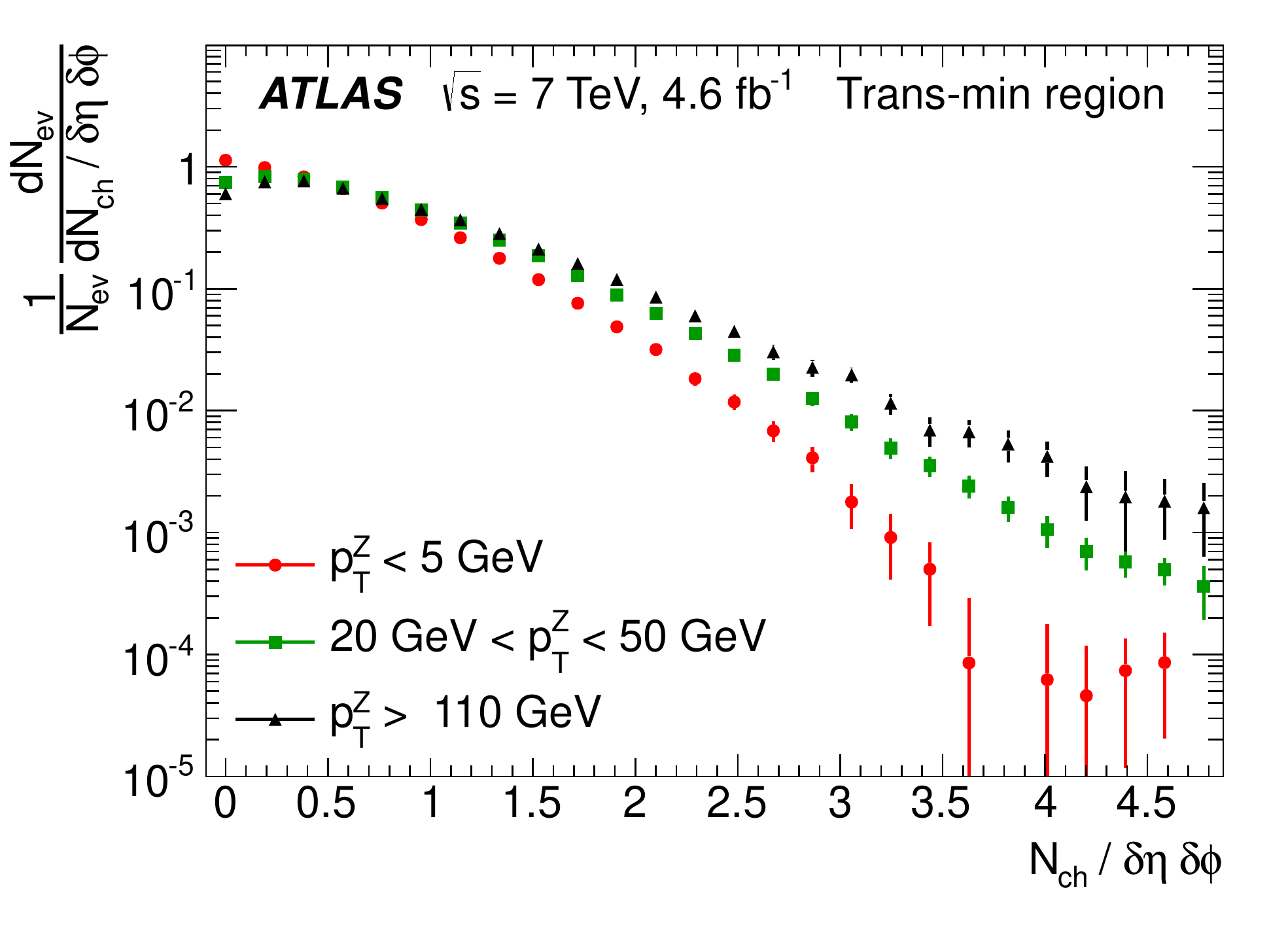}{fig:max-nchg}
   \caption{Distributions of charged particle multiplicity density, \Nchgdetadphi , in three different
  \Z transverse momentum, \ZpT , intervals, in the trans-max (a) and trans-min (b) regions.
 The error bars depict combined statistical and systematic uncertainties.}
   \label{fig:1DnchgCD}
\end{figure*}

\begin{figure*}[tbp]
  \centering
  \vspace{0.5cm}
   \resultsimg{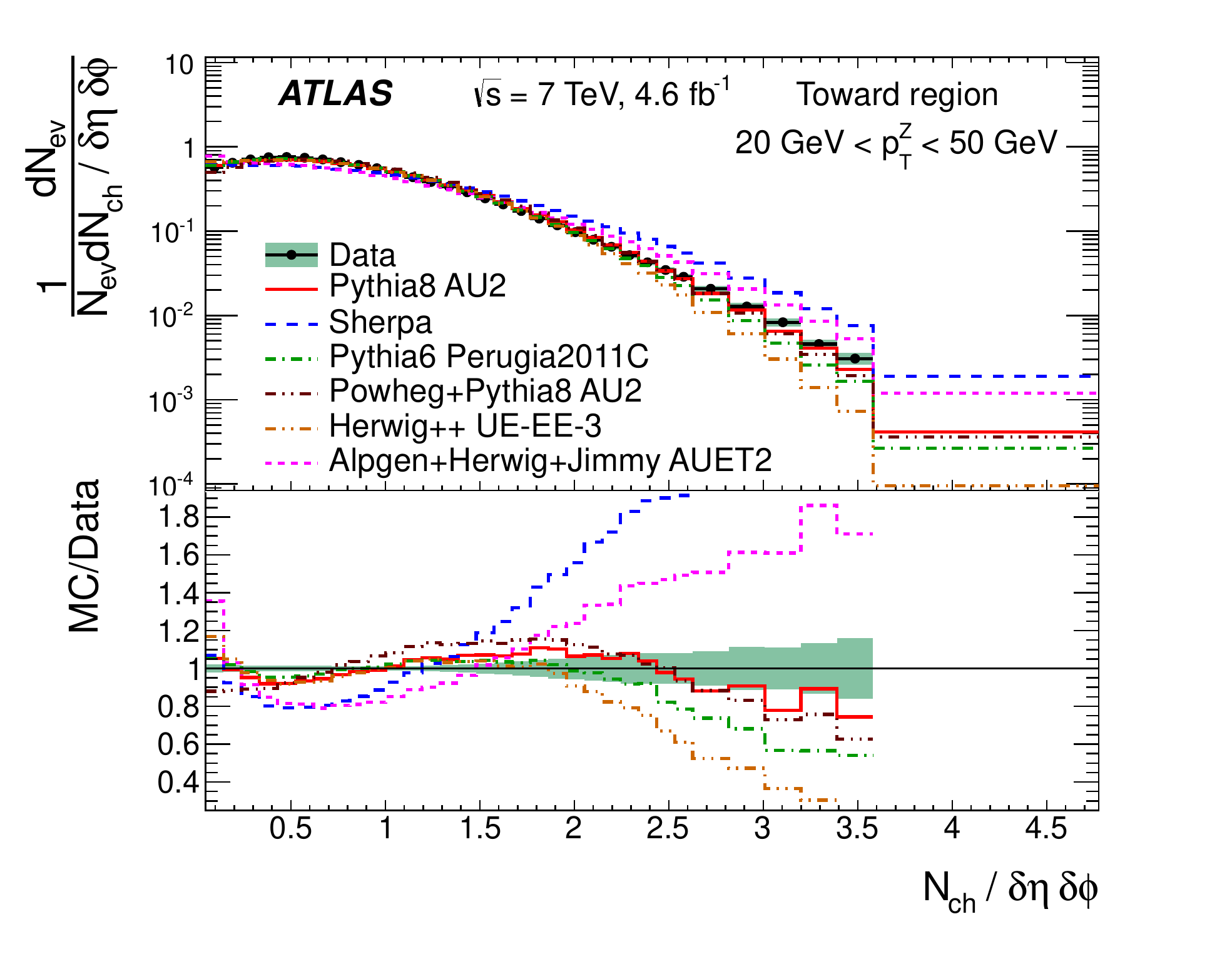}{fig:tow-nchg-mc}\\
   \resultsimg{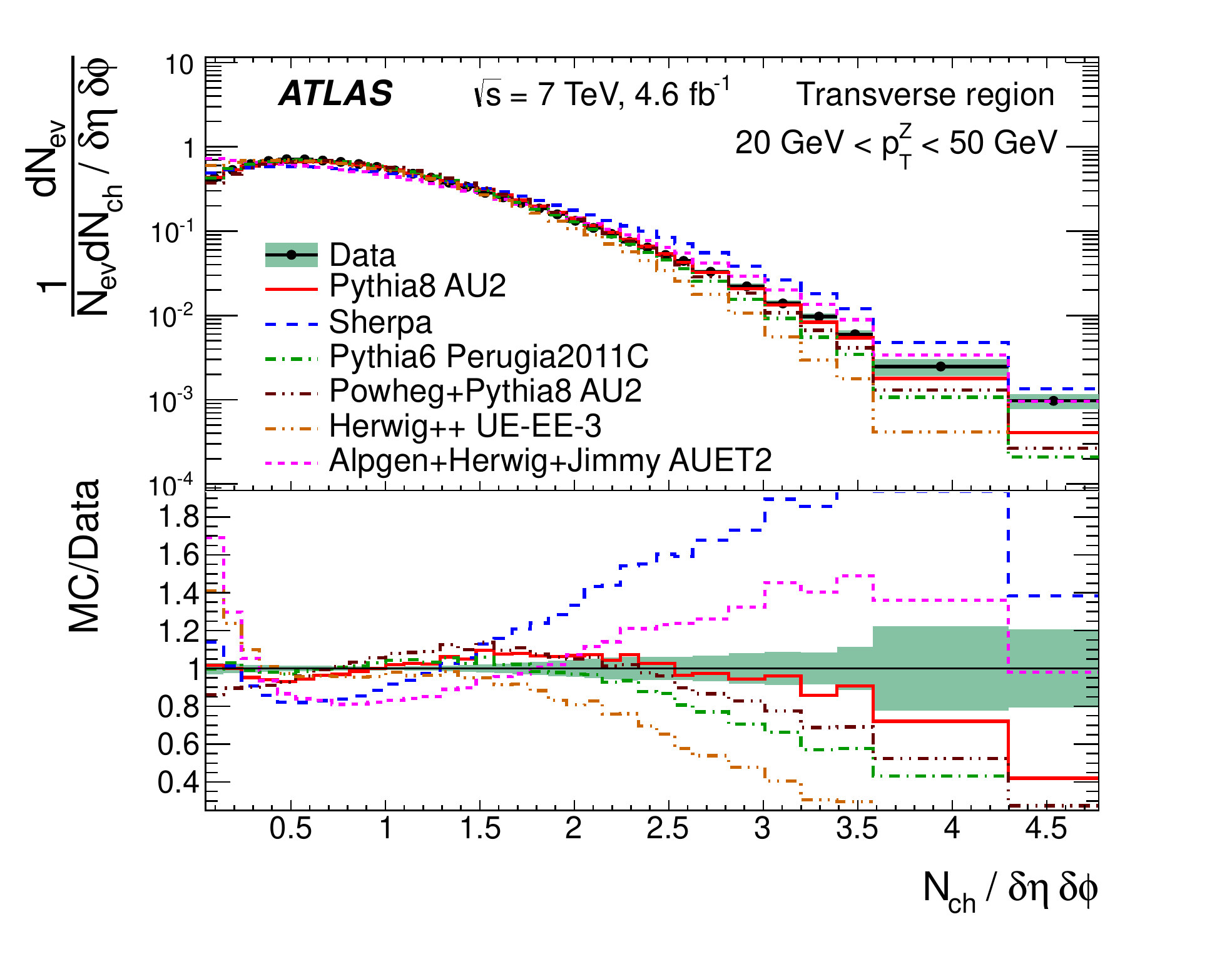}{fig:trv-nchg-mc}
   \caption{Comparisons of data and MC predictions for charged particle multiplicity density, \Nchgdetadphi , for 
    \Z transverse momentum, \ZpT , in the interval $20-50~\GeV$, in the toward (a) and transverse (b) regions. 
    The bottom panels in each plot show the ratio of MC predictions to data.
   The shaded bands represent the combined statistical and systematic uncertainties, while
   the error bars show the statistical uncertainties.}
  \label{fig:mc-1DnchgAB}
\end{figure*}

\begin{figure*}[tbp]
  \centering
  \vspace{0.5cm}
   \resultsimg{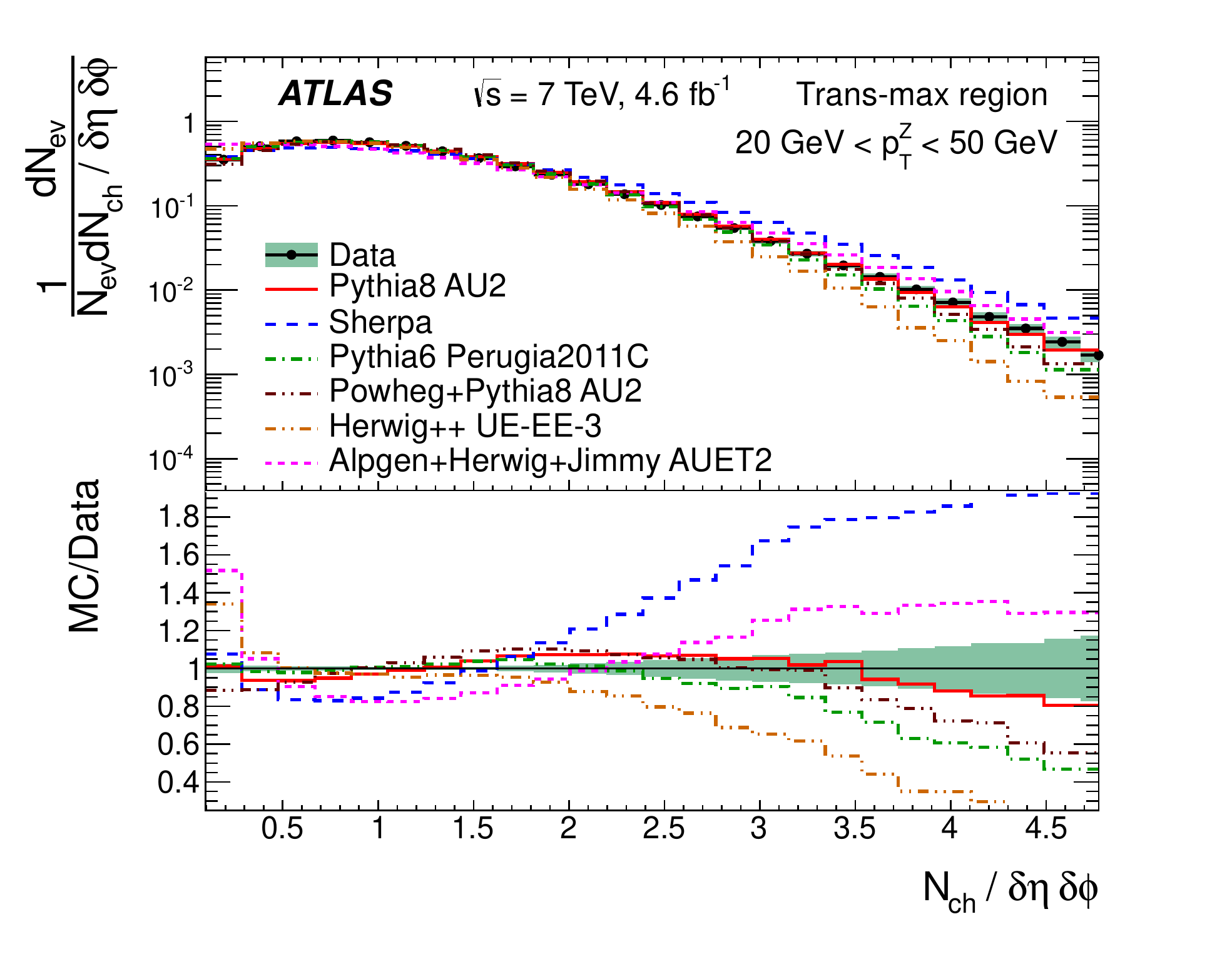}{fig:min-nchg-mc}\\
   \resultsimg{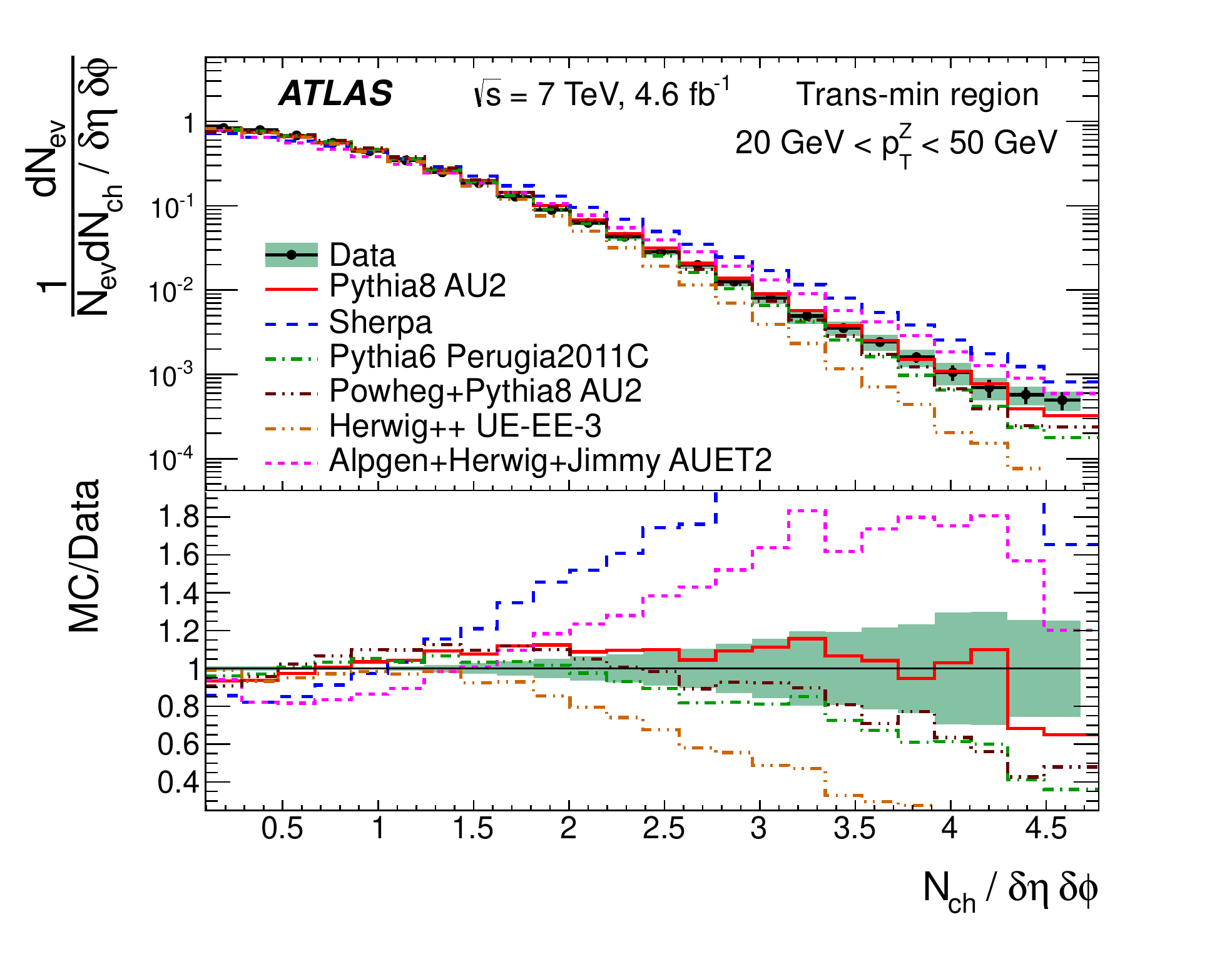}{fig:max-nchg-mc}
   \caption{Comparisons of data and MC predictions for charged particle multiplicity density, \Nchgdetadphi , for 
    \Z transverse momentum, \ZpT , in the interval $20-50~\GeV$, in the trans-max (a) and trans-min (b) regions.
    The bottom panels in each plot show the ratio of MC predictions to data.
   The shaded bands represent the combined statistical and systematic uncertainties, while
   the error bars show the statistical uncertainties.}
  \label{fig:mc-1DnchgCD}
\end{figure*}

The distributions of the charged particle multiplicity density in the four UE regions are
shown in~\FigRef{fig:1DnchgAB} and~\FigRef{fig:1DnchgCD} for the same \ZpT\ intervals used in~\FigRef{fig:1DcptsumAB}
and~\FigRef{fig:1DcptsumCD}, respectively.
The distributions in the transverse, toward and trans-max regions exhibit similar features, with the exception
of the largest multiplicities, which are suppressed in the trans-min region, compared to the trans-max one.
In the trans-min region, 
as for the \pTsumdetadphi distribution, limited dependence on \ZpT\ is observed at low multiplicity.
The suppression of large multiplicities in the trans-min region is more pronounced in the lower \ZpT\ intervals.
The comparison of these multiplicity distributions to various MC models, in the same \ZpT\ interval, between $20 - 50$~GeV,
is shown in~\FigRef{fig:mc-1DnchgAB} and~\FigRef{fig:mc-1DnchgCD} for all the UE regions. 
In contrast to the \pTsumdetadphi distributions, none 
of the MC models, except \pythiaeight , describes the data distributions, in particular for $\Nchgdetadphi  > 2$.

%%%%%%%%%%%%%%%%%%%%%%%%%%%%%%%%%%%%%%%%
\subsection{Average distributions}
\label{sec:pro}

The evolution of the event activity in the four UE regions with the hard scale can be conveniently summarised 
by the average value of the UE observables as a function of \ZpT.

In~\FigRef{fig:pro-ptsum1} the dependence of \dpTsumdetadphi on \ZpT\ is compared 
in different UE regions. The activity levels in the toward and transverse regions are 
both small compared to the activity in the away region.
This difference increases with increasing \ZpT .
The away region density is large due to the presence in most cases of a jet balancing the \Z in \pT . 
The density in the transverse region is seen to be systematically higher
than that in the toward region, which can be explained by the fact that
for high \ZpT\ , additional radiated jets balancing \ZpT affect the transverse
region more than the toward region~\cite{Aad:2013ysa}.
The difference between the three regions disappears at low \ZpT due to the fact that the UE regions
are not well defined with respect to the actual \Z direction.

In~\FigRef{fig:pro-ptsum2}, \dpTsumdetadphi is seen to 
rise much faster as a function of \ZpT\ in the trans-max region than in the trans-min region.
The slowing down of the rise of  \dpTsumdetadphi at high \ZpT\ in the most UE-sensitive toward and trans-min  
regions is consistent with an assumption~\cite{Azarkin:2014cja}
of a full
overlap between the two interacting protons in impact parameter space at high hard scales.

\begin{figure*}[tbp]
  \centering
  \vspace{0.5cm}
   \resultsimg{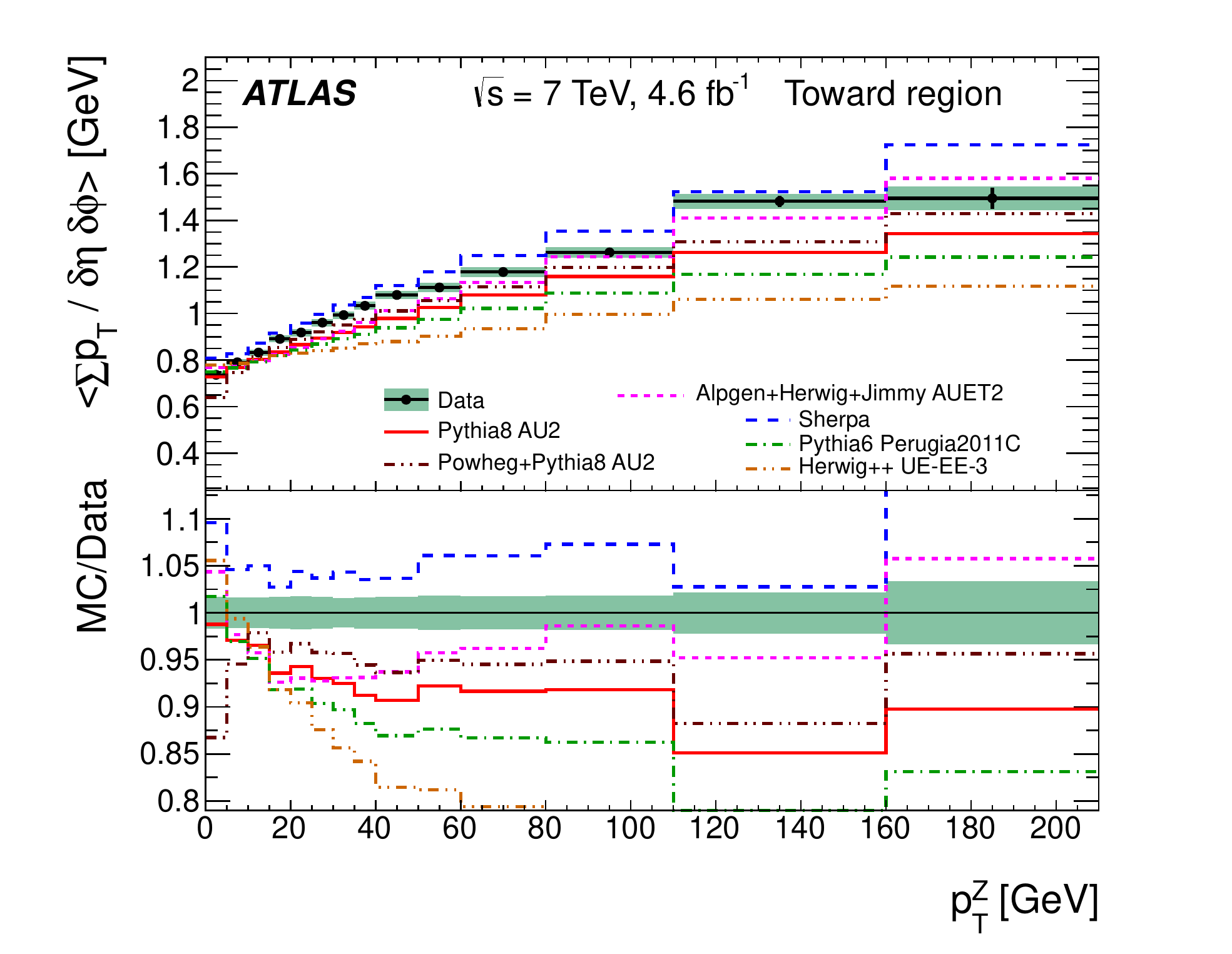}{fig:pro-tow-ptsum}\\
   \resultsimg{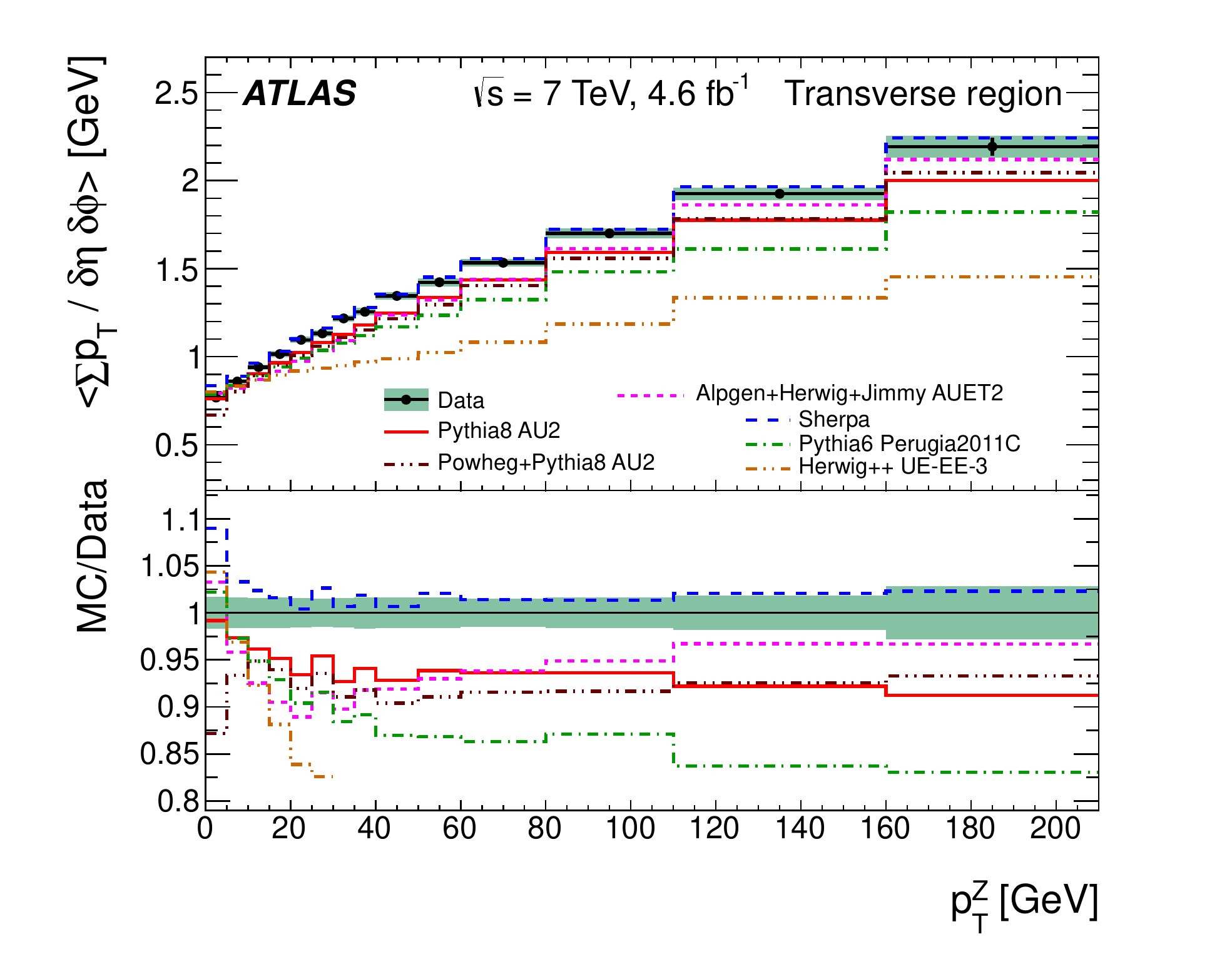}{fig:pro-trv-ptsum}
   \caption{Comparison of data and MC predictions for charged particle scalar \ptsum density average values, \dpTsumdetadphi ,
  as a function of \Z transverse momentum, \ZpT , in the toward (a) and transverse (b) regions. 
  The bottom panels in each plot show the ratio of MC predictions to data.
   The shaded bands represent the combined statistical and systematic uncertainties, while
   the error bars show the statistical uncertainties.}
  \label{fig:sumptDMAB}
\end{figure*}

\begin{figure*}[tbp]
  \centering
  \vspace{0.5cm}
   \resultsimg{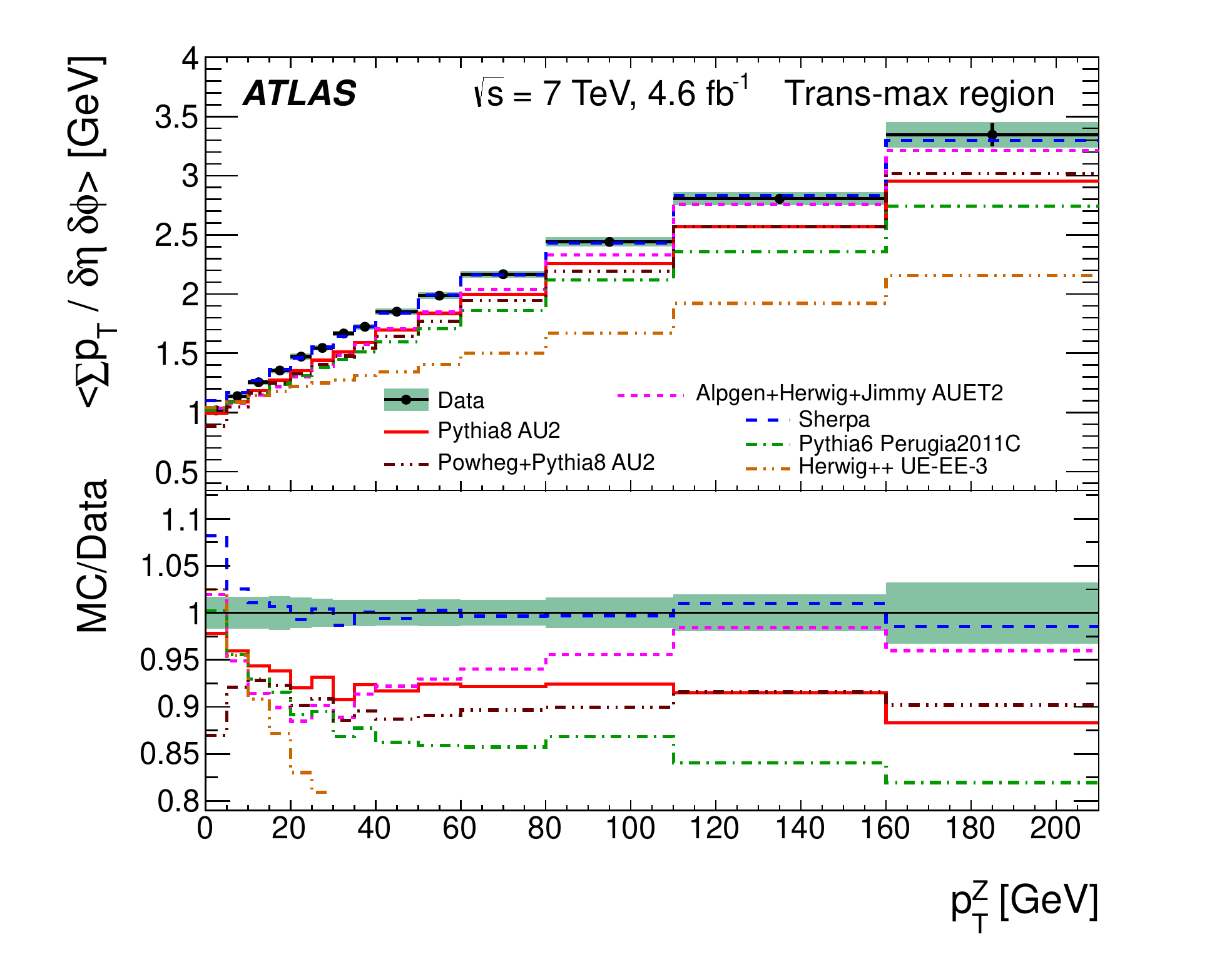}{fig:pro-max-ptsum}\\
   \resultsimg{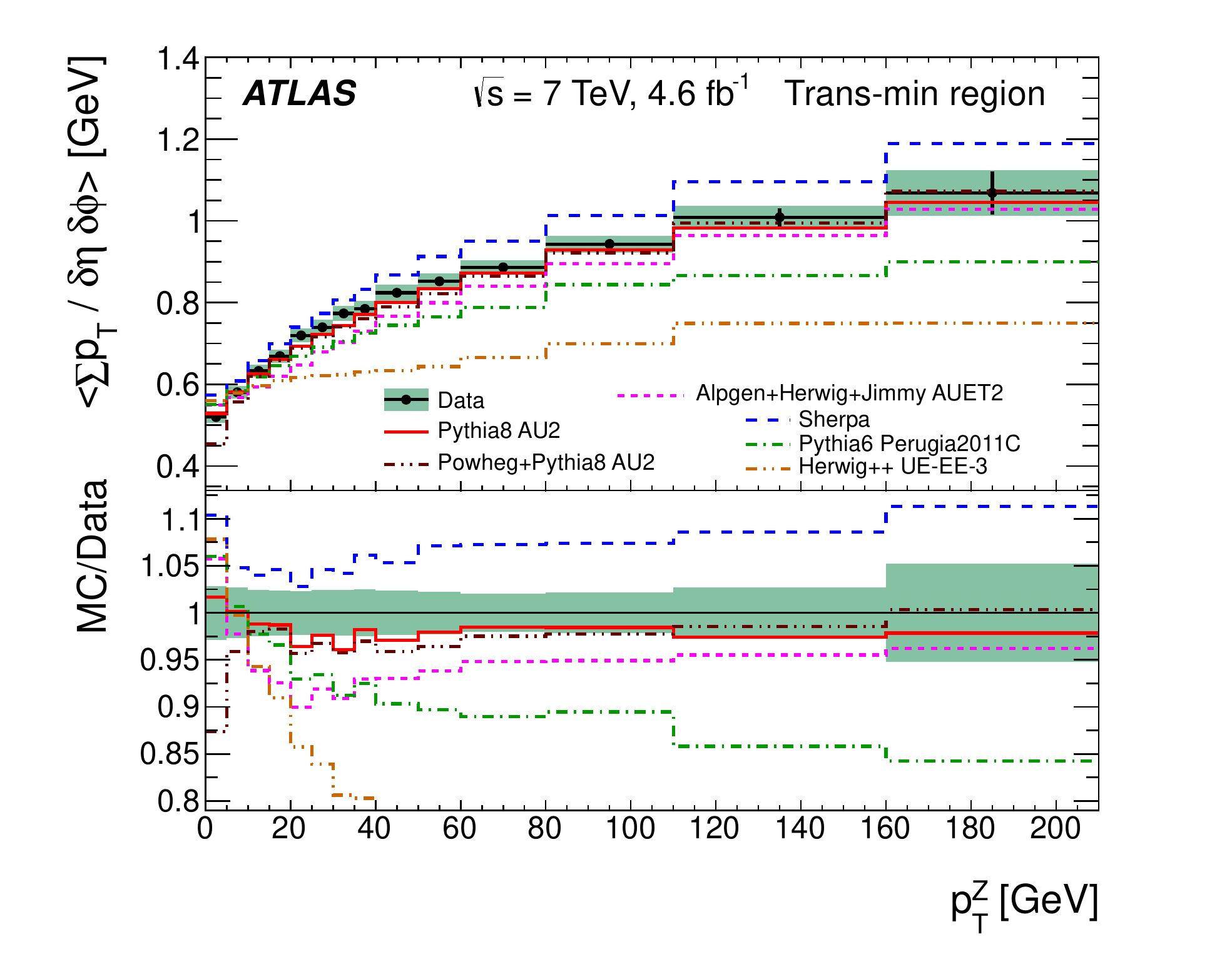}{fig:pro-min-ptsum}
   \caption{Comparison of data and MC predictions for charged particle scalar \ptsum density average values, \dpTsumdetadphi ,
  as a function of \Z transverse momentum, \ZpT , in the trans-max (a) and trans-min (b) regions. 
   The shaded bands represent the combined statistical and systematic uncertainties, while
   the error bars show the statistical uncertainties.}
  \label{fig:sumptDMCD}
\end{figure*}

The comparison of the \dpTsumdetadphi distribution as a function of \ZpT\ with the predictions
of various MC models is shown in~\FigRef{fig:sumptDMAB} and~\FigRef{fig:sumptDMCD} in the UE regions sensitive
to the underlying event characteristics.
For clarity of comparison, the statistically least significant $\ZpT > 210$~GeV bin is omitted.
The variation in the range of predictions is quite wide, although less 
so than for the differential \ptsum distributions. The best description of the transverse and trans-max regions
is given by \sherpa, followed by \pythiaeight, \alpgen and \powheg. The observation that
the multi-leg and NLO generator predictions are closer to the
data than most of the pure parton shower generators suggests that
these regions are affected by the additional jets coming from the hard interaction.
 Jet multiplicities in events with a \Z have been studied by the LHC 
 experiments~\cite{Aad:2013ysa}, and they are well described by \sherpa and \alpgen.

The discrepancy between the \pythiaeight AU2 tune and the \pythiasix Perugia tune
possibly indicates the effect of using LHC UE data for the former in addition to the shower model improvement.
In the trans-min region, which is the most sensitive to the UE, none of the models fully describe the data.
Apart from \herwigpp, and \sherpa, which predicts
a faster rise of \ptsum than observed in data, the other generators model the data better in the
trans-min region than they do in the transverse or trans-max regions.
This possibly indicates that in the LO shower generators the underlying event is well modelled but perturbative jet activity is not.

\begin{figure*}[tbp]
  \centering
  \vspace{0.5cm}
   \resultsimg{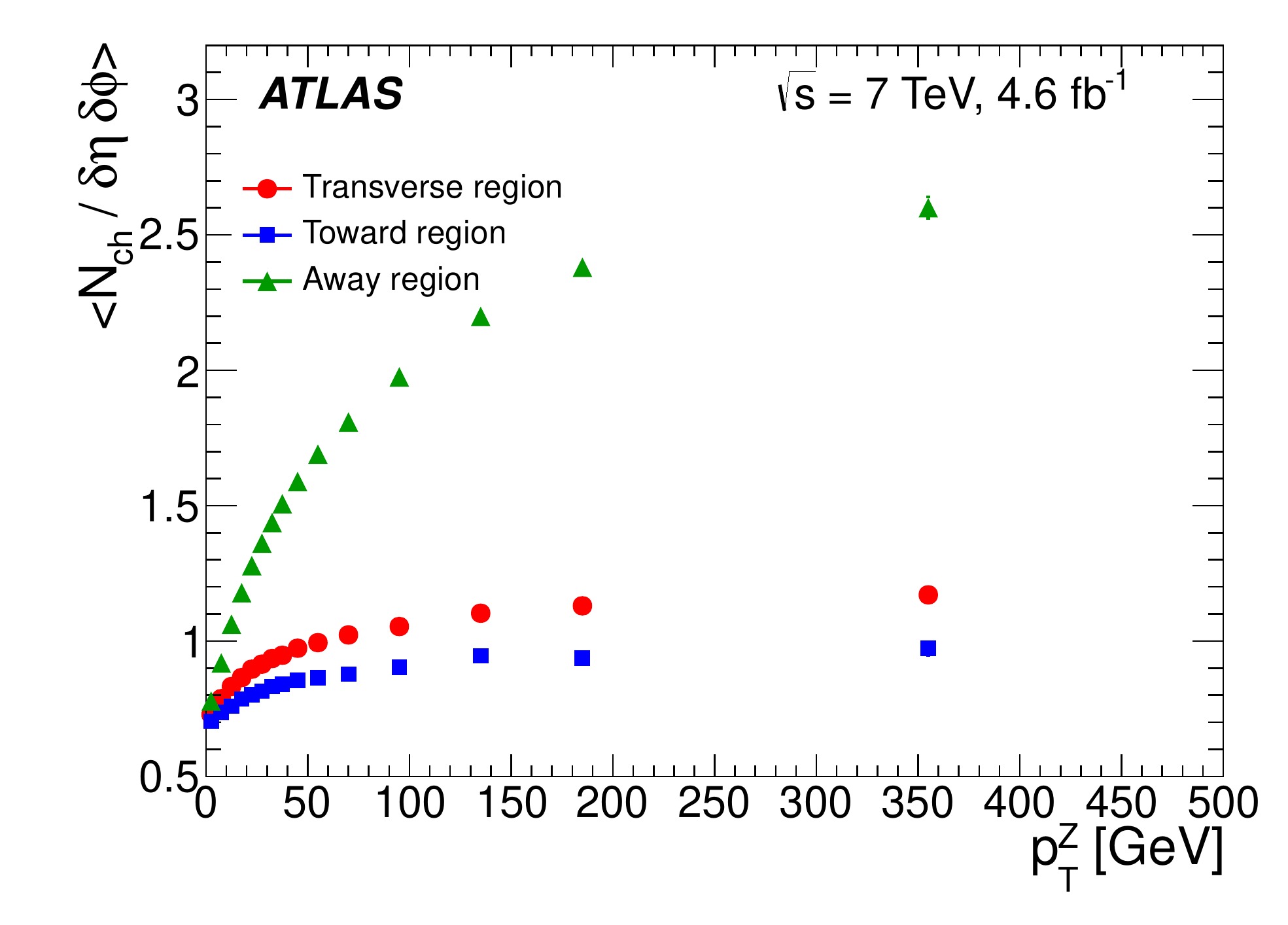}{fig:pro-nchg1}\\
   \resultsimg{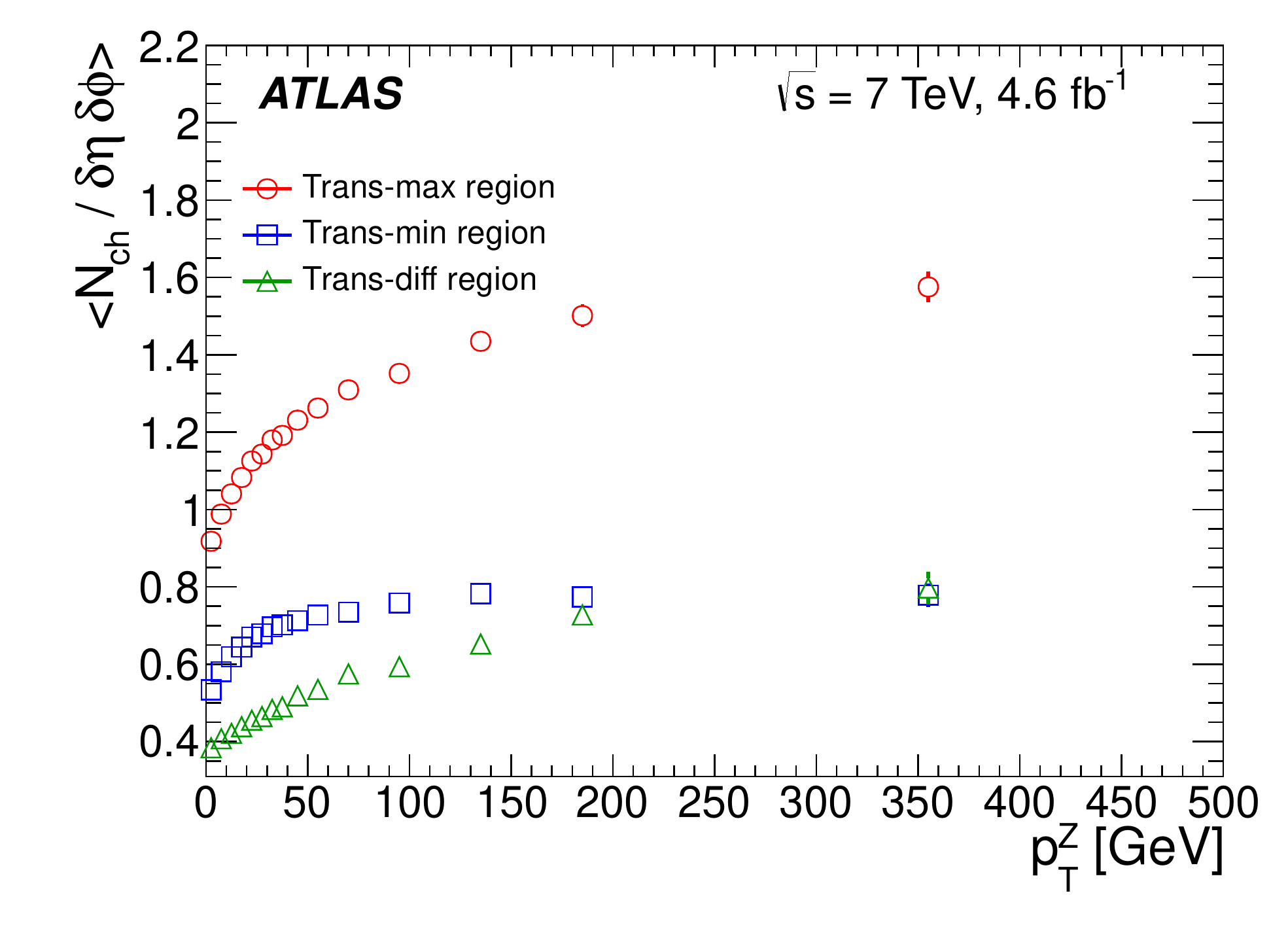}{fig:pro-nchg2}
   \caption{The average values of charged particle multiplicity density, \dNchgdetadphi , as a function of \Z transverse momentum, \ZpT ,
  in the transverse, toward and away regions (a), and in the trans-max, trans-min and trans-diff regions (b).
  The results are plotted at the center of each \ZpT\ bin.
  The error bars depict combined statistical and systematic uncertainties.}
  \label{fig:nchgD}
\end{figure*}

In~\FigRef{fig:nchgD}, \dNchgdetadphi is shown as a function of \ZpT\ 
 in the different UE regions.  The profiles
 behave in a similar way to \dpTsumdetadphi. However, the trans-diff \dNchgdetadphi activity is 
lower than that for trans-min, while for \dpTsumdetadphi ,
it is the other way around. This indicates that the trans-diff region, which is a measure
of extra activity in the trans-max region over the trans-min region, is populated
by a few particles with high transverse momentum, as expected for the leading constituents of jets.

\begin{figure*}[tbp]
  \centering
  \vspace{0.5cm}
   \resultsimg{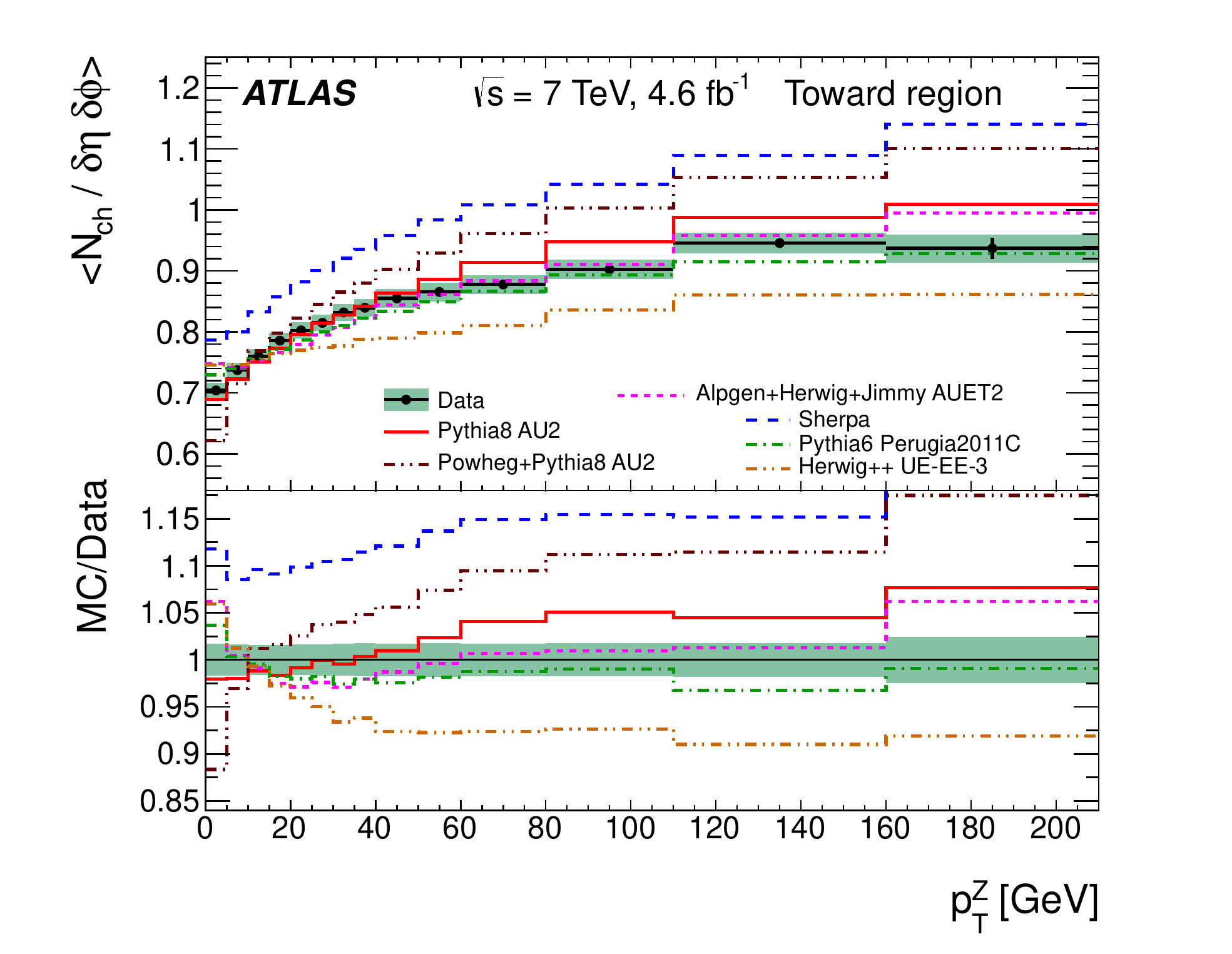}{fig:pro-tow-nchg}\\
   \resultsimg{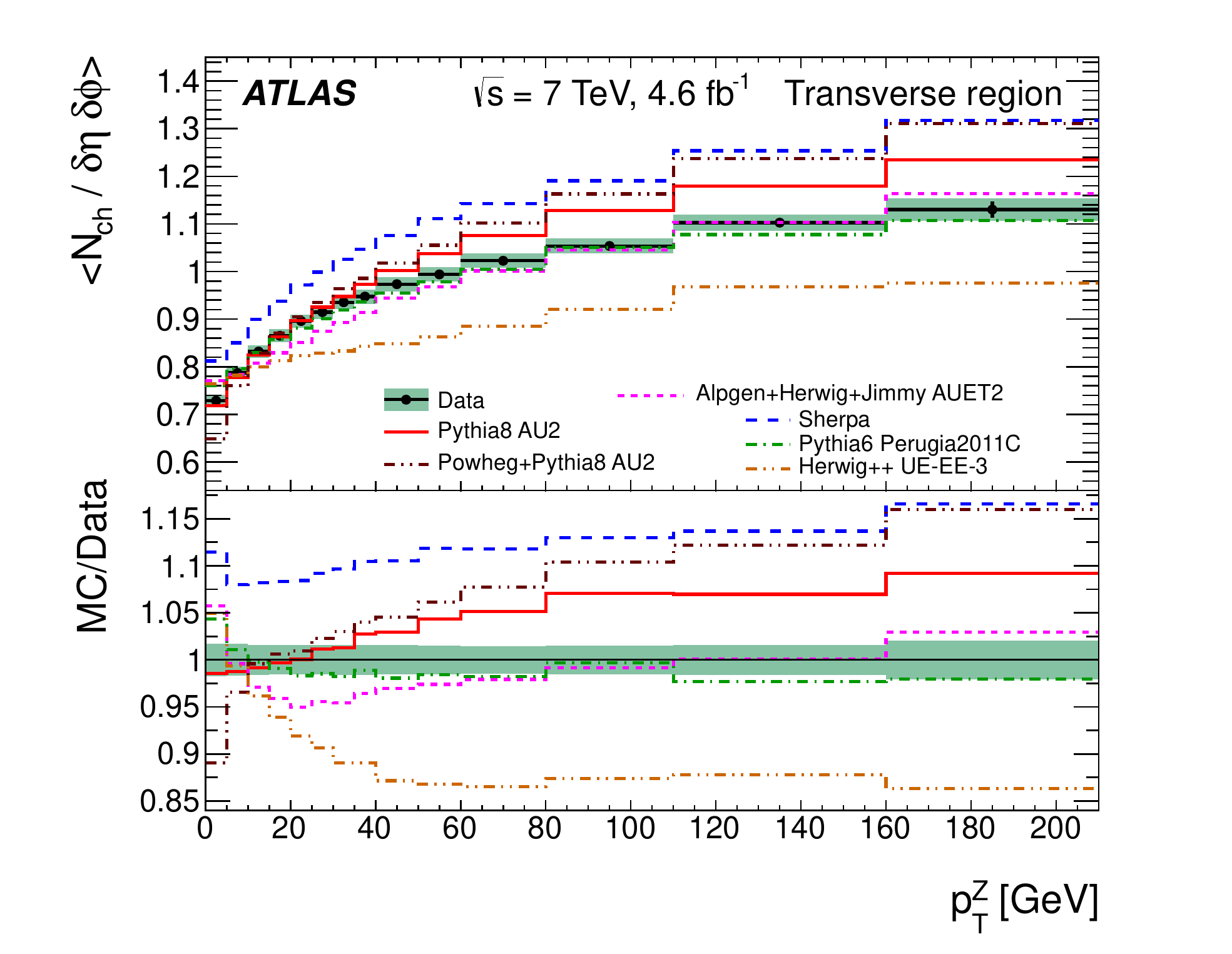}{fig:pro-trv-nchg}
   \caption{Comparison of data and MC predictions for charged particle multiplicity density average values, \dNchgdetadphi ,
  as a function of \Z transverse momentum, \ZpT , in the toward (a) and transverse (b) regions. 
  The bottom panels in each plot show the ratio of MC predictions to data.
   The shaded bands represent the combined statistical and systematic uncertainties, while
   the error bars show the statistical uncertainties.}
  \label{fig:nchgDMAB}
\end{figure*}

\begin{figure*}[tbp]
  \centering
  \vspace{0.5cm}
   \resultsimg{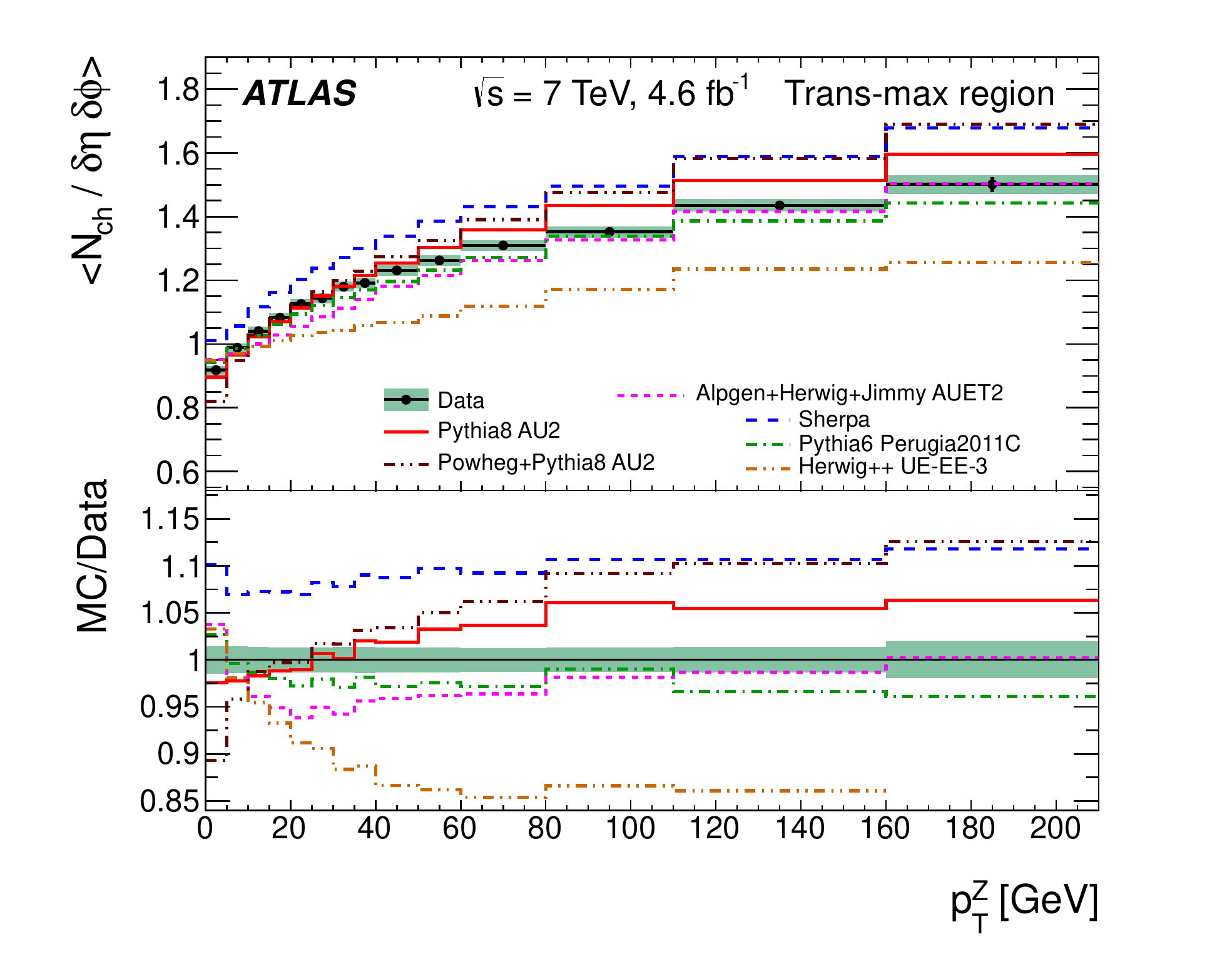}{fig:pro-max-nchg}\\
   \resultsimg{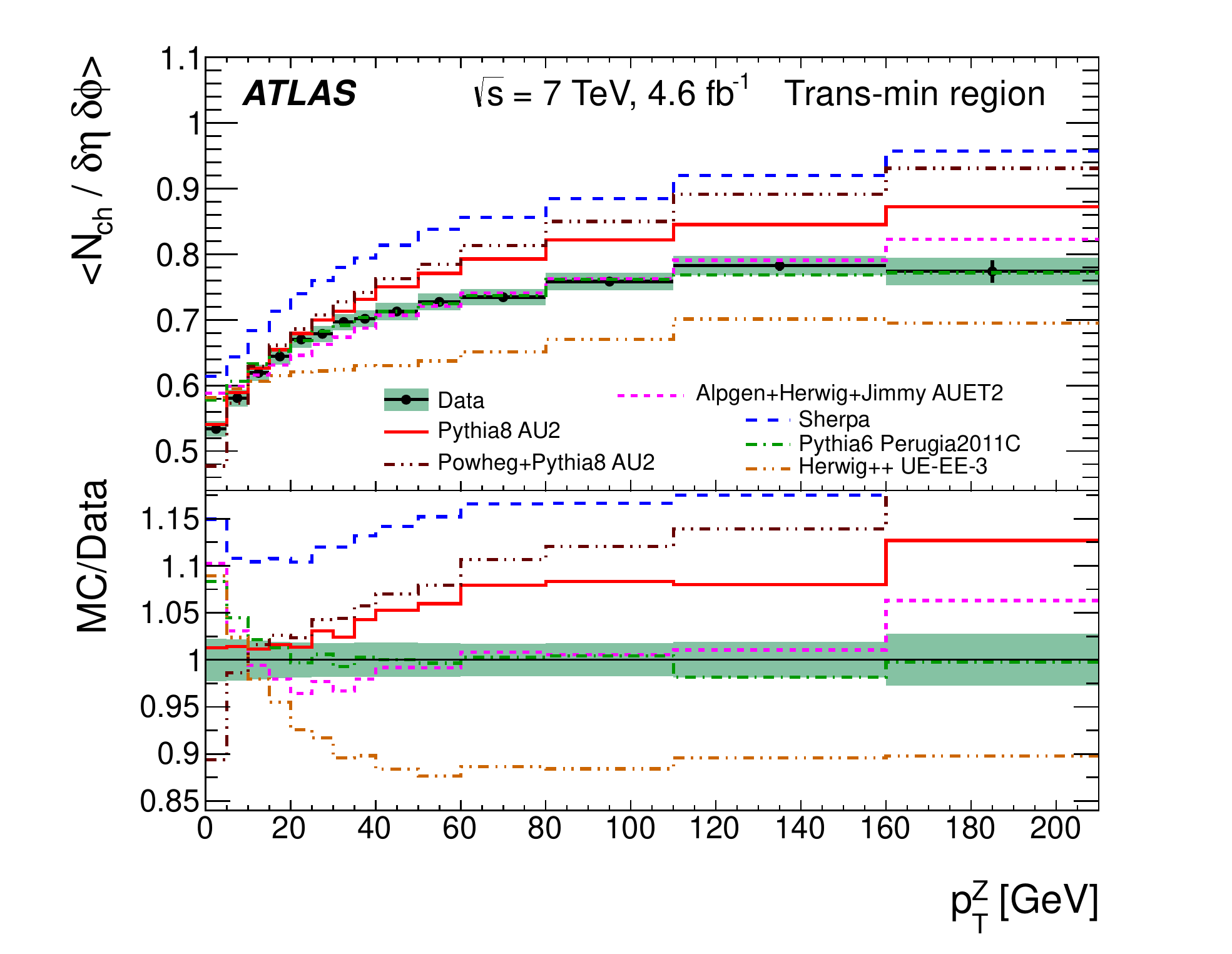}{fig:pro-min-nchg}
   \caption{Comparison of data and MC predictions for charged particle multiplicity density average values, \dNchgdetadphi ,
  as a function of \Z transverse momentum, \ZpT , in the trans-max (a) and trans-min (b) regions. 
  The bottom panels in each plot show the ratio of MC predictions to data.
   The shaded bands represent the combined statistical and systematic uncertainties, while
   the error bars show the statistical uncertainties.}
  \label{fig:nchgDMCD}
\end{figure*}

In~\FigRef{fig:nchgDMAB} and~\FigRef{fig:nchgDMCD}, in which various MC model predictions are compared 
to \dNchgdetadphi as a function of \ZpT , a different pattern
from that of
\dpTsumdetadphi is observed. 
The \pythiasix Perugia 2011C tune and \alpgen 
provide the closest predictions
 in all three regions.
\sherpa , \pythiaeight and
\powheg predict higher average multiplicities, with \sherpa being 
the farthest from the data. On the other hand, \herwigpp mostly underestimates the data.

\begin{figure*}[tbp]
  \centering
  \vspace{0.5cm}
   \resultsimg{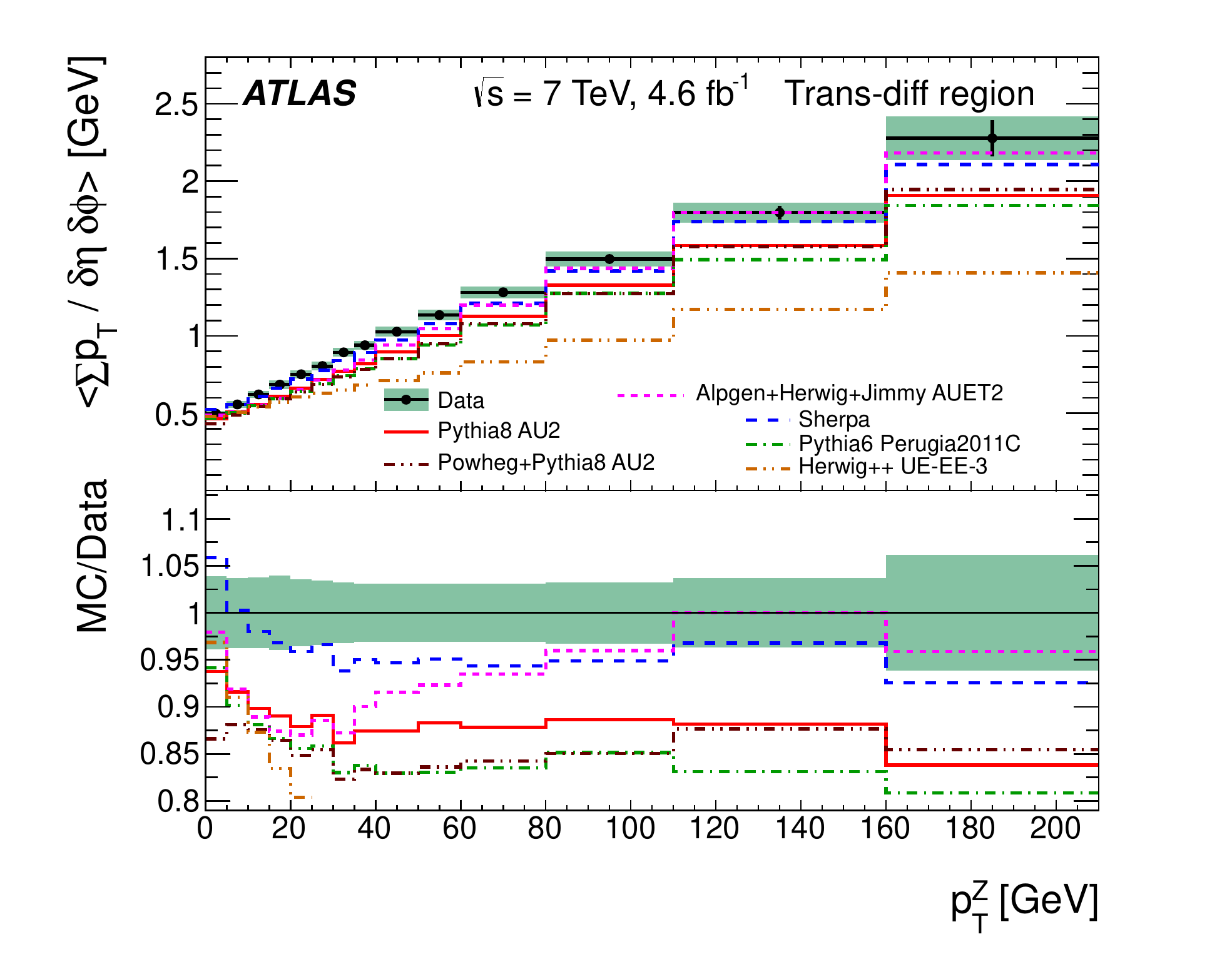}{fig:pro-dif-cptsum}\\
   \resultsimg{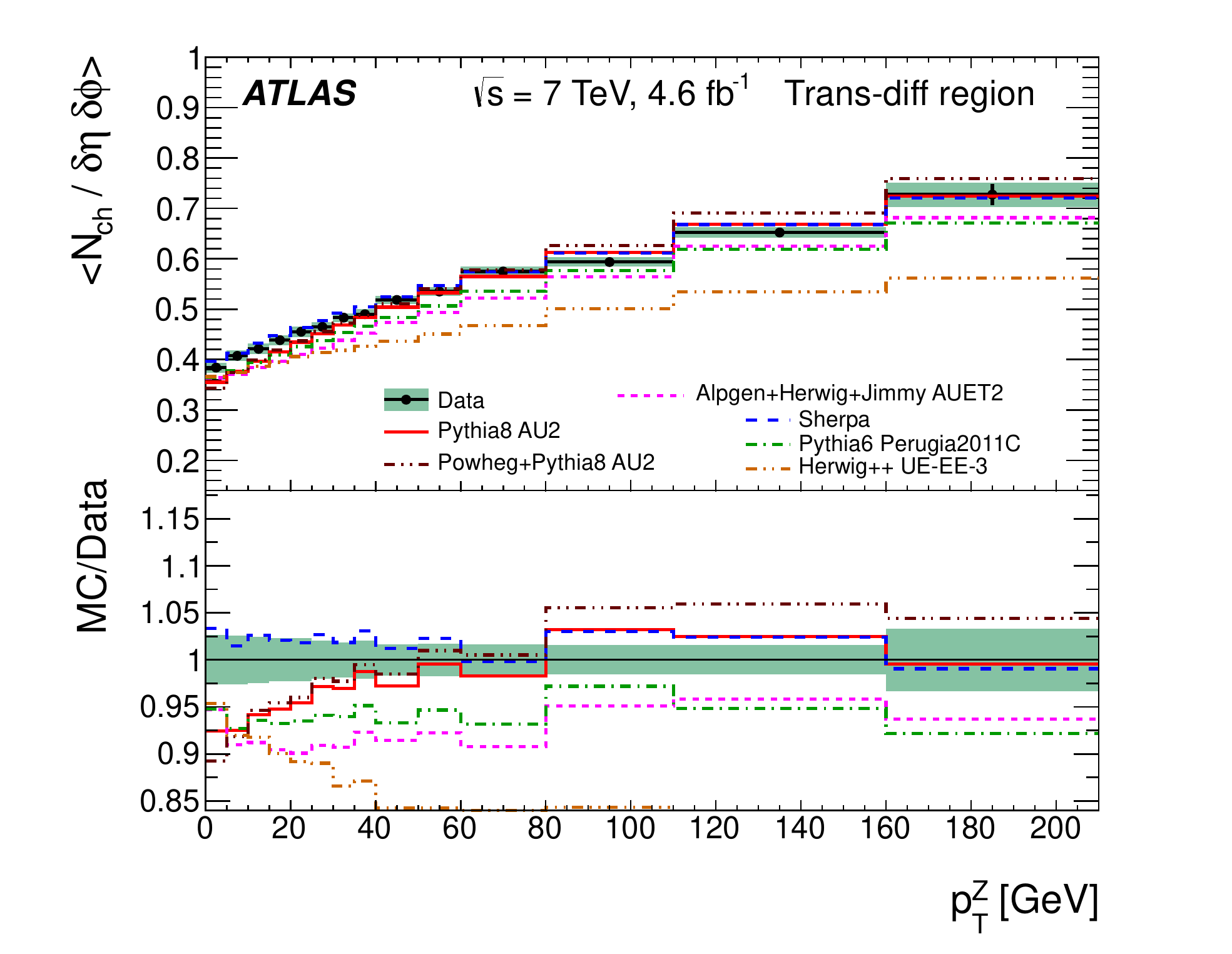}{fig:pro-dif-nchg}
   \caption{Comparison of data and MC predictions for charged particle scalar \ptsum density average values, \dpTsumdetadphi (a),
  and multiplicity average values, \dNchgdetadphi (b) as a function of 
  \Z transverse momentum, \ZpT ,
  in the trans-diff region.
  The shaded bands represent the combined statistical and systematic uncertainties, while
   the error bars show the statistical uncertainties.}
  \label{fig:pro-diff}
\end{figure*}

The \dpTsumdetadphi and \dNchgdetadphi distributions as functions of \ZpT\  
in the trans-diff region are 
compared with the MC model predictions in~\FigRef{fig:pro-diff}.
While all MC models, except for \herwigpp predict the multiplicity fairly well,
only \sherpa and \alpgen predict the \ptsum average values well in certain ranges.
The better modelling of this region by MC models with additional jets
coming from matrix element rather than from parton shower again confirms
that the trans-diff region is most sensitive to the additional radiated jets.

\begin{figure*}[tbp]
  \centering
  \vspace{0.5cm}
   \resultsimg{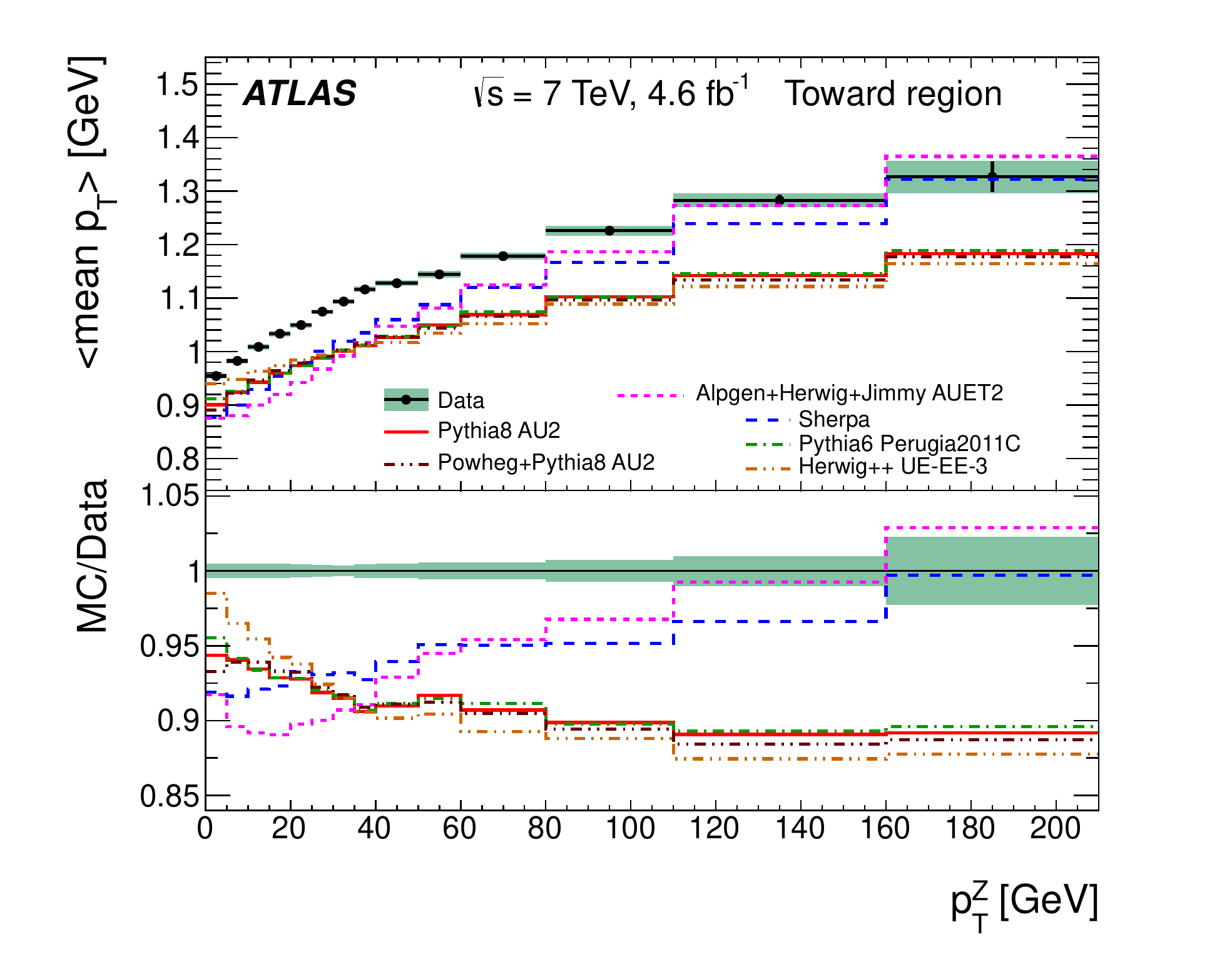}{fig:pro-tow-mpt}\\
   \resultsimg{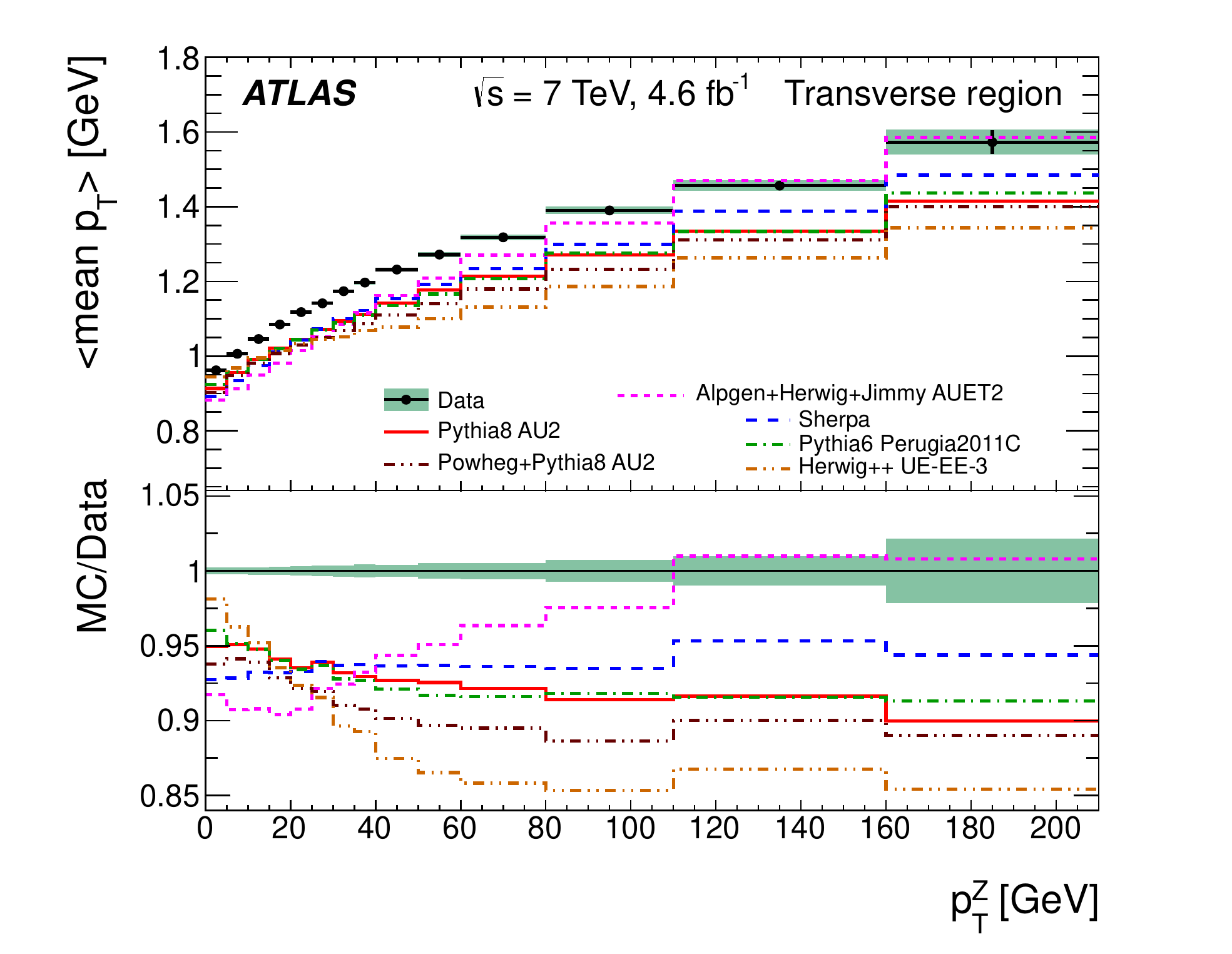}{fig:pro-trv-mpt}
   \caption{Comparison of data and MC predictions for charged particle mean \pT\ as a function of 
  \Z transverse momentum, \ZpT ,
    in the toward (a) and transverse (b) regions.
   The bottom panels in each plot show the ratio of MC predictions to data.
   The shaded bands represent the combined statistical and systematic uncertainties, while
   the error bars show the statistical uncertainties.}
  \label{fig:meanptDMAB}
\end{figure*}

\begin{figure*}[tbp]
  \centering
  \vspace{0.5cm}
   \resultsimg{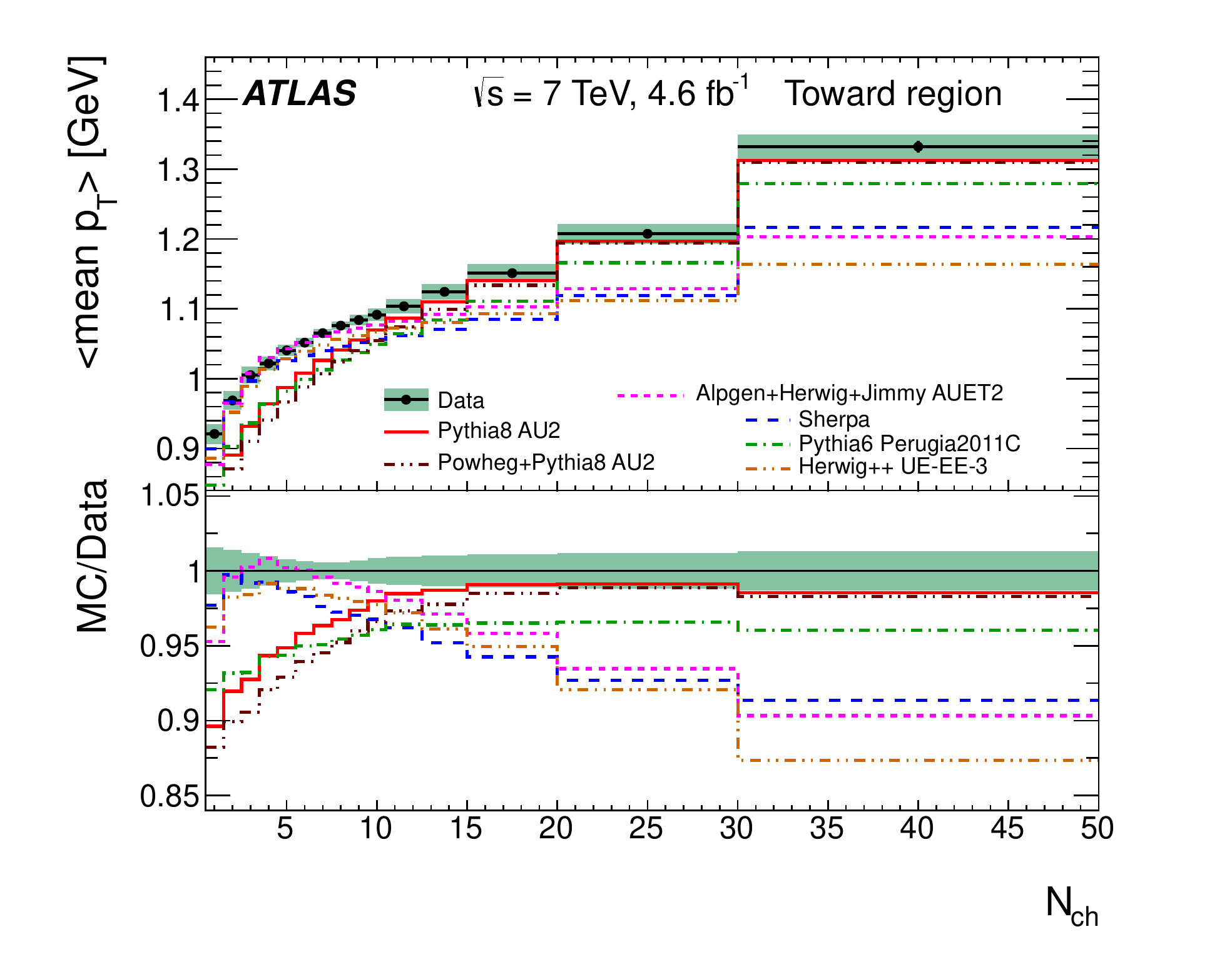}{fig:pro-tow-corr}\\
   \resultsimg{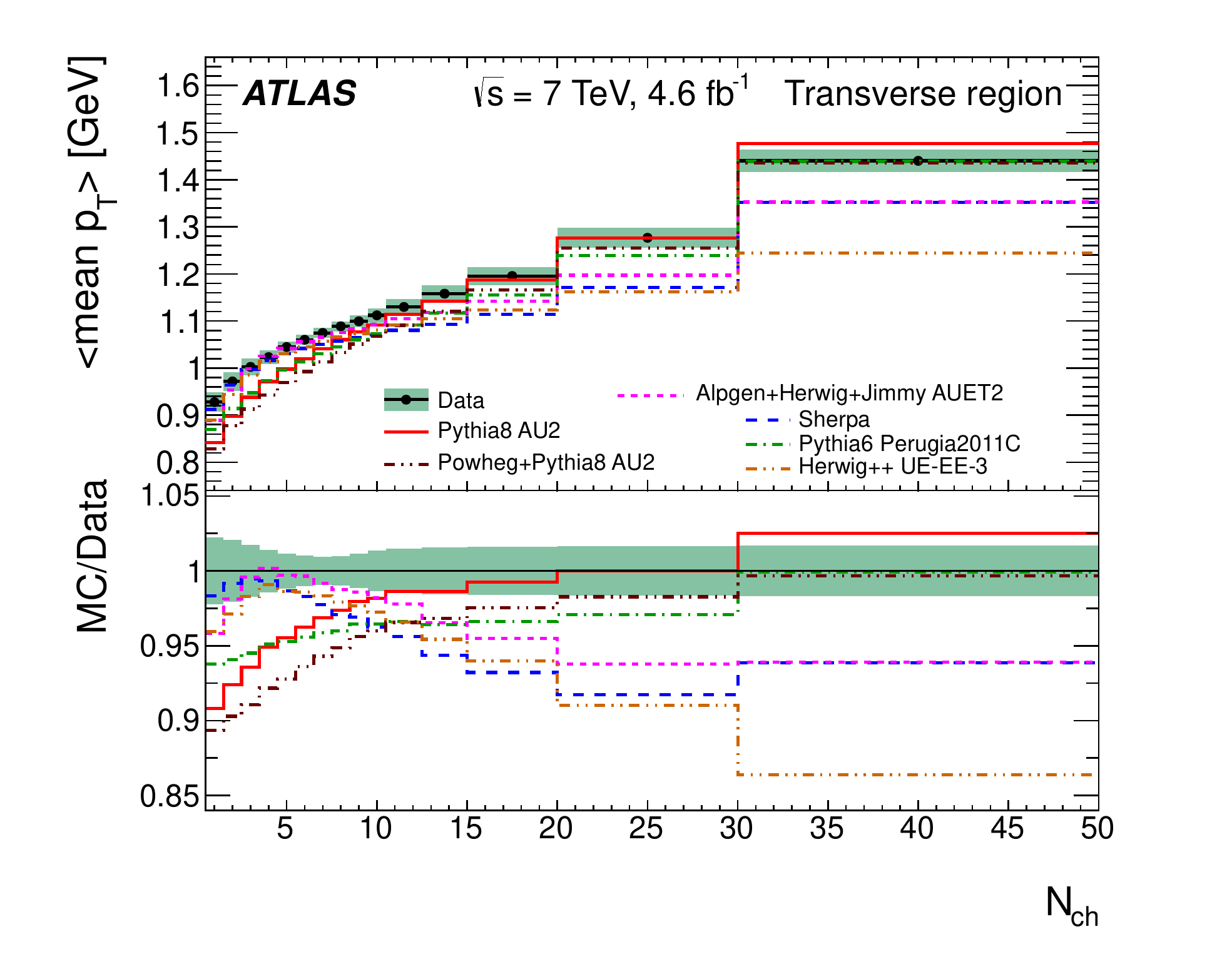}{fig:pro-trv-corr}
   \caption{Comparison of data and MC predictions for charged particle mean \pT\ as a function of 
   charged particle multiplicity, \Nchg ,
   in the toward (a) and transverse (b) regions.
   The bottom panel in each plot shows the ratio of MC predictions to data.
   The shaded bands represent the combined statistical and systematic uncertainties, while
   the error bars show the statistical uncertainties.}
  \label{fig:meanptDMCD}
\end{figure*}

The difficulty of describing the \dpTsumdetadphi and \\
 \dNchgdetadphi average values simultaneously in MC models
is reflected in the comparison of data and MC model predictions for \ptmean in~\FigRef{fig:meanptDMAB}.
The \ptmean as a function of \ZpT\ 
is reasonably described by \alpgen and \sherpa for high \ZpT , while all the other models predict softer spectra.
The correlation of \ptmean with \Nchg , shown in~\FigRef{fig:meanptDMCD}, follows the pattern established by previous
experiments, with a slow increase in mean \pT\ with increasing \Nchg .
 This observable is sensitive to the colour reconnection model in the MC generators.
No MC model is able to predict the full shape in either region. Overall the \pythiaeight
prediction is the closest to the data, followed by \pythiasix and \powheg, although for 
$\Nchg < 5$, all three have much softer distributions than the data. The other models do well in this
low \Nchg region, but are then much lower than the data for high \Nchg .

From all the distributions considered, 
it can be inferred that the jets radiated from the hard scatter will affect the underlying event 
observables and therefore these must be properly reproduced in order to obtain an accurate MC description of the UE.
The UE region least affected by the presence of extra jets is the trans-min region.

%%%%%%%%%%%%%%%%%%%%%%%%%%%%%%%%%%%%%%%%
\subsection{Comparison with other ATLAS measurements}
\label{sec:jet}

The results from this analysis are compared to the results obtained 
when the leading object is either a charged particle~\cite{Aad:2010fh}
or a hadronic jet~\cite{Aad:2014hia}. The underlying event analysis with
a leading charged particle was performed with the early 2010 data,
while the analysis using events with jets utilises the full 2010 dataset.

\begin{figure*}[tbp]
  \centering
  \vspace{0.5cm}
   \resultsimg{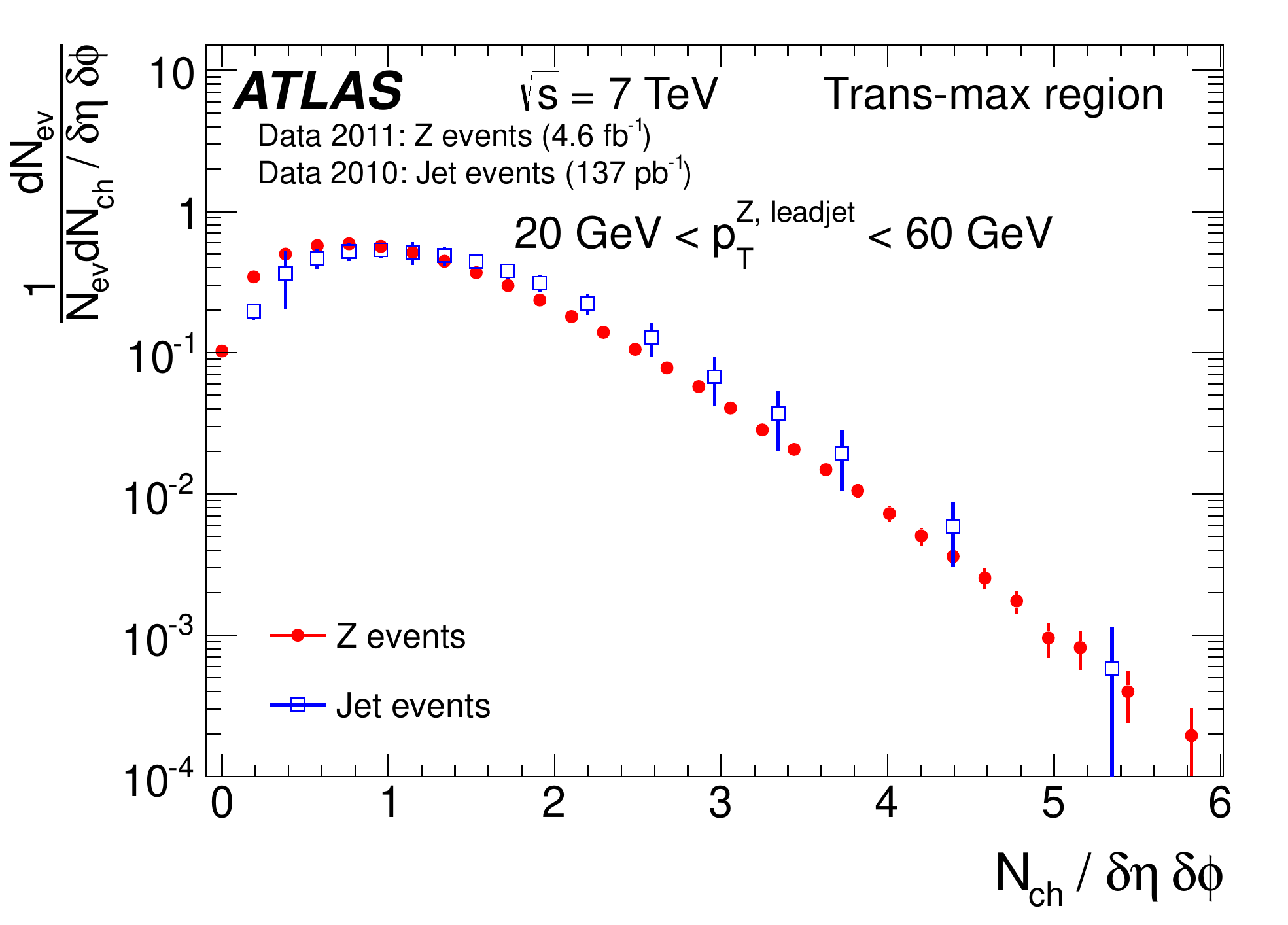}{fig:cmp-mnchg}\\
   \resultsimg{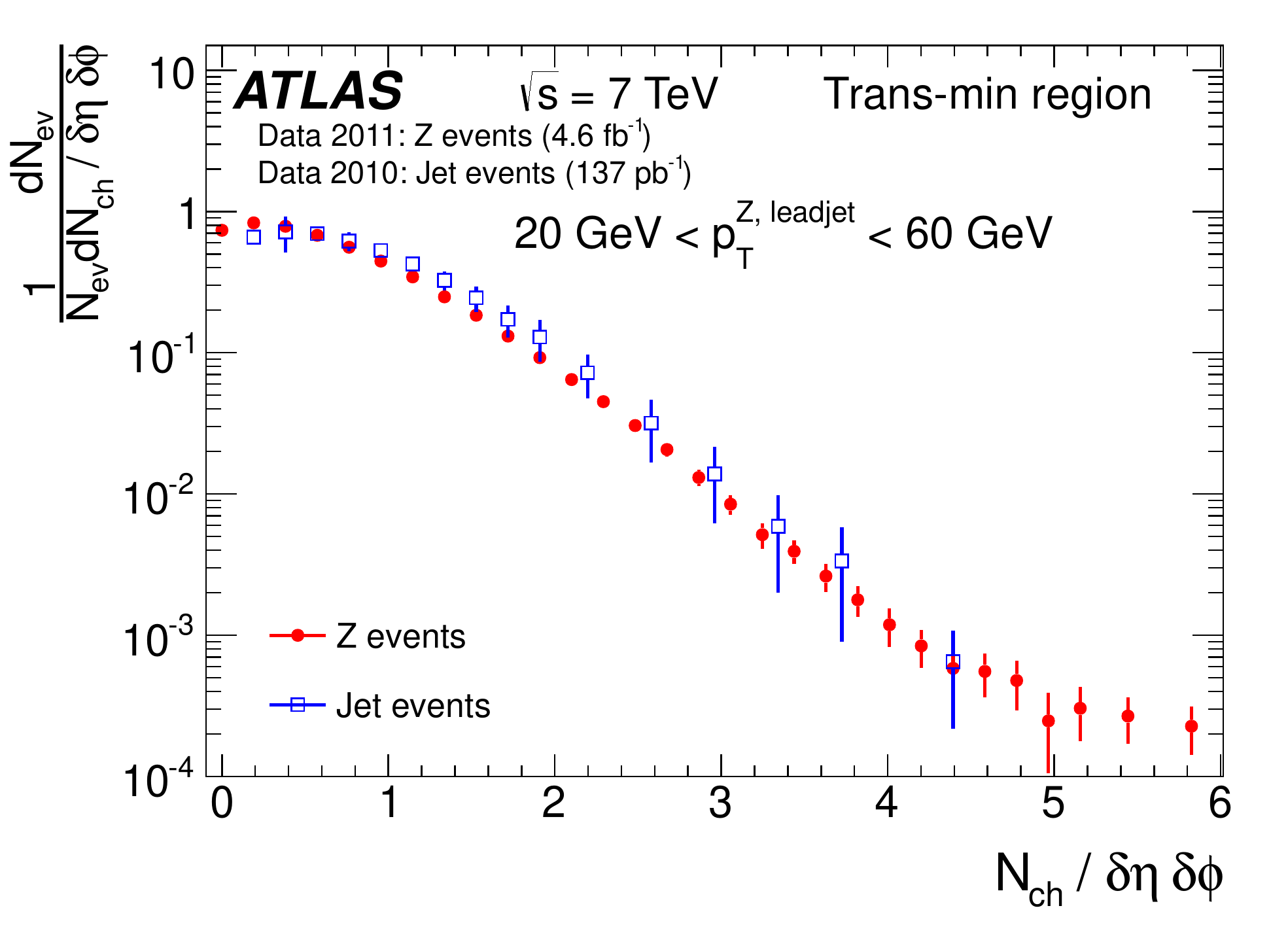}{fig:cmp-hnchg}\\
   \caption{Distributions of charged particle multiplicity density, \Nchgdetadphi ,
   compared between jet and \Z events, respectively in \Z transverse momentum, \ZpT 
   and leading jet transverse momentum , $p_{\textrm{T}}^{\textrm{leadjet}}$ interval between $20-60$~GeV, 
    in the trans-max (a) and trans-min (b) regions. 
  The error bars in each case show the combined statistical and systematic uncertainties.}
  \label{fig:UEcomp_1DAB}
\end{figure*}

\begin{figure*}[tbp]
  \centering
  \vspace{0.5cm}
   \resultsimg{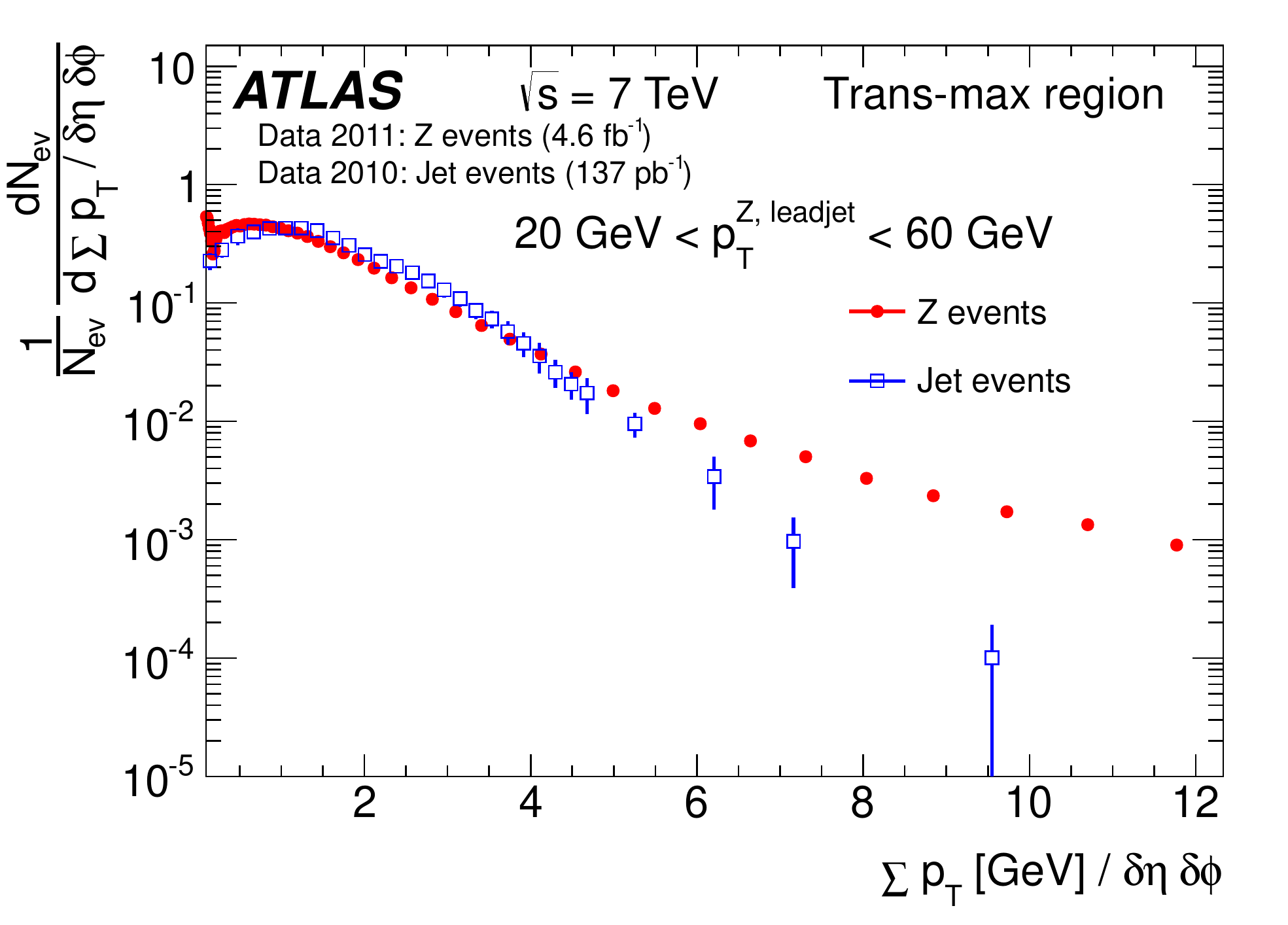}{fig:cmp-mptsum}
   \resultsimg{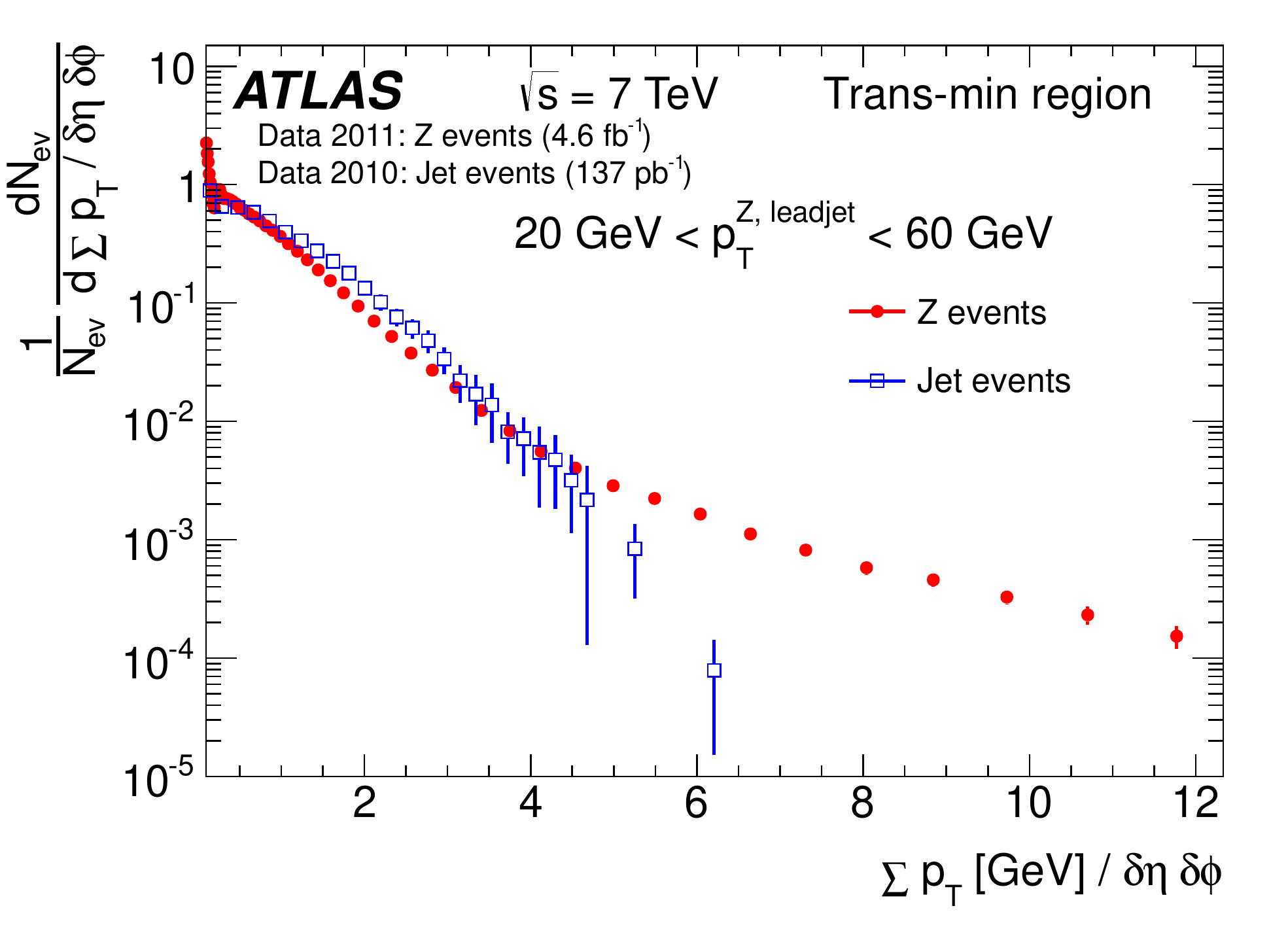}{fig:cmp-hptsum}
   \caption{Distributions of charged particle scalar \pT\ sum  density, \pTsumdetadphi ,
    compared between jet and \Z events, respectively in \Z transverse momentum, \ZpT 
   and leading jet transverse momentum, $p_{\textrm{T}}^{\textrm{leadjet}}$ interval between $20-60$~GeV, 
    in the trans-max (a) and trans-min (b) regions. 
  The error bars in each case show the combined statistical and systematic uncertainties.}
  \label{fig:UEcomp_1DCD}
\end{figure*}

The differential \Nchgdetadphi and \pTsumdetadphi distributions
for leading jet and \Z events are compared in~\FigRef{fig:UEcomp_1DAB} and~\FigRef{fig:UEcomp_1DCD}
for the trans-max and trans-min regions. 
While the \Nchgdetadphi distributions are similar, a clear difference is observed in the high tails of the \pTsumdetadphi distribution,
which are more populated in \Z events than in jet events.
This difference was traced to the definition of the leading object. 
In the case of jets, the accompanying activity can never contain jets with a \pT\ higher than that of the leading jet,
whereas there is no such restriction for \Z events. 
As a test, the average \ptsum was determined for \Z events after rejecting all events in which
at the detector level there was a jet with \pT\ higher than the \ZpT , with jets selected
as in~\cite{Aad:2014hia}. The average was found to be about $20-30\%$ lower than for the standard selection,  and 
the average values in jet and \Z events are in close agreement in this case.

\begin{figure*}[tbp]
  \centering
  \vspace{0.5cm}
   \resultsimg{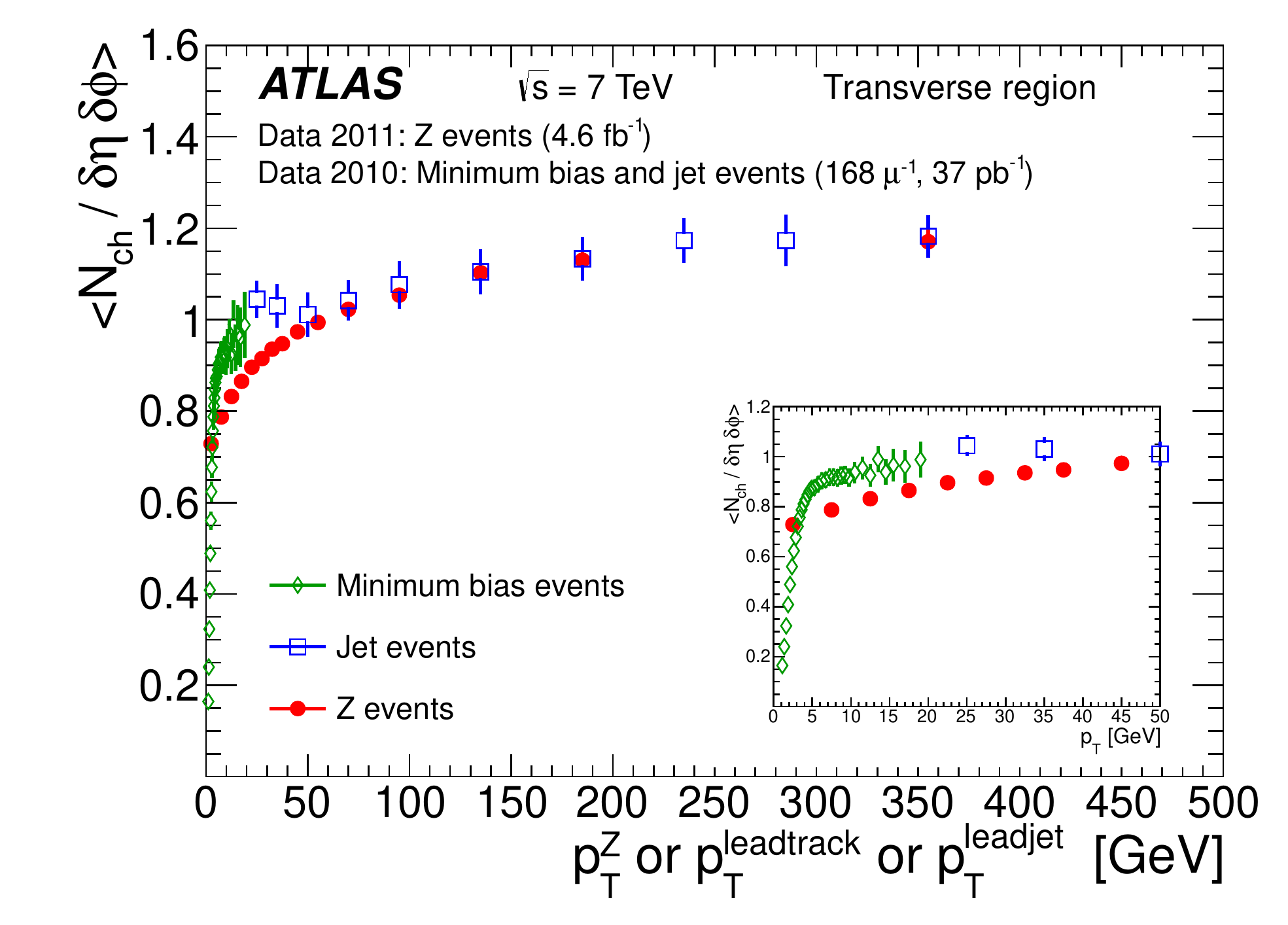}{fig:cmp-nchg}\\
   \resultsimg{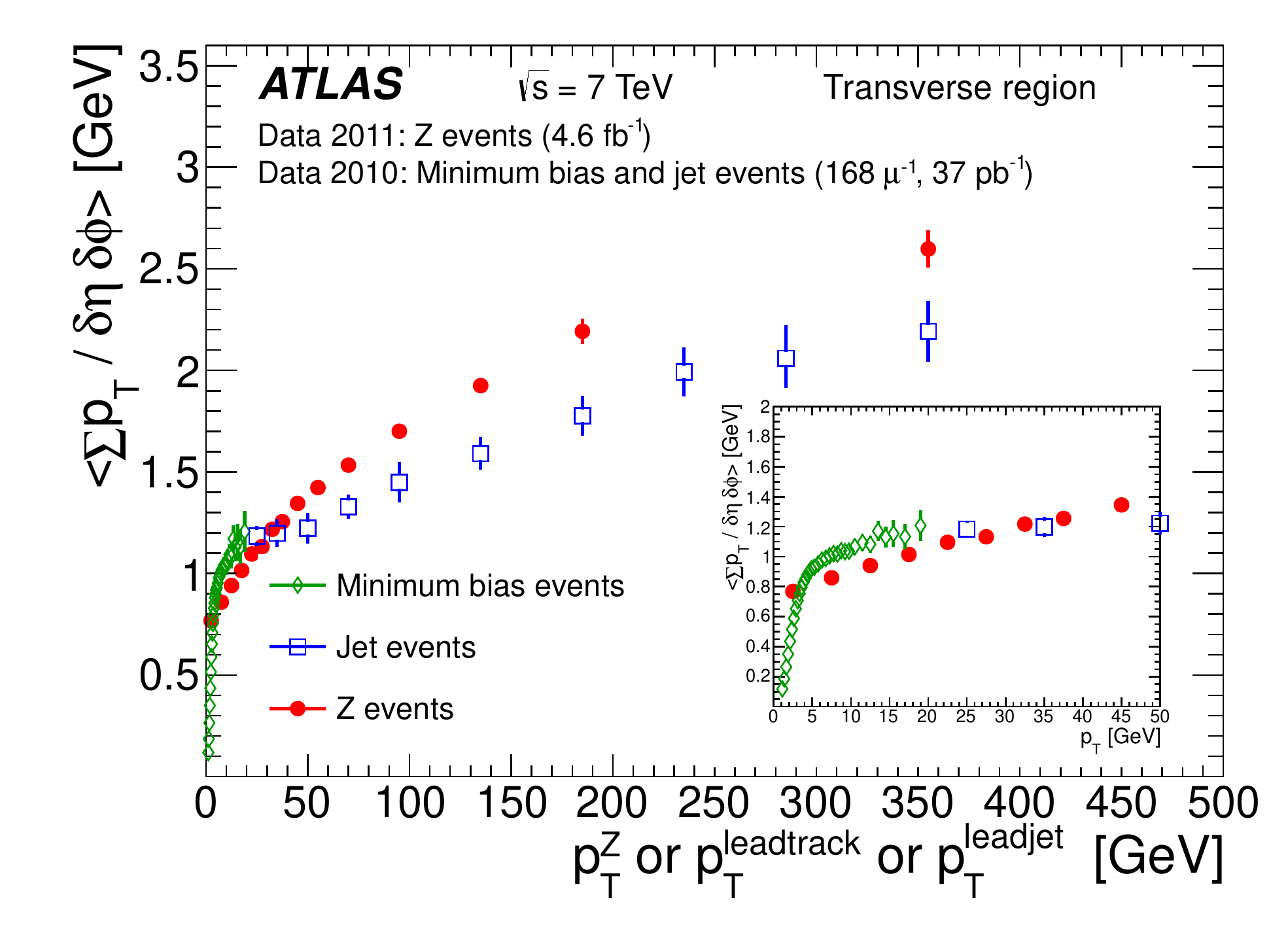}{fig:cmp-ptsum}
   \caption{Charged particle multiplicity average values, \dNchgdetadphi (a), and scalar \ptsum density average values, \dpTsumdetadphi 
  (b), compared between leading charged particle (\textit{minimum bias}), leading jet and \Z events, respectively as
  functions of leading track transverse momentum, $p_{\textrm{T}}^{\textrm{lead}}$,
  leading jet transverse momentum, $p_{\textrm{T}}^{\textrm{leadjet}}$ and 
  \Z transverse momentum, \ZpT , in the transverse region.
  The error bars in each case
  show the combined statistical and systematic uncertainties. The inserts show
  the region of transition between the leading charged particle and leading jet results in more detail.
  }
  \label{fig:UEcompAB}
\end{figure*}

\begin{figure*}[tbp]
  \centering
  \vspace{0.5cm}
   \resultsimg{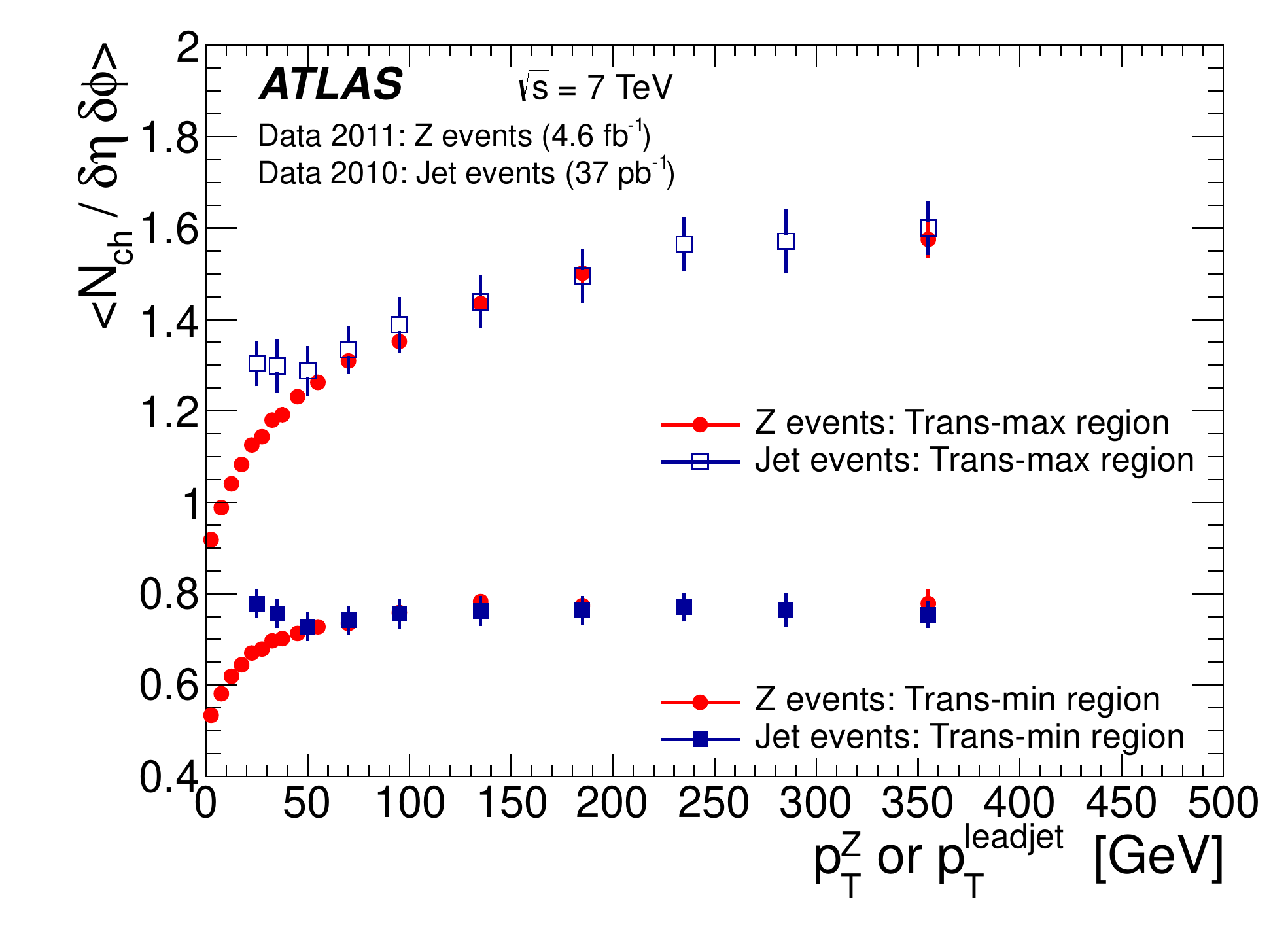}{fig:cmp-maxmin-nchg}\\
   \resultsimg{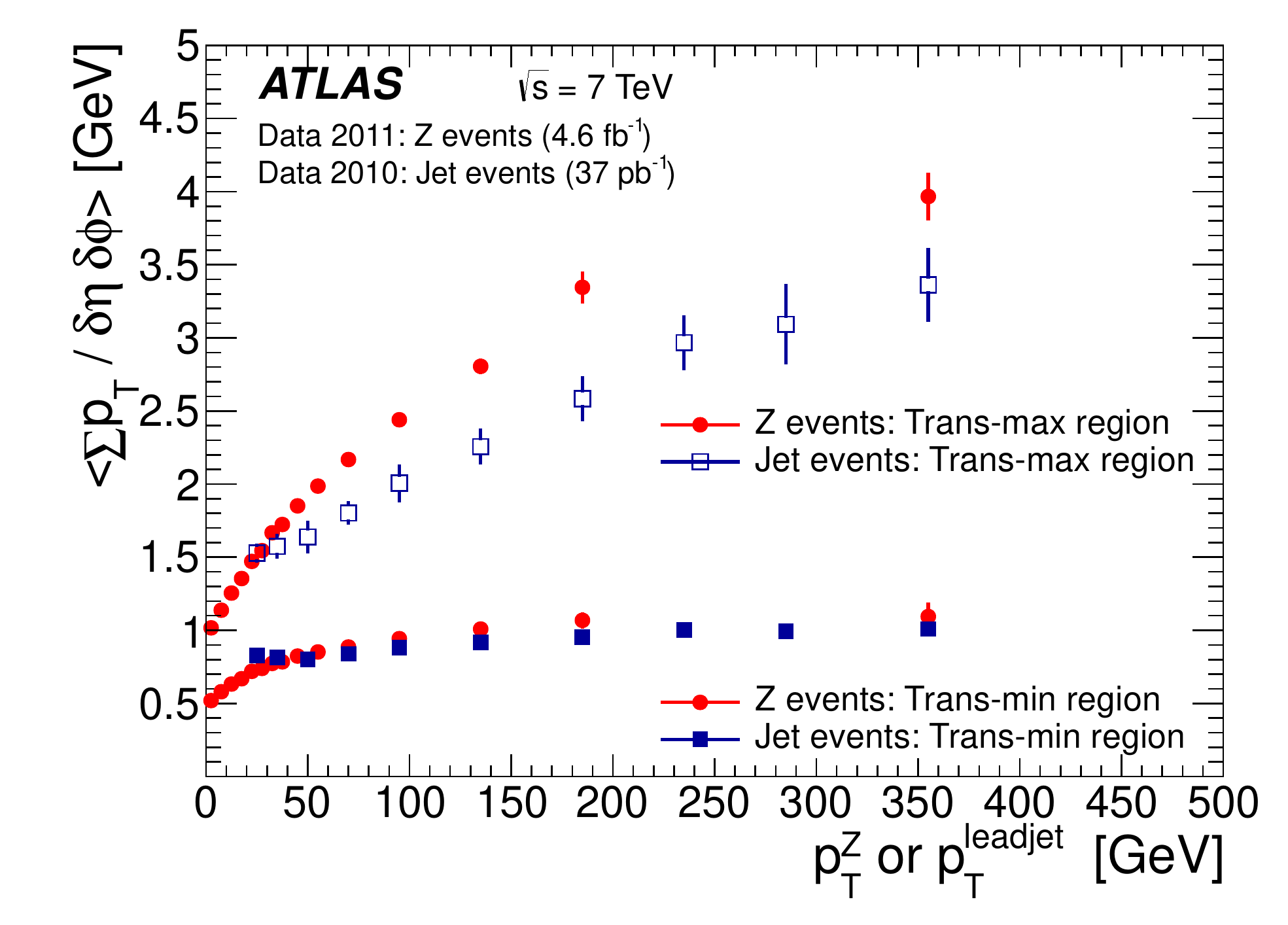}{fig:cmp-maxmin-ptsum}
   \caption{Charged particle multiplicity average values, \dNchgdetadphi (a), and scalar \ptsum density average values, \dpTsumdetadphi 
  (b), compared between leading jet and \Z events, respectively as
  functions of leading jet transverse momentum, $p_{\textrm{T}}^{\textrm{leadjet}}$ and 
  \Z transverse momentum, \ZpT , 
   in the transverse, trans-max and trans-min-regions.
  The error bars in each case
  show the combined statistical and systematic uncertainties.
  }
  \label{fig:UEcompCD}
\end{figure*}

The hard scales used for the analyses are different
and the choice of the main observable used to assess 
the evolution of the underlying event reflects this to a certain extent in the figures.
Nevertheless, certain common 
qualitative features can be observed by comparing \dpTsumdetadphi and \dNchgdetadphi as functions of 
the leading object \pT\ in the transverse region, and also separated into the trans-max/min regions
as shown in~\FigRef{fig:UEcompAB} and~\FigRef{fig:UEcompCD}.
The measurements with a leading jet are complementary to the measurements with a leading track,
and a smooth continuation at $20$~GeV is observed (in~\FigRef{fig:UEcompAB}), corresponding to the
lowest jet \pT\ for which the jet measurement 
could be performed and the highest leading track momentum included in the leading track analysis.
Where the \pT\ of the leading object is less than $50$~GeV, a large difference is observed both
for the \Nchg and \ptsum average values between the jet and \Z measurements in~\FigRef{fig:UEcompAB}; the 
increase of the associated activity as a function of the hard scale \pT\ is very different in
track/jets events from the \Z\ events.

Although the \Nchg density is 
similar in the underlying event associated with a jet to that with a \Z for higher values of the
hard scale ($\geq 50$ GeV), there are residual differences in the average \ptsum densities.
The activity in events with a \Z is systematically higher than that in events with jets.
From the behaviour of the underlying event properties in the trans-max/min regions in~\FigRef{fig:UEcompCD},
this difference originates mostly from the trans-max region, due to selection bias
discussed previously in this section.
The trans-min region is very similar between the two measurements, despite the different hard scales, 
indicating again that this region is least sensitive to the hard interaction and most sensitive to the MPI component.

%%%%%%%%%%%%%%%%%%%%%%%%%%%%%%%%%%%%%%%%
%% Fig:  sumpt slice

%% Fig:  sumpt slice

%%%%%%%%%%%%%%%%%%%%%%%%%%%%%%%%%%%%%%%%%
\FloatBarrier

\begin{figure*}[tbp]
  \centering
  \vspace{0.5cm}
   \resultsimg{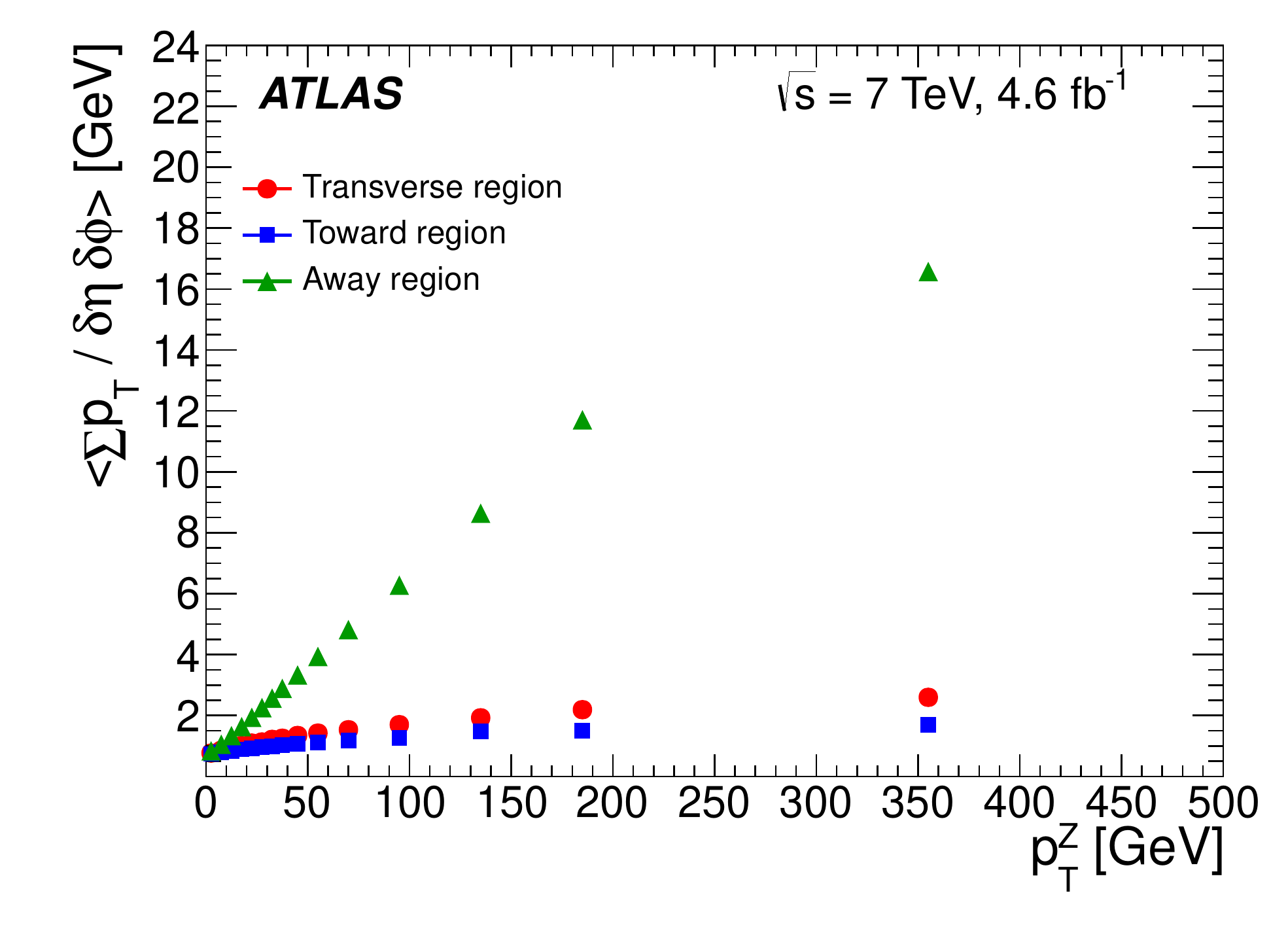}{fig:pro-ptsum1}\\
   \resultsimg{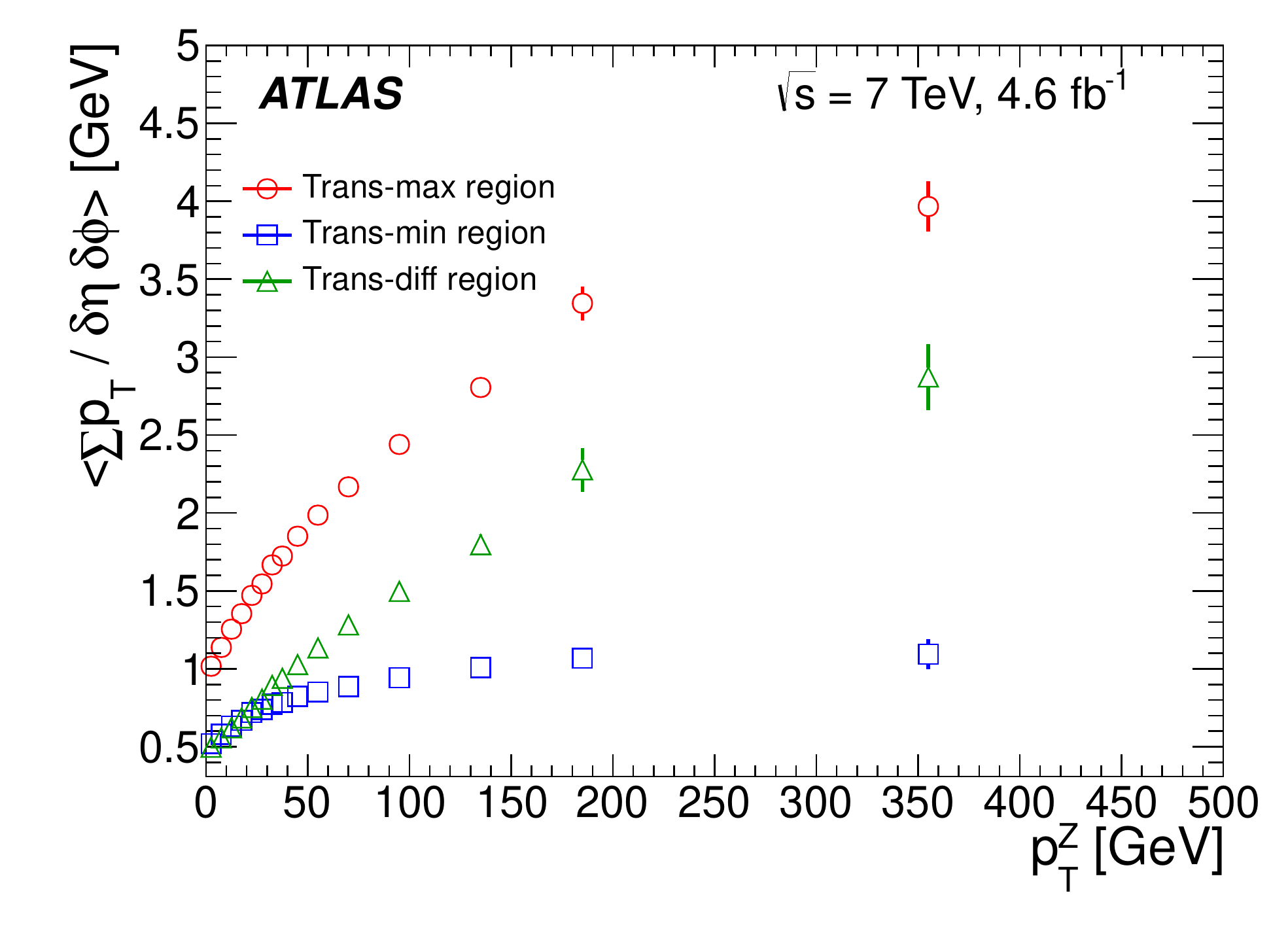}{fig:pro-ptsum2}
   \caption{The average values of charged particle scalar \ptsum density, \dpTsumdetadphi , as a function of \Z transverse momentum, \ZpT ,
  in the transverse, toward and away regions (a), and in the trans-max, trans-min and trans-diff regions (b).
  The results are plotted at the center of each \ZpT\ bin.
  The error bars depict combined statistical and systematic uncertainties.}
  \label{fig:sumptD}
\end{figure*}

%%%%%%%%%%%%%%%%%%%%%%%%%%%%%%%%%%%%%%%
\FloatBarrier

\FloatBarrier

%%%%%%%%%%%%%%%%%%%%%%%%%%%%%%%%%%%%
%         Summary/Conclusions
%%%%%%%%%%%%%%%%%%%%%%%%%%%%%%%%%%%%
\section{Conclusion}
\label{sec:concl}

Measurements sensitive to the underlying event have been presented, 
using an inclusive sample of \Z decays, obtained from a data set collected in 
proton-proton collisions at the LHC corresponding to an integrated luminosity of $4.6\;\invfb$.
The transverse and toward regions with respect to the reconstructed \Z are most sensitive to the underlying event.
The transverse region was further subdivided
into trans-max and trans-min regions on an event-by-event
basis depending on which one had a higher \ptsum; this subdivision
provides additional power to discriminate between the different processeses contributing to the underlying event models.

The results show the presence of a hard component in the \pT distribution of particles, 
presumably originating from extra jet activity associated  with the 
\Z production. It is observed in all the investigated regions, with the trans-min region least affected by it. 
The average underlying event activity increases with
\ZpT , until it reaches a plateau, which is again most prominent in the trans-min region.
The results have been compared to a number of MC models, using several tunes of
commonly used underlying event models.
MC model predictions qualitatively describe the data well, but with
some significant discrepancies, providing precise information sensitive to the 
choices of parameters used in the various underlying-event models.
Careful tuning of these parameters in the future may improve the description 
of the data by the different models in future LHC measurements and studies.

The study of such variables in \Z events provides a probe of the underlying event which is  complementary to that from purely
hadronic events. 
A comparison between them shows similar underlying event activity for the trans-min region.

%%%%%%%%%%%%%%%%%%%%%%%%%%%%%%%%%%%%%%%%%%%%%%%%%%%%%%%%%%%%%%%%%%%%%%%%%%%
%       ATLAS ACKNOWLEDGEMENTS
%%%%%%%%%%%%%%%%%%%%%%%%%%%%%%%%%%%%%%%%%%%%%%%%%%%%%%%%%%%%%%%%%%%%%%%%%%%

\section*{Acknowledgements}

% Acknowledgements for papers with collision data
% Version 14-Oct-2014

We thank CERN for the very successful operation of the LHC, as well as the
support staff from our institutions without whom ATLAS could not be
operated efficiently.

We acknowledge the support of ANPCyT, Argentina; YerPhI, Armenia; ARC,
Australia; BMWFW and FWF, Austria; ANAS, Azerbaijan; SSTC, Belarus; CNPq and FAPESP,
Brazil; NSERC, NRC and CFI, Canada; CERN; CONICYT, Chile; CAS, MOST and NSFC,
China; COLCIENCIAS, Colombia; MSMT CR, MPO CR and VSC CR, Czech Republic;
DNRF, DNSRC and Lundbeck Foundation, Denmark; EPLANET, ERC and NSRF, European Union;
IN2P3-CNRS, CEA-DSM/IRFU, France; GNSF, Georgia; BMBF, DFG, HGF, MPG and AvH
Foundation, Germany; GSRT and NSRF, Greece; ISF, MINERVA, GIF, I-CORE and Benoziyo Center,
Israel; INFN, Italy; MEXT and JSPS, Japan; CNRST, Morocco; FOM and NWO,
Netherlands; BRF and RCN, Norway; MNiSW and NCN, Poland; GRICES and FCT, Portugal; MNE/IFA, Romania; MES of Russia and ROSATOM, Russian Federation; JINR; MSTD,
Serbia; MSSR, Slovakia; ARRS and MIZ\v{S}, Slovenia; DST/NRF, South Africa;
MINECO, Spain; SRC and Wallenberg Foundation, Sweden; SER, SNSF and Cantons of
Bern and Geneva, Switzerland; NSC, Taiwan; TAEK, Turkey; STFC, the Royal
Society and Leverhulme Trust, United Kingdom; DOE and NSF, United States of
America.

The crucial computing support from all WLCG partners is acknowledged
gratefully, in particular from CERN and the ATLAS Tier-1 facilities at
TRIUMF (Canada), NDGF (Denmark, Norway, Sweden), CC-IN2P3 (France),
KIT/GridKA (Germany), INFN-CNAF (Italy), NL-T1 (Netherlands), PIC (Spain),
ASGC (Taiwan), RAL (UK) and BNL (USA) and in the Tier-2 facilities
worldwide.

\FloatBarrier

%%%%%%%%%%%%%%%%%%%%%%%%%%%%%%%%%%%%%%%%%%%%%%%%%%%%%%%%%%%%%%%%%%%%%%%%%%%%%%%
% Bibliography
%%%%%%%%%%%%%%%%%%%%%%%%%%%%%%%%%%%%%%%%%%%%%%%%%%%%%%%%%%%%%%%%%%%%%%%%%%%%%%
%
% Style file to use with mcite.
% Use atlasstyle with just cite.
% \bibliographystyle{epjc}
\bibliographystyle{atlasnote}
\bibliography{refs}

%
%%%%%%%%%%%%%%%%%%%%%%%%%%%%%%%%%%%%%%%%%%%%%%%%%%%%%%%%%%%%%%%%%%%%%%%%%%%%%%
% Author List
%%%%%%%%%%%%%%%%%%%%%%%%%%%%%%%%%%%%%%%%%%%%%%%%%%%%%%%%%%%%%%%%%%%%%%%%%%%%%%
%

\onecolumn
\clearpage
% ATLAS Collaboration author list
% Data extracted on 21-Nov-2014 for paper reference STDM-2011-42
%\documentclass[11pt]{article}
%\usepackage{a4wide}\begin{document}
\begin{flushleft}
{\Large The ATLAS Collaboration}

\bigskip

G.~Aad$^{\rm 84}$,
B.~Abbott$^{\rm 112}$,
J.~Abdallah$^{\rm 152}$,
S.~Abdel~Khalek$^{\rm 116}$,
O.~Abdinov$^{\rm 11}$,
R.~Aben$^{\rm 106}$,
B.~Abi$^{\rm 113}$,
M.~Abolins$^{\rm 89}$,
O.S.~AbouZeid$^{\rm 159}$,
H.~Abramowicz$^{\rm 154}$
% ...
\bigskip
\\
% ...
$^{11}$ Institute of Physics, Azerbaijan Academy of Sciences, Baku, Azerbaijan\\
% ...
$^{84}$ CPPM, Aix-Marseille Universit{\'e} and CNRS/IN2P3, Marseille, France\\
% ...
$^{89}$ Department of Physics and Astronomy, Michigan State University, East Lansing MI, United States of America\\
% ...
$^{106}$ Nikhef National Institute for Subatomic Physics and University of Amsterdam, Amsterdam, Netherlands\\
% ...
$^{112}$ Homer L. Dodge Department of Physics and Astronomy, University of Oklahoma, Norman OK, United States of America\\
$^{113}$ Department of Physics, Oklahoma State University, Stillwater OK, United States of America\\
% ...
$^{116}$ LAL, Universit{\'e} Paris-Sud and CNRS/IN2P3, Orsay, France\\
% ...
$^{152}$ Institute of Physics, Academia Sinica, Taipei, Taiwan\\
% ...
$^{154}$ Raymond and Beverly Sackler School of Physics and Astronomy, Tel Aviv University, Tel Aviv, Israel\\
% ...
$^{159}$ Department of Physics, University of Toronto, Toronto ON, Canada\\
% ...
\end{flushleft}

%\end{document}
% Created with ./xml2latex.py

\end{document}